\documentclass[twocolumn]{aastex631}

\usepackage[nolist]{acronym}
\usepackage{amsmath}
\usepackage{stmaryrd} % for mathbb brackets

\usepackage{CJKutf8}            % Chinese name

\usepackage{multirow}

\usepackage{slashbox}

\usepackage{diagbox}

\begin{document}

\title{Machine Learning-based Search of High-redshift Quasars}

\author{Guangping Ye \begin{CJK*}{UTF8}{gbsn} (叶广平) \end{CJK*}}
\affiliation{Department of Astronomy, Huazhong University of Science and Technology, Wuhan, Hubei 430074, China}
\email{guangping@hust.edu.cn}

\author{Huanian Zhang \begin{CJK*}{UTF8}{gbsn} (张华年) \end{CJK*}}
\affiliation{Department of Astronomy, Huazhong University of Science and Technology, Wuhan, Hubei 430074, China}
\affiliation{Steward Observatory, University of Arizona, Tucson, AZ 85719, USA}
\email{huanian@hust.edu.cn}

\author{Qingwen Wu \begin{CJK*}{UTF8}{gbsn} (吴庆文) \end{CJK*}}
\affiliation{Department of Astronomy, Huazhong University of Science and Technology, Wuhan, Hubei 430074, China}

\begin{abstract}

We present a machine learning search for high-redshift ($5.0 < z < 6.5$) quasars using the combined photometric data from the DESI Imaging Legacy Surveys and the WISE survey. We explore the imputation of missing values for high-redshift quasars, discuss the feature selections, compare different machine learning algorithms, and investigate the selections of class ensemble for the training sample, then we find that the random forest model is very effective in separating the high-redshift quasars from various contaminators. The 11-class random forest model can achieve a precision of 96.43\% and a recall of 91.53\% for high-redshift quasars for the test set. We demonstrate that the completeness of the high-redshift quasars can reach as high as 82.20\%. The final catalog consists of 216,949 high-redshift quasar candidates with 476 high probable ones in the entire Legacy Surveys DR9 footprint, and we make the catalog publicly available. Using MUSE and DESI-EDR public spectra, we find that 14 true high-redshift quasars (11 in the training sample) out of 21 candidates are correctly identified for MUSE, and 20 true high-redshift quasars (11 in the training sample) out of 21 candidates are correctly identified for DESI-EDR. Additionally, we estimate photometric redshift for the high-redshift quasar candidates using random forest regression model with a high precision. %We discover that the known high-redshift quasars may exhibit two subgroups in the color space. This finding could potentially be related to the bimodal distribution of known high-redshift quasars in terms of their number density as a function of redshift. 

\end{abstract}


\section{Introduction} \label{sec:intro}

Quasars in general are driven by a supermassive black hole (SMBH) at the centre of the host galaxy through a process of accretion, and are the brightest non-transient sources of light in the Universe. SMBH activities are a key ingredient of galaxy formation, and are critical to subsequent galaxy evolution. Quasars at $z>5$ are often referred to as high-redshift quasars since they are at the end of the reionization epoch of the Universe, when the vast majority of the Universe's neutral hydrogen has been reionised \citep{Eilers2018, yang2020measurements, bosman2022hydrogen}. High-redshift quasars provide effective probes for the study of galaxy evolution and cosmology, including the evolution of the intergalactic medium \citep[IGM, ][]{Pentericci2002,fan2006constraining,becker2007evolution,bosman2018new,Eilers2018,yang2020measurements} and the circumgalactic medium \citep[CGM, ][]{Zou2021, Davies2023a, Davies2023b, Zou2024}, the formation of early supermassive black holes, the co-evolution of SMBH and their host galaxies \citep{volonteri2012formation}. 

There exist many challenges in searching and verifying high-redshift quasars. On one hand, the number density of high-redshift quasars is very low in the Universe \citep{Zhang2023, Zhang2024}, resulting to a small number of observable high-redshift quasars, which means that the single-object spectroscopic observations on large-aperture telescope of quasar candidates are needed and expensive; on the other hand, the contamination is overwhelming and could be a few orders of magnitude more than signals \citep{barnett2019euclid}. The major contaminations include cool galactic dwarfs with spectral types of M, L, and T \citep{fan2023quasars}, which are similar to those of high-redshift quasars in the color space constructed from optical and near-infrared broad-band photometry. 

The traditional method for searching the high-redshift quasars is the color-cut selection \citep{warren1987first} based on color drop-out, which is caused
by the Ly$\alpha$ break in quasar spectra due to significant IGM
absorption at the wavelength blueward of the Ly$\alpha$ emission
line at the rest frame of the quasar. This method successfully discovers the majority of the currently known high-redshift quasars \citep{fan2006observational,willott2010canada,mortlock2011luminous,venemans2015identification,wu2015ultraluminous,banados2016pan,jiang2016final,mazzucchelli2017physical,mcgreer2018faint,reed2019three,matsuoka2019subaru,matsuoka2019discovery,wang2016survey,wang2018discovery,wang2019exploring,yang2017discovery,yang2019filling,yang2019exploring,yang2020measurements,Ross&Cross2020,yang2021probing,yang2023desi}. The big advantage of this method is that it leads to well-defined selections that are easily reproducible and can be justified with known physics (e.g., the redshift evolution of the Ly$\alpha$ emission through the broadband filters and the drop-out due to neutral IGM absorption). However, color-cut selections might not make use of all the available information in the high-dimensional color space and the two-dimensional color-color diagram might be misleading or biased. And the strict color cuts might result in missing quasars due to the scattering out of the selection regions based on color-color cuts. In contrast to the color-cut selection method, machine learning-based methods can make full use of the color information and construct correlations in the high-dimensional space. Moreover, the large imaging surveys collect data far more than what can be handled manually, machine learning-based automatic methods can easily process those data and can be easily applied to new data. 

So far, there have been many successful examples of machine learning algorithm-based searches for quasars up to redshift $z \sim 6$ quasars \citep{Richards2009,Bovy2011,Jin2019,Khramtsov2019,Wenzl2021}. A variety of supervised machine-learning algorithms have been successfully applied for quasar selections, such as random forest \citep{schindler2017extremely,nakoneczny2019catalog,yeche2020preliminary,He2022}, Support Vector Machines (SVM), XGBoost, and Artificial Neural Networks \citep{schindler2017extremely,Khramtsov2019b,Nakoneczny2021,Wenzl2021}. \cite{Wenzl2021} successfully applied the random forest algorithm to the high-redshift quasars ($4.8 < z < 6.3$) search using the combined dataset of Pan-STARRS DR1 \citep{Chambers2016} and ALLWISE \citep{Wright2010,Mainzer2011}. Although the $y$ band from Pan-STARRS is essential for high-redshift quasars at $z > 6.5$, it is comparably shallow (3$\pi$ stack 5$\sigma$ depth is $\sim$ 21.4), which might not be efficient in capturing faint high-redshift quasars. Here in this study we will use data from Dark Energy Spectroscopic Instrument (DESI) Legacy Imaging Surveys \citep[hereafter referred to as the Legacy Survey;][]{Dey2019OverviewOT}, which is roughly 2 magnitudes deeper than the 3$\pi$ stack 5$\sigma$ depth of $y$ band of Pan-STARRS. We will demonstrate that the constructed features other than the simple photometry provided in the catalog are critical to a successful machine learning model for high-redshift quasars search. %and also that the traditional 2D color-color in high-redshift quasar search could not reflect the feature importance.

The aim of this paper is to search new high redshift quasars from the Legacy Survey data \citep{Dey2019OverviewOT} combined with {\it Wide-field Infrared Survey Explorer} (WISE) all-sky survey data \citep{Wright2010,Mainzer2011} based on machine learning algorithms. % using photometric information of $5<z<6.5$ quasars that have been spectroscopically certified in The Legacy Surveys.
The structure of the paper is organized as follows. In Sec. \ref{sec:Data} we introduce the catalog data we will use in this study and the training sample for the machine learning algorithm. In Sec. \ref{sec:Random Forest} we briefly discuss the various aspects of the machine learning algorithms. In Sec. \ref{sec:Photometric Redshift Estimation} we estimate the photo-z of the high-redshift quasar candidates. In Sec. \ref{sec:High-z candidates and using grism spectra to do some confirmation} we discuss the procedure to obtain the high-redshift quasar candidates, the spectral verification of the candidates and present the final catalog. Finally we summarize. The magnitudes used in this paper are generated based on the AB system after applying Galactic extinction corrections. We use a $\Lambda$CDM cosmology with $\Omega_\Lambda = 0.7$, $\Omega_m = 0.3$ and $H_0 = 70 \text{ km s}^{-1} \text{Mpc}^{-1}$.

\section{Data} \label{sec:Data}
\subsection{Legacy Survey Data} \label{sec: Legacy Survey Data}

\begin{figure}[ht]
\begin{center}
\includegraphics[width = 0.48 \textwidth]{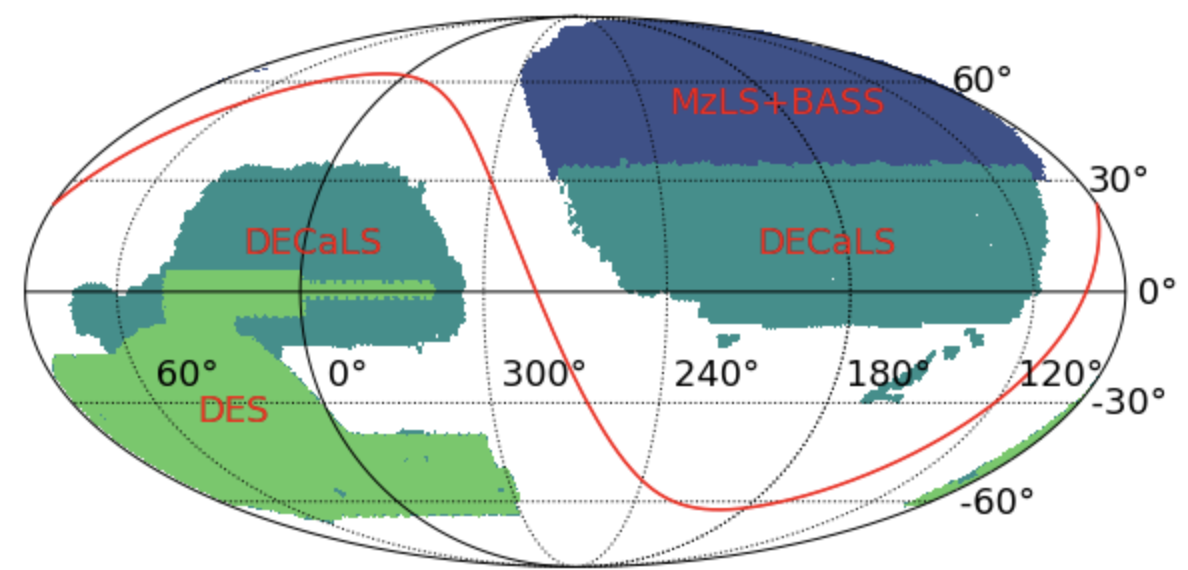}
\end{center}
\caption{ Footprint of the sky covered by the DR9 release of the Legacy Survey \citep{Dey2019OverviewOT}. The Galactic plane is traced by the red curve. }
\label{fig:footprint}
\end{figure}

We use the photometric data from the Data Release 9 (DR9) of the Legacy Survey, which includes observations obtained by the DECam at the CTIO 4 m (DECaLS), an upgraded MOSAIC camera at the KPNO 4 m telescope (MzLS, Mayall z-band Legacy Survey), and the 90Prime camera \citep{Williams2004} at the Steward Observatory 2.3 m telescope (BASS, Beijing-Arizona Sky Survey). Briefly, the Legacy Survey \citep{Dey2019OverviewOT} was initiated to provide targets for the DESI survey drawn from deep, three-band ($g=24.7$, $r=23.9$, and $z=23.0$ AB mag, 5$\sigma$ point-source limits) images, roughly two magnitudes deeper than the Sloan Digital Sky Survey (SDSS) data. The survey covers about 14,000 deg$^2$ of sky visible from the northern hemisphere between declinations approximately bounded by $-$18$^\circ$ and +84$^\circ$. The depth of SDSS \citep{York2000,Abazajian2009,Aihara2011} and Pan-STARRS \citep{Magnier2020}) is insufficient to provide reliable DESI targets. The footprint of DR9 (Figure \ref{fig:footprint}) also includes an additional 6,000 deg$^2$ extending down to $-$68$^\circ$ imaged at the CTIO by the Dark Energy Survey \citep[DES,][]{desCollaboration2016}. Furthermore, the Legacy Survey also includes {\it g, r, z} band images in the North Galactic Cap (NGC) region that overlaps with DES, extending the coverage of the Legacy Survey to over 20,000 square degrees.

\subsection{The Wide-field Infrared Survey Explorer Data} \label{sec: WISE}

In addition to the Legacy Survey data, we also take advantage of the WISE data release, providing infrared photometry of four bands at central wavelength of 3.4, 4.6, 12, and 22 $\mu$m ({\it W1, W2, W3, W4}) over the entire sky \citep{Wright2010,Mainzer2011}. The {\it W1, W2, W3} and {\it W4} data in the Legacy Survey catalog are based on forced photometry in the unWISE stacked images (including all imaging through year 7 of NEOWISE-Reactivation) at the locations of the Legacy Survey optical sources. For our selection process we only focus on the {\it W1} (3.4 $\mu$m) and $W2$ (4.6 $\mu$m) photometry with limiting magnitudes of 19.8, 19.0 in AB magnitude (Vega: 17.1, 15.7), respectively. The conversion between AB and Vega magnitude for {\it W1} and {\it W2} is {\it W1}$_{\rm AB}$ = {\it W1}$_{\rm Vega}$ + 2.699 and {\it W2}$_{\rm AB}$ = {\it W2}$_{\rm Vega}$ + 3.339, respectively.

\subsection{Training sample} 
\label{sec:Training sample}

\begin{figure*}[ht]
\begin{center}
\includegraphics[width= 1. \textwidth]{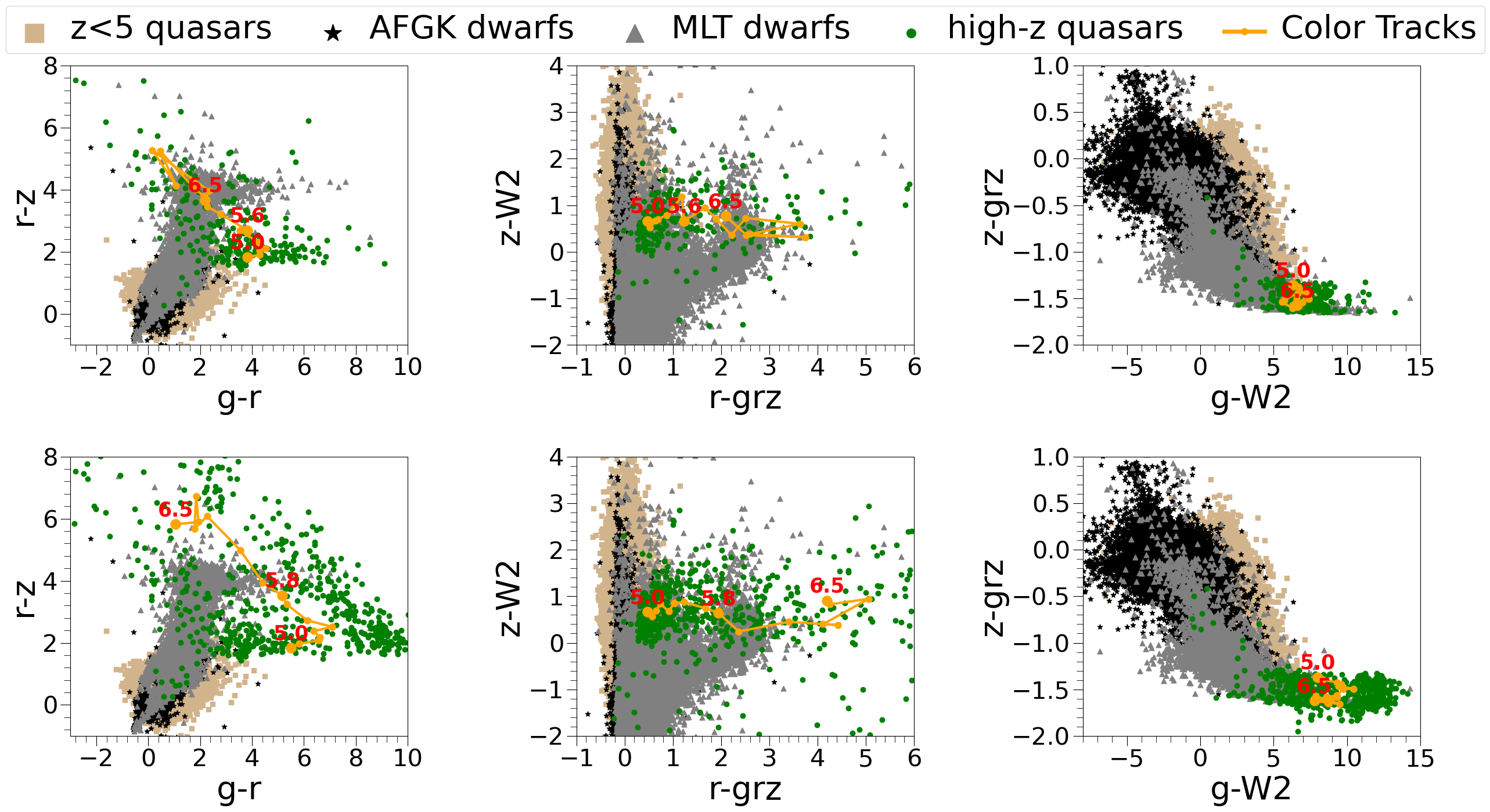}
\end{center}
\caption{ The color-color distribution of the signal and contaminators in the training sample. The top panels are based on the dataset without missing values (only 252 known high-redshift quasars), while the bottom panels are based on all the data in the dataset (including 588 high-redshift quasars), where the missing values of high-redshift quasars are filled in by regression modelling, see Sec. \ref{sec:missingvalue} for detail. The tan square represents quasars with redshifts below 5, the black star stands for AFGK stars, the grey triangle is for MLT dwarfs, and the green dot represents the known high-redshift quasars. The orange dots and lines in each figure are the redshift trajectories of the known high-redshift quasars in the color space.} 
\label{fig:colorcut}
\end{figure*}

We use supervised machine learning algorithms to perform the classification, therefore the training sample is critical to construct a good and reliable model. Any bias in the training sample will propagate to the new data, resulting in highly unreliable outputs. In order to construct a representative and unbiased training sample, we collect the spectra-confirmed signal (high-redshift quasars at $5.0 < z < 6.5$) and all possible contaminators which populate the same or similar color space (eg. $r-z$, $z-W1$, $z-W2$) because of similar absorption features in the optical and near-infrared bands for both the signal and the contaminators. For quasars at $z \sim 5$, Lyman series absorption systems begin to dominate in the {\it r} band and Ly$\alpha$ emission moves to {\it i} or {\it z} band, resulting in redder $r-z$ color. However, as the redshift increases, most $z > 5$ quasars enter the MLT dwarfs locus in $g-r$/$r-z$ color-color diagram (as shown in the upper-left subplot of Figure \ref{fig:colorcut}). At $z \gtrsim 5.7$, Ly$\alpha$ emission line moves into the $z$ band and the $g, r, i$ bands become the drop-out bands, resulting in very red $r-z$ color and comparably red $z-W1$ color. And the infrared photometry $W1$ and $W2$ (constructed colors such as $z-W1$, $z-W2$, $W1-W2$) are useful for all redshifts. The medium-redshift quasars with redshift just below 5, the quasars at lower redshift, the MLT dwarfs, the different types of stars also share similar color-space with the high-redshift quasars, consisting of the major contaminants for the signal of high-redshift quasars.

We then obtain the photometry data from both Legacy Survey ({\it g, r, z} bands) and WISE ({\it W1, W2} bands) for the spectra-confirmed high-redshift quasars ($5.0 < z < 6.5$), and the spectra-confirmed contaminators, which include the MLT brown dwarfs, AFGK stars, medium-redshift quasars ($3.5 < z < 5.0$), low-redshift quasars ($1.5 < z < 3.5$), and very low-redshift quasars ($0 < z < 1.5$). Although there are a number of quasars at $z > 6.5$, we will not include those since the Ly$\alpha$ emission of those quasars is already beyond the {\it z} band coverage of the Legacy Survey. We also do not include O- and B-type stars, because they are far from the high-redshift quasars in color space \citep{Wenzl2021}. The distant compact early-type galaxies at intermediate or higher redshift, exhibiting very red $r-z$ color, might be a possible type of contaminants, which will be further discussed in Sec. \ref{sec:selection}.

\begin{table*}[t]
\centering
\begin{tabular}{cccccccccccc}
\hline
\multicolumn{1}{l|}{} 
& \multicolumn{4}{c|}{quasars}                           & \multicolumn{7}{c}{star}                              \\ \hline
& vlow-z      & low-z         & mid-z       & high-z     & T     & L     & M     & A     & F     & G     & K     \\
redshift              
& {[}0,1.5{]} & {[}1.5,3.5{]} & {[}3.5,5{]} & {[}5,6.5{]} 
& --    & --    & --    & --    & --    & --    & --    \\
number                
& $2.9\times 10^5$       & $4.2\times 10^5$         & 9,431       & 588        & 1,024 & 4,488 & $1.6\times 10^5$ & $1.2\times 10^5$ & $5.8\times 10^4$ & $2.7\times 10^5$ & $6.3\times 10^4$ \\ \hline
\end{tabular}
\caption{\label{tab:trainingsample}The number of spectra-confirmed high-redshift quasars as well as various categories of spectra-confirmed contaminators in the training sample.} 
\end{table*}

\begin{figure}[ht!]
\begin{center}
\includegraphics[width = 0.48 \textwidth]{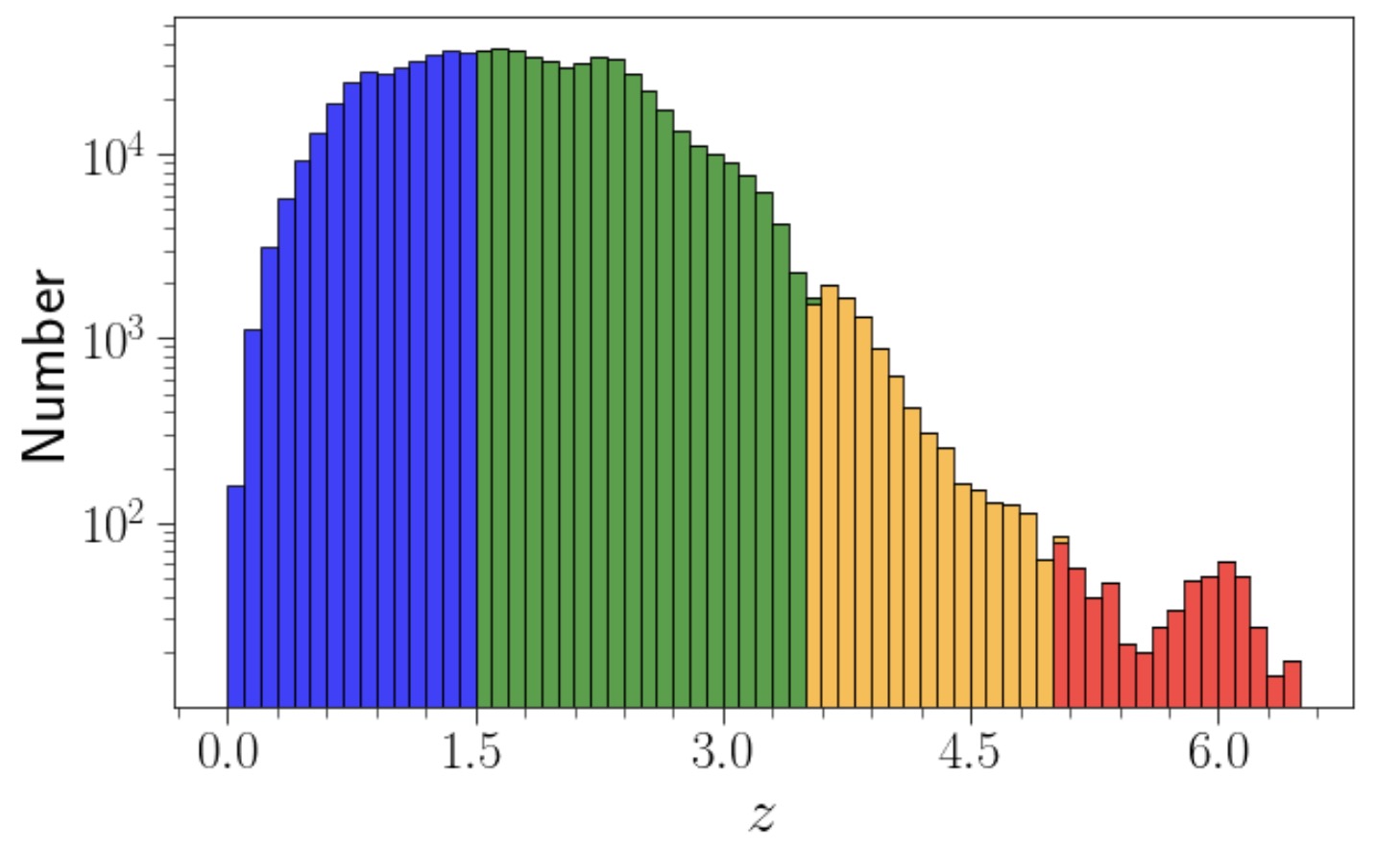}
\end{center}
\caption{The redshift distribution of the quasars in the training sample. The different colors represent the divided categories.}
\label{fig:quasar_distribution}
\end{figure}

The sample of high-redshift quasars is extracted from the Supplementary Database\footnote{\href{https://www.annualreviews.org/content/journals/10.1146/annurev-astro-052920-102455\#supplementary\_data}{SupplementaryDatabase.org/}} \citep{fan2023quasars}, which are constructed from previous observations \citep{wang2016survey,wang2019exploring,yang2020measurements,yang2021probing}, and the DESI spectra-confirmed quasars at $z > 5$\footnote{\href{https://cdsarc.cds.unistra.fr/viz-bin/cat/J/ApJS/269/27\#/browse}{DESIHighzQSO.org/}} \citep{yang2023desi}. 
The MLT brown dwarfs are extracted from both UltracoolSheet\footnote{\href{http://bit.ly/UltracoolSheet}{UltracoolSheet.org/}} \citep{Best2020}
and SDSS DR16 Database\footnote{\href{https://www.sdss.org/dr16}{sdss.org/dr16}} \citep{Ahumada2020}. 
The quasars at $z < 5$ are extracted from SDSS quasars catalog DR16\footnote{\href{https://data.sdss.org/sas/dr16/eboss/qso/DR16Q/}{data.sdss.org/sas/dr16/eboss/qso/DR16Q}}\citep{Lyke2020}. 
The AFGK stars are obtained from the SDSS DR16 Database \citep{Ahumada2020}. %Extended sources like galaxies can be easily separated by morphology from the train sample whose morphological types are all Point Spread Function (PSF), therefore our training sample does not include galaxies. 
To obtain a training sample as representative as possible, we remove the entries whose {\it z} band magnitude is either too bright (top 1\%) or too faint (last 1\%) for each category, which are not representative among the whole training sample. The number of instances for each category in the training sample is shown in Table \ref{tab:trainingsample} and the redshift distribution of the quasars is shown in Figure \ref{fig:quasar_distribution}. Note that the data size for each class in the training sample is not equal, which means that we are dealing with the imbalanced classification. This issue will be further discussed in detail in Sec. \ref{sec:imbalance}.

The Legacy Survey catalog data also provides fluxes at different aperture radius for each band, calling `apflux', which is useful because it provides more features for the machine learning model and essential to discriminate the signal from extended but compact sources like distant compact galaxies. The radius of the apertures are [0.5, 0.75, 1.0, 1.5, 2.0, 3.5, 5.0, 7.0] arcseconds for the {\it g, r, z} bands and [3, 5, 7, 9, 11] arcseconds for the {\it W1, W2} bands. We will construct apflux ratios as the machine learning features, with a detailed description  in Sec. \ref{sec:Data Features Selection}.
%{\bf For each of these two sets of bands ({\it g, r, z} and {\it W1, W2}), we separately perform the following steps: 1) For apfluxes in the same band, we construct the apflux ratios by taking the ratio of neighboring radii. 2) For apfluxes in different bands, we construct the apflux ratios by taking the ratio of the same radius. Moreover, for apflux ratios at different radius across different bands between {\it g, r, z} and {\it W1, W2} bands, we only construct the apflux ratios between the {\it z} band and the {\it W1} band because we believe that such apflux ratios, where the aperture radii differ significantly, are not very effective for training our model. The absence of these apflux ratios in Table \ref{tab:FeatureImportance} effectively supports our argument.}
%{\bf The apflux ratio involving {\it W1, W2} is calculated at different radius across different bands between {\it g, r, z} and {\it W1, W2} bands because the aperture radius are quite different.} 
Moreover, the combined fluxes {\it grz} and {\it W}, which are constructed from {\it g, r, z} fluxes and {\it W1, W2} fluxes, respectively, are useful in quasars search according to a recent study \citep{yeche2020preliminary}. The definitions of the combined fluxes of {\it grz} and {\it W} and the conversion between flux and magnitude are:

    \begin{equation}
	\label{eqn:grz}
 	flux_{grz} = (flux_{g} + 0.8*flux_{r} + 0.5*flux_{z})/2.3
    \end{equation}	

    \begin{equation}
	\label{eqn:w}
	flux_{W} = 0.75*flux_{W1} + 0.25*flux_{W2}
    \end{equation}	

    \begin{equation}
	\label{eqn:mag_flux}
	mag = 22.5 - 2.5*lg(flux)
    \end{equation}

The measurements we will use in the machine learning model include: {\it g, r, z, grz, W1, W2, W} fluxes and magnitudes, as well as apfluxes. Some selected color-color distributions of the training sample are shown in Figure \ref{fig:colorcut}. The top panels are based on the data set with no missing values of all measurements for high-redshift quasars (252 sources), and for all background sources. The bottom panels are based on the entire dataset, where the missing values of high-redshift quasars have been filled in. The process of filling in the missing values is described in detail in the following Sec. \ref{sec:missingvalue}. As is shown clearly that some colors such as {\it g-r, g-W2, r-grz} are effective in separating high-redshift quasars from the contaminators. We would expect that the machine learning algorithm would significantly enhance the performance in separating the high-redshift quasars from the contaminators when making full use of all color information. It is also obvious that the MLT brown dwarfs and the mid-z quasars are the major contaminators since they overlap with the high-redshift quasars to a greater extent in the color space.
 
\subsection{Imputation of Missing Values}
\label{sec:missingvalue}

Currently, the number of spectra-confirmed quasars at $z > 5$ is 727, of which 602 quasars are in the Legacy Survey footprint. Because of the color drop-out, a large number of quasars (335) have nearly zero or even negative fluxes in the {\it g, r} bands, and a small fraction of quasars (29) also have nearly zero or negative flux measurements in the {\it W1} and {\it W2} bands due to the low sensitivity of WISE survey. There are no well-defined magnitudes when converting from zero/negative fluxes into magnitudes, which we call ``missing values". The missing value issue highly limits the high-redshift quasars sample since the machine learning algorithm could not deal with missing values. Therefore, special considerations and techniques may be required to address this issue to ensure accurate analysis and interpretation of the available data.

To fully utilize the limited photometric information available for the known high-redshift quasars, we first employ common imputation methods to handle missing values in the training dataset. One common imputation method is to fill in zeros for the missing values, but it will highly bias the classification model because of the extreme color constructed from the filled-in zeros. Moreover, a zero AB mag has a physical meaning, so we will not discuss it here. The other common imputation methods include: 1) mean imputation, replacing missing values with the mean value of the respective feature; 2) LS limit imputation, substituting missing values with the limiting magnitudes of the Legacy Survey; 3) future imaging survey limit imputation(LR limit), adopting the 5$\sigma$ point-source depth of 27.5 AB mag from the Large Synoptic Survey Telescope \citep[LSST, ][]{LSST} for {\it g, r, z} bands \citep{LSST2019} and of 24 from
the Roman Space Telescope \citep[RST, ][]{RST2015} for {\it W1, W2} bands \citep{Akeson2019} to replace the corresponding missing values, which will be further discussed in Sec. \ref{sec:future}; 4) random forest (RF) imputation, training a random forest regression model to generate reasonable values for missing values; 5) Multiple Imputation by Chained Equations \citep[MICE, ][]{Azur2011} forest, generating imputed values by iterative random forest model on multiple independent divisions of dataset. To validate the feasibility of the five methods mentioned above, we compare the distributions of the imputed values with the complete dataset without missing values and assess the differences.

Here we briefly introduce MICE. MICE is based on the random forest algorithm, and it constructs multiple different imputed datasets randomly to examine the uncertainty and other effects caused by missing values. Each dataset undergoes independent imputation processes, where the imputed values for each feature in the dataset are predicted using a random forest regression model trained on the non-missing dataset. The above process is repeated to update the imputed values. This iterative process does not stop until the average of the imputed values for each feature tends to converge. One big advantage of MICE is its ability to automatically select features related to the imputation variable without requiring manual specification. It also calculates weights for each variable to optimize the variable selection and model fitting during the imputation process. By combining the predictions from multiple random forest models, MICE provides accurate and reliable imputation results. 

\begin{figure}[ht!]
\begin{center}
\includegraphics[width = 0.48 \textwidth]{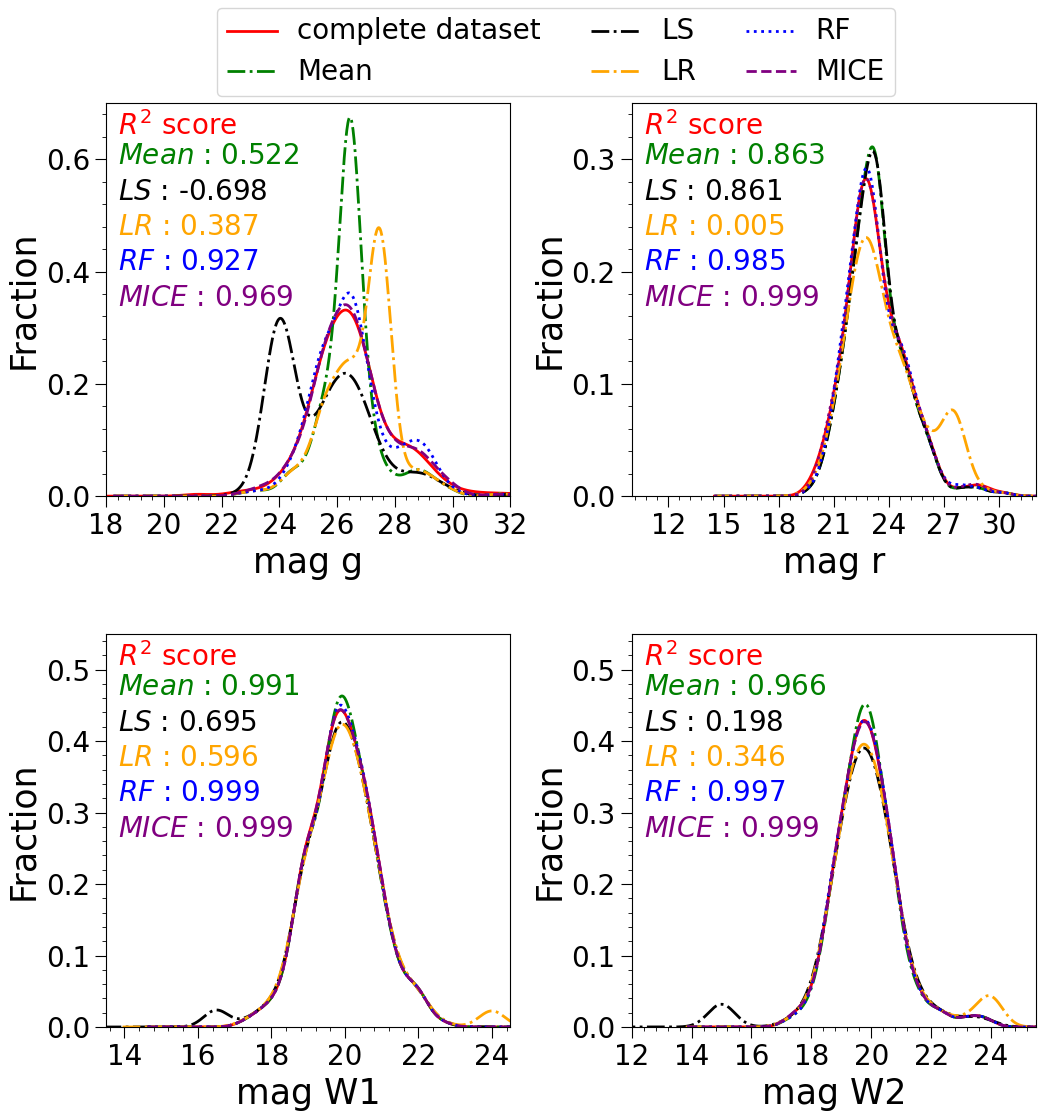}
\end{center}
\caption{ The distribution of {\it g, r, W1, W2} magnitudes using various imputation methods. The red solid line represents the density distribution of the corresponding feature in the complete dataset. The green, black, yellow, blue and purple lines represent the density distributions of the respective features in the datasets imputed using the Mean, LS limit, LR limit, RF and MICE methods. }
\label{fig:Imputa_rf}
\end{figure}

We determine the best imputation method by comparing the density distributions of the magnitudes between the imputed dataset and the complete dataset. We use $R^2$ to represent the differences between the imputed datasets and the complete dataset for these features. $R^2$ assesses how well the model explains variable changes (0 to 1, closer to 1 for better fit). Figure \ref{fig:Imputa_rf} illustrates the density distributions of the four magnitudes with missing values for the complete dataset as well as the five imputed datasets. For the {\it g} magnitude with more than half missing values (55.6\%), the RF and MICE methods can achieve a $R^2$ value of $>$ 0.90, while the results of the other methods are poor. It is easily seen that MICE performs significantly better than the RF model. For the {\it r, W1, W2} magnitudes with fractions of missing values of 12.29\%, 2.33\%, 3.82\%, respectively, the performance of RF and MICE are similar and are much better than the other methods. Note that the $R^2$ criteria could not reflect the effectiveness of the imputation method using future imaging survey limits, and we will further discuss it in Sec. \ref{sec:future}.

Overall, MICE is an effective approach that leverages the power of random forest regression models for missing value imputation. It is very effective in that it automates feature selection and enhances the imputation process through weighted variables. The combination of multiple random forest models enables MICE to provide accurate and reliable imputation results. Therefore, we will adopt the MICE method to perform imputation in the following discussion. The robustness of the MICE imputation will be discussed in the following discussions.

\section{Machine Learning} \label{sec:Random Forest}

In the past decades, many large surveys such as SDSS \citep{York2000}, Legacy Survey \citep{Dey2019OverviewOT}, DESI \citep{DESI1, DESI2} collected tons of data that is challenging to analyze. And the approach using automatic methods like machine learning is both efficient and easily reproducible. Here we briefly introduce the various aspects of the machine learning technique.

\subsection{Metrics} \label{sec:Metrics}

We already introduce the training sample in Sec. \ref{sec:Training sample}. Here we will introduce the other aspects of the machine learning technique. To evaluate the performance of a classification model, we typically consider the evaluation metrics including precision, recall, accuracy, f1 score, and the area under the Receiver Operating Characteristic curve (ROC AUC). In the evaluation process, the comparison between the predicted labels from the machine learning algorithms and the true labels yields four outcomes: true positives (TP), true negatives (TN), false positives (FP), and false negatives (FN), from which we can obtain the desired evaluation metrics.

Precision refers to the fraction of true positives in total positives predicted, defined as:

\begin{equation}
	\label{eqn:precision}
	precision = \frac{TP}{TP + FP} 
\end{equation}	

Recall refers to the ratio of positive class instances correctly identified by a classifier, defined as:

\begin{equation}
	\label{eqn:recall}
	recall = \frac{TP}{TP + FN} 
\end{equation}	

The f1 score metrics combines precision and recall. In fact, the f1 score is the harmonic mean of the two. A high f1 score symbolizes a high precision as well as a high recall. It is defined as:

\begin{equation}
	\label{eqn:f1}
        f1 = \frac{2 \times {precision} \times {recall}}{{precision} + {recall}}
\end{equation}

The f1 score belongs to a unified formula $f_{\beta}$ where $\beta = 1$. $f_{\beta}$ is defined as:

\begin{equation}
	\label{eqn:fb}
        f_{\beta} = \frac{(1 + \beta^2)({precision} \times {recall})}{(\beta^2 \times {precision}) + {recall}}
 \end{equation}	

On datasets with balanced class quantities, the f1 score performs well. For highly imbalanced dataset, the Adjusted f-score (AGF) is an improvement upon the $f_{\beta}$ score \citep{Akosa2017}. The AGF considers all elements of the original confusion matrix, making it a more equitable evaluation metrics for the classification of the minority class. It is defined as:

\begin{equation}
	\label{eqn:agf}
        AGF = \sqrt{f_{\beta=2} \times inv f_{\beta=0.5}}
 \end{equation}
where the $f_{\beta=2}$ score weights recall more than precision, the $inv f_{\beta=0.5}$ refers to the $f_{\beta=0.5}$ score calculated using a confusion matrix obtained by inverting the labels, where TP is replaced by TN, and then substituted into $f_{\beta}$.

%Accuracy refers to the ratio of correctly classified instances by a classifier to the total number of all instances, defined as:

%\begin{equation}
%	\label{eqn:acc}
%	accuracy = \frac{TP + TN}{TP + TN + FP + FN} 
%\end{equation}	

In principle, the precision, recall, f1 as well as AGF is least affected by the imbalanced data sample. And it is almost impossible to tune the desired metrics for all classes for a multi-class classification task. Therefore, we will focus on the metrics of precision, recall, f1 and AGF for the signal of high-redshift quasars in the following discussions, because the correct identification of the high-redshift quasars from enumerous backgrounds is our focus in this study. Although there will be confusion between the classes of contaminants, such as the confusion between the different types of stars, they are not the focus of this study. But we will discuss the weighted metrics of all other classes except the high-redshift quasars for completeness. 

%\subsection{Fiducial Model}

\subsection{Feature Selection} 
\label{sec:Data Features Selection}

The feature selection, the appropriate algorithm and the choices of class ensemble is essential to a good machine learning model, all of which will be discussed in the following. Many studies \citep{schindler2017extremely, nakoneczny2019catalog, yeche2020preliminary, Wenzl2021, He2022} have demonstrated that random forest is very effective in separating signals from contaminators, therefore we will adopt the 11-class random forest classifier as our fiducial model to discuss the feature selection, and the 11 classes are shown in Table \ref{tab:trainingsample}. And then we will compare the different machine learning algorithms and finally we will discuss whether the different class ensembles will help improve the model performance.

The measurements we will use in this study include dereddened magnitudes, dereddened fluxes, and aperture fluxes in the {\it g, r, z, grz, W1, W2}, and {\it W} bands. The aperture radius are [0.5, 0.75, 1.0, 1.5, 2.0, 3.5, 5.0, 7.0] arcseconds for the {\it g, r, z} bands, and are [3, 5, 7, 9, 11] arcseconds for the {\it W1} and {\it W2} bands. The apfluxes are essential to remove extended but compact sources like galaxies, which will be further discussed in Sec. \ref{sec:selection}. %We also construct new fluxes of {\it grz} and {\it W} and calculate the corresponding magnitude based on the above known fluxes, as discussed in Sec. \ref{sec:Training sample}. 
The features used in the machine learning algorithm are constructed from magnitudes (colors), dereddened fluxes (flux ratios) and aperture fluxes (flux ratios). The magnitude features include {\it g, r, z, W1, W2, grz, W}. The color features are constructed from the subtraction of two magnitudes. The flux ratios are constructed from the division of two fluxes, which is the same as the construction of color features. The apflux ratios are constructed as follows: 1) for apfluxes in the same band, we construct the apflux ratios by taking the ratio of aperture fluxes at adjacent radius indices; 2) for apfluxes across different bands, we construct the apflux ratios  by taking the ratio of aperture fluxes at the same radius index across adjacent bands for the first 5 radius for both optical ({\it g, r, z}) and the infrared ({\it W1, W2}) bands. For the extra three radius in the optical bands ({\it g, r, z}), the apflux ratios are constructed by the ratio of aperture fluxes at the same radius across different bands.

The apflux ratios can be represented by a general formula: $ap\_{[a]}\_{[i]}/ap\_{[b]}\_{[j]}$, where [a] and [b] represent indices for the photometric bands ({\it g, r, z, W1, W2}). And the indices [i] and [j] representing the aperture radius, range from 1 to 8 for the optical bands and range from 1 to 5 for the infrared bands. The conditions are: 

\begin{enumerate}
    \item \( [a] = [b] \) (the same band) and \( [i] - [j] = 1 \) (adjacent radius indices).
    \item \( [a] - [b] = -1 \) (adjacent bands) and \( [i] = [j] \) (the same radius index).
\end{enumerate}

There are 7 magnitudes, 21 colors or flux ratios and 55 apflux ratios in total. To fully understand the contribution to improving the performance of the machine learning algorithm, we investigate the following subsamples with different feature subsets:

\begin{itemize}
	\item FeatureSet-A (7): {\it g, r, z, W1, W2, grz, W} magnitude features. 

	\item FeatureSet-B (28): magnitude (7) and color (21) features.
	
	\item FeatureSet-C (83): magnitude (7), color (21) and apflux ratio (55) features.
	
	\item FeatureSet-D (83): flux (7), flux ratio (21) and apflux ratio (55) features.
	
\end{itemize}

FeatureSet-D is the mirror of FeatureSet-C and the big advantage of FeatureSet-D is that there is no missing value issue. We train the random forest classification model on the above four different datasets separately. In the model training, we use {\it RandomizedSearchCV} to find the best set of hyperparameters and we split the entire training sample into ``training set" and ``test set". Table \ref{tab:precision} presents the evaluation scores of the four feature sets for the class of high-redshift quasars. We also use cross-validation (CV) to avoid overfitting on the training sample, and the fold number is set to be 10. In Table \ref{tab:precision}, the ``Val" column represents the average results of ten equally folds of the cross-validation datasets. We adopt the standard deviation within the ten folds as the error. The ``Test" column shows the results of the models for the test set.

\begin{table*}[ht]
\centering
\begin{center}
\begin{tabular}{|c|cc|cc|cc|ll|}
\hline
& \multicolumn{2}{c|}{FeatureSet-A(7)} 
& \multicolumn{2}{c|}{FeatureSet-B(28)} 
& \multicolumn{2}{c|}{FeatureSet-C(83)} 
& \multicolumn{2}{c|}{FeatureSet-D(83)} 
\\ \hline
& \multicolumn{1}{c|}{Val}   & Test 
& \multicolumn{1}{c|}{Val}   & Test  
& \multicolumn{1}{c|}{Val}   & Test  
& \multicolumn{1}{c|}{Val}   & Test  
\\ \hline
precision 
&  \multicolumn{1}{c|}{0.90\(\pm 0.04\)}  & 0.89 
&  \multicolumn{1}{c|}{0.90\(\pm 0.05\)}  & 0.94  
&  \multicolumn{1}{c|}{0.92\(\pm 0.04\)}  & 0.96  
&  \multicolumn{1}{c|}{0.93\(\pm 0.03\)}  & 0.95          
\\ \hline
recall    
&  \multicolumn{1}{c|}{0.81\(\pm 0.04\)}  & 0.85 
&  \multicolumn{1}{c|}{0.91\(\pm 0.03\)}  & 0.92  
&  \multicolumn{1}{c|}{0.89\(\pm 0.05\)}  & 0.92  
&  \multicolumn{1}{c|}{0.86\(\pm 0.07\)}  & 0.87      
\\ \hline
f1        
&  \multicolumn{1}{c|}{0.85\(\pm 0.03\)}  & 0.87 
&  \multicolumn{1}{c|}{0.91\(\pm 0.03\)}  & 0.93  
&  \multicolumn{1}{c|}{0.91\(\pm 0.03\)}  & 0.94  
&  \multicolumn{1}{c|}{0.89\(\pm 0.04\)}  & 0.91  
\\ \hline
AGF        
&  \multicolumn{1}{c|}{0.91\(\pm 0.02\)}  & 0.93 
&  \multicolumn{1}{c|}{0.95\(\pm 0.01\)}  & 0.96  
&  \multicolumn{1}{c|}{0.95\(\pm 0.02\)}  & 0.96  
&  \multicolumn{1}{c|}{0.93\(\pm 0.03\)}  & 0.94  
\\ \hline
\end{tabular}
\end{center}
\caption{\label{tab:precision}
The precision, recall, f1 and AGF score of the high-redshift quasars for the four different feature sets. }
\end{table*}

The comparison between FeatureSet-A and B tells us that the addition of 21 color features significantly improve the performance for the validation data set. This also indicates that in high-dimension color space, the high-redshift quasars are separable from the contaminators. When incorporating the additional 55 apflux ratio into the models, we find that the performance is also enhanced for the test set. Additionally, we use pure flux values from different bands and their ratios as features in FeatureSet-D. Comparing the performance of FeatureSet-C and FeatureSet-D in the table, it can be observed that their precision are very similar, with FeatureSet-C having slightly higher recall, f1 and AGF. The consistency of all evaluation metrics between FeatureSet-C and FeatureSet-D as shown in Table \ref{tab:precision} demonstrates the robustness of our imputation method. %{\bf These models on FeatureSet-C and FeatureSet-D are named the ``mag model" and the ``flux model", respectively.}

Table \ref{tab:FeatureImportance} presents top 20 features ranked by the importance for FeatureSet-C. We can see that the color features align with the trends shown in Figure \ref{fig:colorcut}. As can be seen clearly, the newly introduced features of {\it grz} and {\it W} magnitudes have comparably high importance, demonstrating that these additional features indeed play an important role in improving the model performance. We notice that the most important feature, {\it z-W2}, is not effective in separating the high-redshift quasars from the contamination shown in Figure \ref{fig:colorcut}, which will be further discussed in detail in Sec. \ref{sec:Discussion of Random Forest performance}.

\begin{table}[]
	\centering
	\begin{tabular}{lc}
		\toprule
		Feature                             & \multicolumn{1}{l}{Importance {[}\%{]}} \\
		\toprule
		$\text{z}-\text{W2}$      	 & 6.14                                      \\
		$\text{z}-\text{W1}$      	 & 5.59                                     \\
		$\text{z}-\text{W}$    	     & 5.59                                       \\
		$\text{g}-\text{r}$    	     & 4.94                                      \\
		$\text{g}-\text{z}$	         & 4.83                                      \\	
		$\text{g}-\text{grz}$        & 4.58                                  \\
		$\text{z}-\text{grz}$        & 4.41  
                    \\
		$\text{r}-\text{z}$          & 4.17            
                \\		
		$\text{W}-\text{grz}$        & 3.42                                     \\	
		$\text{r}-\text{grz}$     	 & 3.32                                  \\				
		$\text{r}-\text{W}$      	 & 3.15                                       \\	
		$\text{g}-\text{W}$          & 3.08                                       \\
		$\text{W2}-\text{grz}$       & 2.91                                 \\
		$\text{W1}-\text{grz}$     	 & 2.67                                       \\		
		$\text{apflux\_W1\_1}-\text{apflux\_W2\_2}$     	                                   & 2.54                                       \\
		$\text{r}-\text{W2}$ 		 & 2.54                                       \\
		$\text{g}-\text{W1}$ 		 & 2.31                                       \\
		$\text{r}-\text{W1}$ 		 & 2.25                                       \\
		$\text{g}-\text{W2}$ 		 & 2.21                                       \\
		$\text{apflux\_r\_5}-\text{apflux\_z\_5}$     	                                       & 1.98                                       \\
		\toprule
	\end{tabular}

	\caption{\label{tab:FeatureImportance}The top 20 feature in terms of their importance ranking provided by the random forest classification model trained on FeatureSet-C.} 
\end{table}

\subsection{Comparison of Different Algorithms} \label{sec:Comparison of Different Algorithms}

There are many machine learning algorithms with their own strength and weakness, here we compare the performance of different machine learning algorithms on separating high-redshift quasars from contaminators. The classification algorithms used in this study are the following:

\begin{enumerate}

\item \texttt{K-Nearest Neighbors (KNN)}: A non-parametric algorithm that forms a model by placing the data from the training set in the high-dimensional feature space and makes predictions based on the similarity between the data point and its $k$ nearest neighbours of the training set in the high-dimension feature space. The hyper-parameters in this algorithm include {\it number of neighbors ($n\_neighbors$), weights ($w$)} and {\it distance metric ($p$)}. The distance metric can be either Manhattan distance or Euclidean distance, used to measure the similarity.

\item \texttt{Decision Tree (DT)}: A non-parametric supervised learning that constructs a tree-like classification or regression structure by continuously dichotomising features based on a discrete set of values. In these tree structures, leaves represent class labels and branches represent conjunctions of features that lead to those class labels. The hyper-parameters in this algorithm include {\it criterion for split quality, maximum depth of trees, minimum number of samples required to split an internal node, minimum number of samples required at each leaf node, maximum number of features to consider for each split}, etc.

\item \texttt{Random Forest}: An integrated approach that performs classification or regression by combining multiple decision tree models, where each tree is constructed using a random subset of the training data and a random subset of features. This randomness helps to reduce the overfitting issue and to improve generalisation. The hyper-parameters of this algorithm include {\it number of estimators} and the ones of the decision tree algorithm. 

\item \texttt{Light Gradient Boosting Machine (LGBM)}: A gradient boosting algorithm that makes predictions by iteratively training multiple decision trees. Each decision tree is trained based on the residuals of the previous tree to gradually reduce the prediction error. The {\it learning rate} in LGBM algorithm is critical to the model performance. Other hyper-parameters are similar to the ones in random forest classifier.

\item \texttt{Gaussian Naive Bayes (GNB)}: A plain Bayesian probabilistic classification algorithm assuming that features are independent of each other and that their conditional probabilities follow a Gaussian distribution. The {\it prior probability ($p$)} is the most important hyper-parameter of this algorithm. By default, $p$ is estimated from the proportion of samples in each category in the training data.

\end{enumerate}

We apply the above algorithms to FeatureSet-C and obtain the corresponding precision, recall, f1 and AGF scores of the high-redshift quasars for these models, as shown in Table \ref{tab:Different Algorithms}. The maximum value for each evaluation metrics is highlighted. It can be seen clearly that the random forest model is the best one for all four evaluation metrics. Then we will adopt the random forest algorithm in the following discussion. 

Note that FeatureSet-C is the best feature set selected via the random forest model, but it might not be the best feature set for other algorithms. Here we investigate whether FeatureSet-A or FeatureSet-B on other algorithms have better performance than FeatureSet-C. We find that all metrics on FeatureSet-A for other algorithms are worse than those on FeatureSet-B. We present the best metrics using FeatureSet-B for the other four algorithms, we have the best precision (\(0.88 \pm 0.05\), from KNN), best recall (\(0.84 \pm 0.06\), from LGBM), best f1 score (\(0.83 \pm 0.05\), from LGBM) and best AGF (\(0.83\pm 0.05\), from LGBM). It is obvious that the results obtained on FeatureSet-B for these four algorithms are all lower than the corresponding metrics on Featureset-C. Therefore, we conclude that Featureset-C is the best feature set not only for the random forest algorithm, but also for other classification algorithms considered in this work.

\begin{table}[]
	\centering
		\begin{tabular}{|c|c|c|c|c|c|}
			\hline
		& Precision  	    & Recall 		 	& f1 	 & AGF   
                \\ \hline
KNN 
& 0.88\(\pm 0.07\)  
& 0.60\(\pm 0.09\)  
& 0.71\(\pm 0.07\)  
& 0.80\(\pm 0.05\)  
\\ \hline
DT       
& 0.85\(\pm 0.04\)  
& 0.80\(\pm 0.05\)  
& 0.82\(\pm 0.03\)
& 0.90\(\pm 0.02\) 
\\ \hline		
RF 		
& \textbf{0.92\(\pm 0.04\)}  
& \textbf{0.89\(\pm 0.05\)} 
& \textbf{0.91\(\pm 0.03\)}  
& \textbf{0.95\(\pm 0.02\)}
\\ \hline
LGBM                
& 0.85\(\pm 0.05\)  
& 0.84\(\pm 0.07\) 
& 0.84\(\pm 0.04\) 
& 0.92\(\pm 0.03\) 
\\ \hline
GNB    
& 0.39\(\pm 0.04\) 
& 0.68\(\pm 0.07\)  
& 0.49\(\pm 0.05\)
& 0.77\(\pm 0.03\) 
\\ \hline
												
		\end{tabular}
  
	\caption{\label{tab:Different Algorithms}Evaluation scores for different classification algorithms based on FeatureSet-C for the validation dataset.}
\end{table}

\subsection{Class Selection} \label{sec:Class Selection}

We adopt the 11-class classification model in the above discussion, and the 11 classes are shown in Table \ref{tab:trainingsample}. In order to investigate whether further reducing the number of classes in the training sample can improve the model's performance, we will now discuss four scenarios in which certain classes from FeatureSet-C are merged. The four scenarios are the following:

\begin{itemize}
	\item 11-class classification ($P_{11}$): 
		\begin{itemize}
		\item $P_{11}$: vlowz, lowz, midz, highz quasars and M, L, T, A, F, G, K dwarfs.
		\end{itemize}
	
	\item 4-class classification ($P_4$): 
		\begin{itemize}
			\item $P_4^0$: vlowz, lowz, midz quasars
			\item $P_4^1$: highz quasars
			\item $P_4^2$: M, L, T dwarfs
			\item $P_4^3$: A, F, G, K stars 
		\end{itemize}		
	\item 3-class classification ($P_3$):
 		\begin{itemize}
			\item $P_3^0$: vlowz, lowz, midz quasars
			\item $P_3^1$: highz quasars
			\item $P_3^2$: M, L, T, A, F, G, K dwarfs
   		\end{itemize}		
	\item 2-class classification ($P_2$):
 		\begin{itemize}
			\item $P_2^0$: vlowz, lowz, midz quasars and M, L, T, A, F, G, K dwarfs
			\item $P_2^1$: highz quasar
		\end{itemize}		

\end{itemize}

We apply the above four scenarios to FeatureSet-C and obtained four new datasets, called $P_{11}$, $P_4$, $P_3$, $P_2$. In $P_4$, $P_3$, $P_2$, the reason why we can combine vlowz, lowz, midz quasar samples is that they are in the period when the cosmic reionisation epoch is already over. In $P_4$ scenario, we separate the M, L, T dwarfs from other stars because they are the main contaminants of high-redshift quasars and they are quite different from other types of stars in the optical and near-infrared bands. M, L, T dwarfs are grouped together because of the similarity between them. A, F, G, K dwarfs are also grouped together because they are relatively far away from the high-redshift quasars on the color-color diagrams, as can be seen from Figure \ref{fig:colorcut}. 

\begin{table*}[]
	\centering
        \begin{tabular}{|c|ccc|ccc|ccc|ccc|}
        \hline
        \multirow{3}{*}{} & 
        \multicolumn{3}{c|}{$P_{11}$} & 
        \multicolumn{3}{c|}{$P_4$}    &
        \multicolumn{3}{c|}{$P_3$}    & 
        \multicolumn{3}{c|}{$P_2$}                              \\ \cline{2-13} &
        \multicolumn{1}{c|}{Val}     &
        \multicolumn{2}{c|}{Test}    &
        \multicolumn{1}{c|}{Val}     &
        \multicolumn{2}{c|}{Test}    &
        \multicolumn{1}{c|}{Val}     &
        \multicolumn{2}{c|}{Test}    &
        \multicolumn{1}{c|}{Val}     &
        \multicolumn{2}{c|}{Test}                         \\ \cline{2-13} &
        \multicolumn{1}{c|}{high-z}     & 
        \multicolumn{1}{c|}{high-z} & others & 
        \multicolumn{1}{c|}{high-z}     & 
        \multicolumn{1}{c|}{high-z} & others & 
        \multicolumn{1}{c|}{high-z}     & 
        \multicolumn{1}{c|}{high-z} & others & 
        \multicolumn{1}{c|}{high-z}     & 
        \multicolumn{1}{c|}{high-z} & others \\ \hline
precision         
& \multicolumn{1}{c|}{0.92\(\pm 0.04\)} & \multicolumn{1}{c|}{0.96}         & 0.86          
& \multicolumn{1}{c|}{0.94\(\pm 0.04\)} & \multicolumn{1}{c|}{0.97}         & 0.99          
& \multicolumn{1}{c|}{0.94\(\pm 0.03\)} & \multicolumn{1}{c|}{0.97}         & 0.99          
& \multicolumn{1}{c|}{0.97\(\pm 0.03\)} & \multicolumn{1}{c|}{0.96}         & 1.00          \\ \hline
recall            
& \multicolumn{1}{c|}{0.89\(\pm 0.05\)} & \multicolumn{1}{c|}{0.92}         & 0.87          
& \multicolumn{1}{c|}{0.86\(\pm 0.04\)} & \multicolumn{1}{c|}{0.91}         & 0.99          
& \multicolumn{1}{c|}{0.86\(\pm 0.03\)} & \multicolumn{1}{c|}{0.92}         & 0.99          
& \multicolumn{1}{c|}{0.79\(\pm 0.04\)} & \multicolumn{1}{c|}{0.90}         & 1.00          \\ \hline
f1             
& \multicolumn{1}{c|}{0.91\(\pm 0.03\)} & \multicolumn{1}{c|}{0.94}         & 0.85          
& \multicolumn{1}{c|}{0.89\(\pm 0.02\)} & \multicolumn{1}{c|}{0.94}         & 0.99          
& \multicolumn{1}{c|}{0.90\(\pm 0.02\)} & \multicolumn{1}{c|}{0.95}         & 0.99          
& \multicolumn{1}{c|}{0.87\(\pm 0.03\)} & \multicolumn{1}{c|}{0.93}         & 1.00          \\ \hline
AGF               
& \multicolumn{1}{c|}{0.95\(\pm 0.02\)} & \multicolumn{1}{c|}{0.96}         & 0.91          
& \multicolumn{1}{c|}{0.93\(\pm 0.02\)} & \multicolumn{1}{c|}{0.96}         & 0.99          
& \multicolumn{1}{c|}{0.93\(\pm 0.01\)} & \multicolumn{1}{c|}{0.96}         & 0.99          
& \multicolumn{1}{c|}{0.91\(\pm 0.02\)} & \multicolumn{1}{c|}{0.95}         & 0.95          \\ \hline
\end{tabular}
	\caption{\label{tab:ClassSelection}
The precision, recall, f1 and AGF of the high-z class in the validation and test datasets for the four scenarios based on FeatureSet-C. In addition, we present the average metrics weighted by the number of instances of each class  for all other classes in the ``others" column for the test set.}
\end{table*}

We train the random forest classification model on the above four datasets. The precision, recall, f1 and AGF score of the high-redshift quasars for the four scenarios are presented in Table \ref{tab:ClassSelection}. Meanwhile, we also present the  average metrics weighted by the number of instances of each class  for all other classes except the high-z class for the test set. It is clearly seen that the precision and recall are consistent within error for the different scenarios identified by the number of classes in the dataset. This might indicate that there is certain correlation among the different contaminators, which is demonstrated by the weighted metrics of all other classes. The relatively stable precision indicates that the sample of high-redshift quasars is clean and possesses inherent features that is distinguishable from other contaminants, ensuring high precision and recall regardless of the number of classes in the classification models. As we reduce the complexity of the classification model by fewer classes via combining certain classes, the weighted metrics should get better
because the confusion between classes that we combine gets ignored, which is consistent with the results shown in Table \ref{tab:ClassSelection}.

Based on the statistics in Table \ref{tab:ClassSelection}, we decide to pick the 11-class model for two reasons. First of all, the 11-class model would provide us the full picture of contamination within the different classes, expecially the contamination for the high-redshift quasars. Secondly, we find that the strict boundary between ``mid-z" and ``high-z" redshift bins introduces confusion between these two classes, which should not be a concern. The inspection on those ``mid-z" quasars, which are classified as ``high-z" quasars, tells us that the redshift for those ``mid-z" quasars are between 4.84 and 4.98, right around the redsfhit boundary of 5. Further inspection on the spectra of these ``mid-z'' quasars indicates that they bear a striking resemblance to the typical spectra of high-redshift quasars. This implies that these ``mid-z" quasars which are predicted as the ``high-z" class are not actually contaminants of the ``high-z" quasars. Therefore, the precision of the ``high-z" class for the 11-class model has been underestimated. If we exclude these ``mid-z" quasars, the precision of of the ``high-z" class for the 11-class model would reach 0.99.

\subsection{Important feature: {\it i-band photometry}} \label{sec:Important feature: i magnitude}

For quasars with redshifts between 5 and 6.5, the Ly$\alpha$ emission line is shifted to the wavelength range of 7296 \AA \ $-$ 9120 \AA. However, the Legacy Survey {\it g, r} bands does not cover this wavelength range, and the {\it z} band only has a small coverage in this wavelength range. Then the {\it i} band, whose wavelength coverage mostly overlaps with this wavelength range, is essential for high-redshift quasar ($5 < z < 6.5$) searching . Fortunately, the Legacy Survey DR10 not only includes images in the {\it g, r}, and {\it z} bands from DECaLS, but it also contains DECam observations for {\it g, r, i}, and {\it z} bands  from several non-DECaLS surveys, primarily the Dark Energy Survey, the DELVE Survey \citep{delve}, and the DeROSITA Survey\footnote{\href{https://noirlab.edu/science/programs/ctio/instruments/Dark-Energy-Camera/DeROSITAS}{DeROSITAS.org/}}. These surveys mainly cover the southern sky (Declination $\le$ 32.375$^\circ$). The area covered in the {\it i} band for number of passes greater than 3 is 13,024 square degrees, while the total area covered in the {\it g, r, i}, and {\it z} bands jointly is 9,923 square degrees.

Because the combined sky coverage of the {\it g, r, i}, and {\it z} bands is approximately half of that for the {\it g, r, z} bands, we divide the entire dataset into the one that includes the {\it i} band data (FeatureSet-i) and the one that does not include the {\it i} band data (FeatureSet-non-i). The feature construction is the same as above except that the FeatureSet-i dataset includes all i band related measurements. We apply MICE to fill in missing values as discussed in Sec.\ref{sec:missingvalue} and we train models using the 11-class 
random forest classification algorithm on these two datasets. We evaluate the models on the validation set and test set to assess the {\it i} band importance.

\begin{table*}[]
	\centering
		\begin{tabular}{|c|cc|cc|l|c|c|}
			\hline
& \multicolumn{2}{c|}{FeatureSet-i}   
& \multicolumn{2}{c|}{FeatureSet-non-i} &  
& Feature               
& Importance {[}\%{]} \\ \cline{1-5} \cline{7-8} 
& \multicolumn{1}{c|}{Val}   & Test 
& \multicolumn{1}{c|}{Val}   & Test  &  
& $\text{g}-\text{i}$   & 5.40               
\\ \cline{1-5} \cline{7-8} 
precision 
& \multicolumn{1}{c|}{0.94\(\pm 0.03\)}  & 0.97 
& \multicolumn{1}{c|}{0.91\(\pm 0.04\)}  & 0.92  &  
& $\text{z}-\text{W2}$  & 5.37               
\\ \cline{1-5} \cline{7-8} 
recall    
& \multicolumn{1}{c|}{0.88\(\pm 0.04\)}  & 0.87 
& \multicolumn{1}{c|}{0.87\(\pm 0.04\)}  & 0.89  &  
& $\text{r}-\text{z}$   & 4.99                
\\ \cline{1-5} \cline{7-8} 
f1        
& \multicolumn{1}{c|}{0.91\(\pm 0.03\)}  & 0.92 
& \multicolumn{1}{c|}{0.89\(\pm 0.03\)}  & 0.90  &  
& $\text{g}-\text{r}$ & 4.92  
\\ \cline{1-5} \cline{7-8} 
AGF
& \multicolumn{1}{c|}{0.94\(\pm 0.02\)}  & 0.94 
& \multicolumn{1}{c|}{0.94\(\pm 0.02\)}  & 0.95  &  
& $\text{g}-\text{z}$ & 4.64
\\ \hline
		\end{tabular}
	\caption{\label{tab:precision-i-mag} The precision, recall f1 and AGF of the dataset including {\it i} band measurements (FeatureSet-i) and not including {\it i} band measurements (FeatureSet-non-i). We also present the first five most importance features from FeatureSet-i.}
\end{table*}

Table \ref{tab:precision-i-mag} shows the evaluation scores of validation and test sets for the two datasets (FeatureSet-i v.s. FeatureSet-non-i), as well as the first few most important features of the models trained on FeatureSet-i. Compared to FeatureSet-non-i, which does not include features related to the {\it i} band, the models trained on FeatureSet-i achieve higher precision on the test set, while the other evaluation scores are similar. This indicates that the {\it i} band can improve the performance of our model in searching for high-redshift quasar to some extent. The columns of ``Feature" and ``Importance" display the top-ranked features and the corresponding importance in the models trained on FeatureSet-i. It can be observed that the feature {\it g-i} has high importance, which reflects the significance of {\it i} band related features in our model.

In this section, we already demonstrate that the {\it i} band and its corresponding features are indeed playing an important role in the classification model for high-redshift quasars. However, due to the limited amount of {\it i} band data, our model is ultimately trained on the data provided by Legacy Survey DR9. In future work, we will consider incorporating photometric data from more bands to obtain additional features and further improve the model's performance.

\subsection{Discussion of Random Forest performance} \label{sec:Discussion of Random Forest performance}

\begin{figure*}
\begin{center}
\includegraphics[width = 0.95 \textwidth]{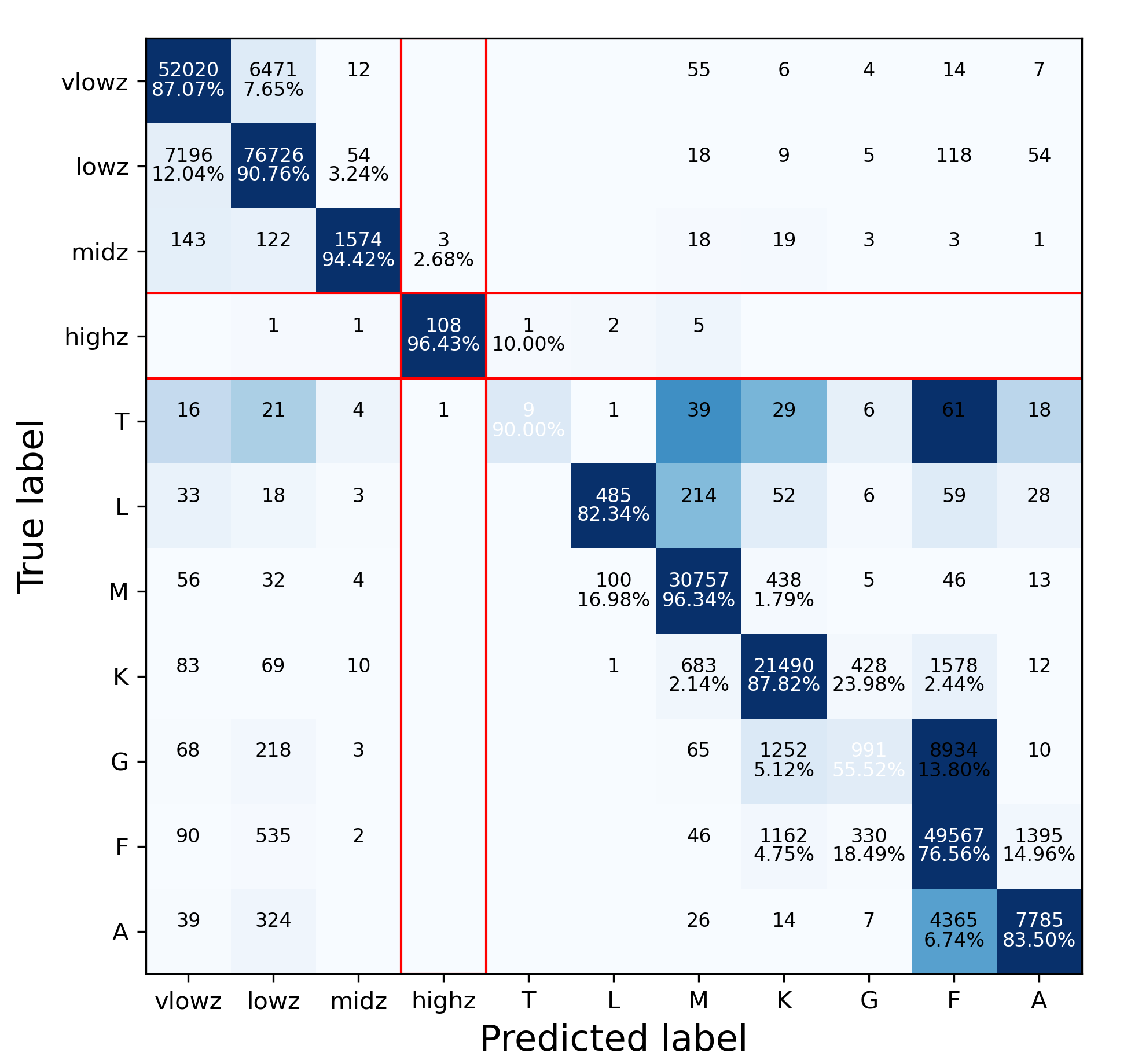}
\end{center}
\caption{Confusion matrix of the 11-class random forest classifier trained on FeatureSet-C for the test set. Each column in each row represents the predicted class counts for the corresponding true class. The percentages on the diagonal represent the precision of each class, and the shades of color represent the recall of each class. The precision of the high-redshift quasars is as high as 96.43\%. The recall of the high-redhisft quasars can achieve 91.53\%, which means that approximately 10\% of the high-redshift quasars are incorrectly predicted as other types. }
\label{fig:Confusion Matrix}
	\end{figure*}

After the discussion on feature selections, the comparison of different classification algorithms and class selections of the dataset, we find that the 11-class random forest classification model on the FeatureSet-C and D yields the best performance for separating the high-redshift quasars from the contaminators. Figure \ref{fig:Confusion Matrix} shows the confusion matrix of the model using FeatureSet-C for the test set, where each row represents the true class and each column represents the predicted class. The percentages in each square indicate the fraction of the corresponding class in each column. Squares with a percentage of less than 1\% are not displayed. For the diagonal elements, these percentages represent the precision of the corresponding class. 

For the class of high-redshift quasars, the precision achieves as high as 96.43\% and the recall can also reach 91.53\%. The precision of the high-redshift quasars for this model is significantly enhanced compared to the previous work \citep{Wenzl2021}. The high precision ensures that the final high-redshift quasar candidates have much larger probability to be true high-redshift quasars, which significantly improves the efficiency of the future spectra verification. The recall is also significantly enhanced compared to the previous work \citep{Wenzl2021}, which means much less true high-redshift quasars will be missed. The average precision and recall weighted by the number of instances of each class for all other classes except the signal can also reach 86\% and 87\%, respectively, which are also significantly higher than those presented in \cite{Wenzl2021}.

As can be seen that the main contaminators to the high-redshift quasars are M, L, and T dwarfs, which agrees with the studies using traditional color-cuts, because M, L, and T dwarfs exhibit many absorption features in the optical and near-infrared bands, making them to have similar colors as the high-redshift quasars in those bands. To further improve the classification model, on one hand, we need to expand the training sample size, especially the high-redshift quasars, L and T dwarfs. On the other hand, considering the inclusion of photometric data from additional bands, such as {\it i}, and {\it y} bands, which are not available in Legacy Survey DR9 but will be available in future imaging surveys, may provide us with useful features to distinguish the high-redshift quasars from the contaminators. 

For quasars with $z < 5$, they all have $\sim$ 90\% precision, indicating that each class has its own intrinsic characteristics that make them distinguishable. However, there is also significant contamination among them. On one hand, this is because they are in the post-reionization era of the universe, where the IGM is already mostly ionized, resulting in color similarities among these quasars. On the other hand, our rough setting of the redshift boundaries might lead to misclassification of quasars around the redshift boundary into other classes.

For M, L, and T dwarfs, the contamination among them is significant. This is because they all belong to subtypes of brown dwarfs, which are differentiated based on the spectra \citep{Best2020,Ahumada2020}, and photometric data lacks the precision of spectroscopic data. We find that a significant portion of L and T dwarfs are misclassified as other classes, which is due to the high similarity in the optical spectra and comparably small sample size. Moreover, because of the limitation of the sample size, the color space for L and T dwarfs are not fully spanned and not representative enough to separate them. The issues for both M dwarfs and quasars with $z < 5$ are highly mitigated. As for other type stars, although there is serious contamination among them, we have not find any significant impact on the predictions of the high-redshift quasars. 

\begin{figure}
\begin{center}
\includegraphics[width = 0.48 \textwidth]{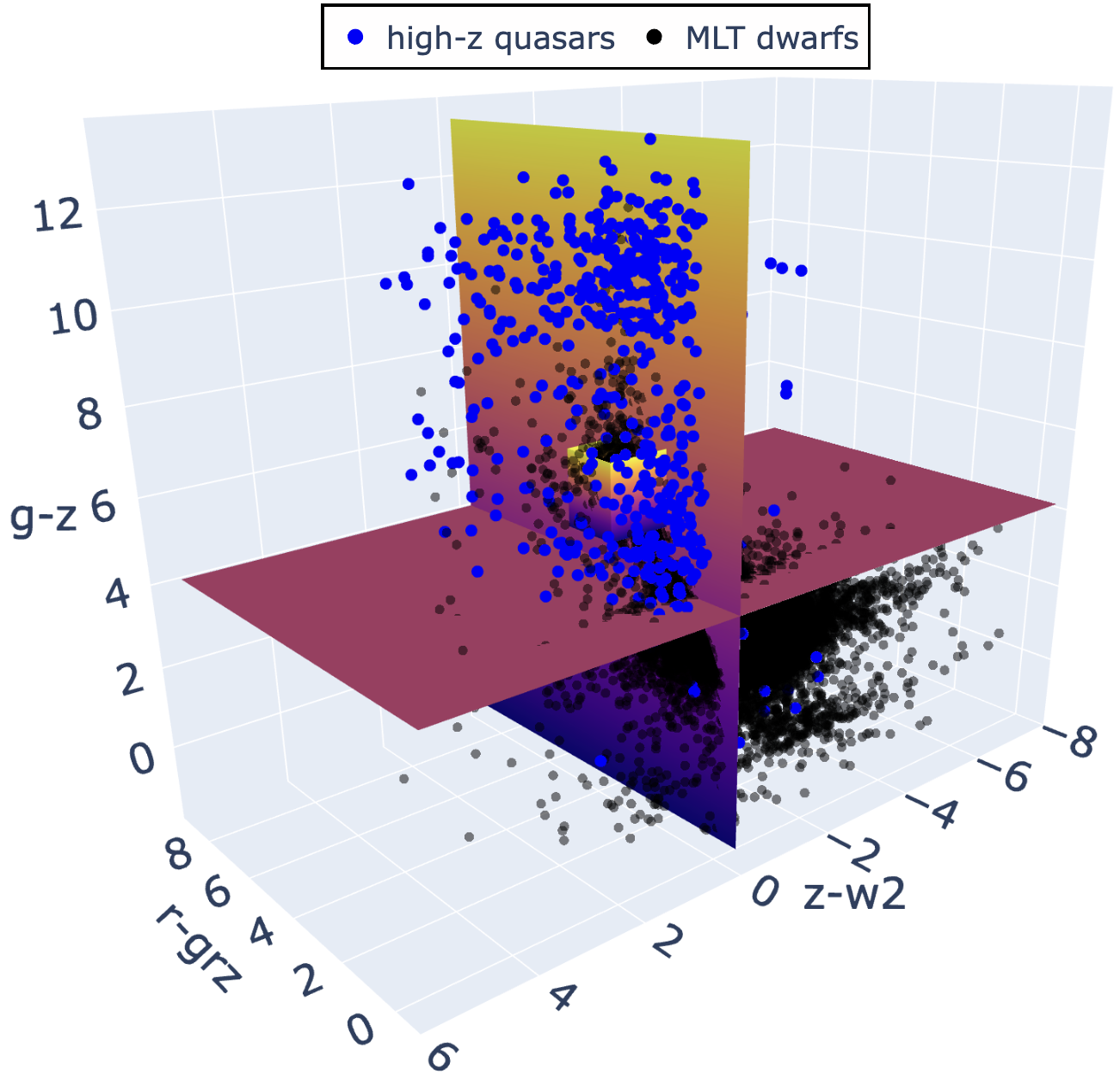}
\end{center}
	\caption{The distribution of the high-redshift quasars and the main contaminators, MLT dwarfs, in the three-dimensional color space spanned by the $z-W2$, $g-z$, and  $r-grz$ features. It can be observed that the most important feature $z- W2$ from the random forest model is very effective to separate the high-redshift quasars from the contaminators in the three-dimensional color space, while $z-W2$ is not effective in the traditional two-dimensional color space.}
 \label{fig:colorcut_3d}
\end{figure}

We notice that the top-ranked feature of {\it z-W2} in Table \ref{tab:FeatureImportance} contradicts to the color-color diagram presented in Figure \ref{fig:colorcut}. Here we investigate this issue in detail. It is known that the color-color diagram presented in Figure \ref{fig:colorcut} is the projection of the high-dimensional color space. Then the projection on certain direction might diminish the separation between the high-redshift quasars and the contaminators. In order to fully resolve this issue, we construct a three-dimensional (3-D) color space from the top ranked features presented in Table \ref{tab:FeatureImportance}. Figure \ref{fig:colorcut_3d} illustrates the color diagram of high-redshift quasars and the main contaminant sources, MLT dwarfs, in the 3-D color space of {\it z-W2}, {\it g-z}, and {\it r-grz}. It can be observed that while there was significant overlap in the 2-D color space of {\it z-W2} and {\it r-grz}, these sources exhibit clear separation in the 3-D color space. This demonstrates that the signal and the contaminators can be better distinguished in high-dimensional color space, which is not evident in the projected 2-D color space.
 
\subsection{Discussion of the imbalanced issue}
\label{sec:imbalance}

We notice that the number of instances in each class in the training sample is not balanced, as shown in Table \ref{tab:trainingsample}. Here we are discussing the impact of the imbalanced data sample on the random forest classification model. The ratio of the most numerous class (low-z quasars) to the least abundant class (high-z quasars) is as high as {$700:1$}. %Most other class also have hundreds of times the number of instances as the high-z class, it is evident that our dataset is highly imbalanced.
Typically, classification models trained on the extremely imbalanced datasets tend to favor the majority class, making it easier for the minority classes to be incorrectly classified as the majority. This skews the model's classification performance for the minority class, more specifically, it results in a lower recall for the minority class, which also 
indirectly affects the precision of the minority class.

There are multiple approaches to deal with the imbalanced datasets. For example, the random forest classification algorithm offers a {\it $class\_weight$} parameter to adjust the weights of each class based on the number of instances in the input dataset, with the goal of giving the classes that are less frequent a higher weight. This approach aims at balancing the dataset by ensuring that the minority classes are not overwhelmed by the majority classes. There are two modes (``balanced” and ``balanced\_subsample”) to adjust the class weights. The ``balanced” mode uses the values of labels to automatically adjust weights inversely proportional to class frequencies in the input data. The ``balanced\_subsample” mode is the same as ``balanced” except that weights are computed based on the bootstrap sample for every tree grown. %We train RF models with this parameter on Featureset-C. 

%\Huanian{You should further mention ``balanced" and ``balanced-subsample".} %As shown in Table~\ref{tab:balance}, the ``$class\_weight$" column indicates the evaluation scores of these two models on the test data set using Featureset-C for the high-z class. The corresponding scores are consistent with each other. Compared to the ``RF(no resample)" column, there is a slightly decrease in precision for both models, while the recall remains unchanged.

The two methods mentioned above do not involve any reprocessing of the training samples but rather assigning different weights to different classes to achieve balanced datasets. There are methods that directly alter the number of instances for each class in the training sample to balance the dataset, and the methods can be further divided into two categories. One approach is to reduce the entire training sample according to the minority class, which is called ``under-sampling". And the other approach is generating more training sample according to the majority class, which is called ``over-sampling".

\begin{table*}[ht]
\centering
\begin{tabular}{c|c|cc|ccc|ccc}
\hline
%\diagbox{}{Methods}
\multirow{2}{*}{\resizebox{3.5cm}{!}{\diagbox{\Huge{\bf Metrics}}{\Huge{\bf Methods}}}}
& RF (Imbalance)
& \multicolumn{2}{c|}{class\_weight}                  
& \multicolumn{3}{c|}{Under-sampling} 
& \multicolumn{3}{c}{Over-sampling} 
\\
%\diagbox{metric}{}
&                 
& BAL 
& \multicolumn{1}{c|}{BAL\_SUB} 
& BRF        
% & ENN        
& NM
& OSS       
& ROS      
& SMOTE     
& ADASYN     
\\ \hline
precision 
& 0.96               
& 0.87         & 0.88           
& 0.42         & 0.67           & 0.86          
& 0.89         & 0.88           & 0.88           \\
recall    
& 0.92                
& 0.93         & 0.92           
& 0.98         & 0.98           & 0.96          
& 0.92         & 0.95           & 0.93           \\
f1        
& 0.94               
& 0.90         & 0.90           
& 0.59         & 0.80           & 0.91    
& 0.91         & 0.91           & 0.90          \\
AGF        
& 0.96               
& 0.96         & 0.96           
& 0.88         & 0.95           & 0.97    
& 0.96         & 0.97           & 0.96          
\\ \hline
\end{tabular}
\caption{\label{tab:balance} The evaluation metrics for the RF model on the imbalanced sample without resampling and for the RF models on the balanced samples for the test set.}
\end{table*}

The commonly used under-sampling algorithms include: Balanced Random Forest \citep[BRF,][]{Chen2004}, Near Miss \citep[NM,][]{Mani2003}, and One Sided Selection \citep[OSS,][]{Kubat1997}. The main difference among the algorithms lies in the way they undersample the training sample. BRF draws a bootstrap sample from the minority class and sample with replacement the same number of samples from the majority class. NM uses the KNN algorithm to remove majority class samples near the minority class. OSS rejects the majority class samples around the minority class and near the decision boundary, and those that have minimal impact on the model. We adopt these three under-sampling methods to reprocess the training sample. %keeping the number of samples in the high-z class unchanged, while resampling each majority class to match the number of samples in the high-z class.
%As shown in Table~\ref{tab:balance}, the ``Under-sampling" column represents the results obtained using these three algorithms. It is clearly seen that those algorithms sacrifices precision while achieving high recall, which is contradicting to our goal of achieving high precision while sacrificing recall. %It can be observed that the best precision among them is lower by up to 10 percentage points compared to the results obtained from the imbalanced training set. Although their recalls all show an improvement of approximately 5 percentage points, our focus is primarily on the precision compared to the recall of the high-z class.

The commonly used over-sampling algorithms include: Random Over Sampler (ROS), Synthetic Minority Oversampling Technique \citep[SMOTE,][]{Chawla2011}, and Adaptive Synthetic Sampling \citep[ADASYN,][]{He2008}. The approaches to over sample the training sample are different for these algorithms. ROS randomly samples and duplicates the minority class with replacement. SMOTE synthesizes new samples between the nearest neighbors of minority class samples in the feature space, while ADASYN focuses on synthesizing new samples between the nearest neighbors of minority class samples misclassified by the KNN classifier. We adopt these three algorithms to reprocess the training sample. %As shown in Table~\ref{tab:balance}, the ``Over-sampling" column represents the results obtained using these three methods. Their results are very similar to those of the ``$class\_weight$" column. Compared to the ``RF(no resampling)" results, the precisions in the ``Over-sampling" column are lower, while the recalls are basically the same.

Table \ref{tab:balance} shows the performance for the RF model without resampling and the RF models on the balanced samples for the test set. 
For the $class\_weight$ balanced methods, it can be observed that setting the $class\_weight$ parameter has little impact on recall but reduces precision compared to the results on imbalanced samples. %This could be due to assigning higher weights to each sample of the high-z class, causing some leaf nodes with originally low proportions of high-z class predictions to shift from non-high-z class to high-z class predictions. This introduces more sources of contamination incorrectly classified as high-z class, and this parameter setting does not significantly alter predictions for leaf nodes already predicted as high-z class. 
For the under-sampling balanced methods, the precision of both BRF and NM methods decreases significantly, while both recalls show a significant improvement. This is because the under-sampling methods reduces the number of the majority samples significantly, leading to the loss of important information from the contaminators and shrinking in the color space. This issue is particularly obvious for the BRF method which is based on random under-sampling. %This causes the decision boundary between high-z and other contamination sources to lean more towards including the high-z class. Furthermore, we can see the results of the OSS method, which ensures a minimal decrease in precision while slightly improving recall. This is quite 
Surprisingly, we find that the OSS method returns comparably good precision and excellent recall. %as our under-sampling method requires the number of instances from all classes to be comparable to the number of instances from the high-z class (around 500). 
Despite such a small training set, we are able to obtain comparably good models with balanced dataset. This also reflects the fact that the high-redshift quasars possess inherent features such that they are separable from the contaminators in the high-dimensional color space. For the over-sampling balanced methods, the metrics obtained by these three methods are quite consistent.
Compared to the metrics on the imbalanced dataset, they show little change in recall but a decrease in precision. Although the performance of the over-sampling balanced methods is excellent, one issue about these methods is that the over-sampled instances  might even not exist or might be biased in the color space. %The balancing of the training set by the three oversampling methods, ROS, SMOTE, and ADASYN, essentially means providing weights for the minority class within the training set, which is why the results obtained from these oversampling techniques are very similar to those of ``$class\_weight$". The difference lies in the fact that SMOTE and ADASYN synthesize new data to enhance the weight of the minority class. Consequently, they extend the decision boundaries in the feature space, allowing the previously isolated minority class instances to be better classified.  However, this may introduce too many falsely classified instances, leading to a decrease in precision for the high-z class. This is because the inclusion of more isolated high-z instances results in an improvement in the recall of the high-z class.

In conclusion, most of the methods discussed for handling imbalanced datasets do not show significant differences in all four metrics (precision, recall, f1, AGF) compared to those on the imbalanced dataset. Additionally, these balanced methods often prioritize on enhancing the recall of the high-z class while sacrificing its precision. However, what we are more concerned about is the precision of the high-z class, which is more crucial for correctly classifying a vast amount of data with unknown label using our model. Therefore, we will still adopt the 11-class random forest classification model on our original training sample.

% \Huanian{Add a table to show the results for the above algorithms.}

\subsection{Future Imaging Surveys}
\label{sec:future}
There are multiple complementary imaging surveys on the way, which will be the frontier for high-redshift quasars searching. On one hand, there will be more confirmed high-redshift quasars in the near future, expanding the size of training sample significantly. On the other hand, those deep imaging surveys not only provide much better images with high-quality, but also do they provide many more features for both high-redshift quasars searching and photo-z estimation. For example, the Chinese Space Station Telescope \citep[CSST,][]{Zhan2011} will cover $\sim$ 17000 square degrees of the sky and has a wavelength coverage of $2600-10000$ \AA \ with broad-band filters of near-ultraviolet (NUV), {\it u, g, r, i, z, y}. The Roman Space Telescope 
\citep[RST,][]{RST2015} has a wide wavelength coverage of $0.48-2.3$ microns with 8 broad-band filters, whose near-infrared images will be a lot better than the WISE {\it W1, W2} images. The 5$\sigma$ AB magnitude limits could reach $\sim$ 24\footnote{\href{https://roman.gsfc.nasa.gov/science/WFI\_technical.html}{romanWFItechnical.org/}} \citep{Akeson2019} 
for filter F184 and F213 (central wavelength are 1.84 and 2.31 $\mu$m, respectively) even with only 55 seconds exposure. The Euclid Space Telescope \citep[EST,][]{Euclid} will deliver data over 15,000 square degrees of the sky with a wavelength range of $0.95-2.02$ microns. The Large Synoptic Survey Telescope \citep[LSST,][]{LSST} will deliver the best images in 6 filters for over 18,000 square degrees sky coverage, and the 5$\sigma$ point-source depth of the coadded maps can reach 27.5 AB magnitude for $r$-band \citep{LSST2019}. The near- or mid-infrared photometric data from those wide-field imaging surveys will significantly improve the performance of high-redshift quasars searching.

Furthermore, those future imaging surveys could reach much deeper detection limit, which might resolve the missing values issue for some of the high-redshift quasars. Here we investigate the imputation method in which the missing values are replaced with the $5\sigma$ point-source depth 
from future imaging surveys (refer to Sec. \ref{sec:missingvalue} for details). The evaluation scores (precision: 0.96, recall: 0.91, f1: 0.93, and AGF: 0.95) of the imputation method using future imaging survey limits on the test set are very close to those using the MICE imputation method (precision: 0.96, recall: 0.92, f1: 0.94, and AGF: 0.96). This indicates that the future imaging surveys are effective to identify the high-redshift quasars. When incorporating additional features other than the {\it g, r, z, W1, W2} bands from future imaging surveys, one could imagine how effective our random forest classification model will be.

%\Huanian{Run the RF model on Featureset-C using the future survey limits.}

\section{Photometric Redshift Estimation} \label{sec:Photometric Redshift Estimation}

Generally, spectroscopic redshift is challenging to obtain for a large number of high-redshift quasar candidates, then the photometric redshift could provide us a good estimation to study various distance-related aspects of the high-redshift quasars. The photometric redshift estimation algorithms can be divided into two categories: template fitting and machine learning methods. Template fitting involves establishing a relationship between photometric magnitudes or fluxes and spectroscopic redshifts using a series of spectral energy distribution (SED) templates. For example, EAZY \citep{Brammer2008} is an algorithm based on template fitting, in which the default template set, as well as the default functional forms of the priors are from semianalytical models. However, machine learning methods seek to find the relationship between photometric information and spectroscopic redshifts based on spectra-confirmed samples. Algorithms that are commonly used include k-nearest neighbors \citep{Ball2007,Zhang2013,Zhang2019,Han2021,Curran2021}, random forest regression \citep{Carliles2010,Carrasco2013,schindler2017extremely,Zhang2019,Wenzl2021,Li2022,Zeraatgari2024}, and CatBoost \citep{Li2022}. Here we are using the random forest regression algorithm to establish a model to estimate the photometric redshift of the high-redshift quasars and introducing the evaluation metrics of the photometric redshift estimation.

The evaluation of photometric redshift is often based on the ratio of the number where the absolute value of $\Delta z = z_{\rm spec} - z_{\rm photo}$ is less than some threshold, $e$, to the total number, as shown in Equation \ref{eqn:Ze}. \cite{Bovy2012} conducted statistical analyses on the spectroscopic redshifts of previously verified quasars and the photometric redshifts derived from their optical and near-infrared photometric data. %The typical threshold values, $e$, are 0.1, 0.2, and 0.3, used to assess the accuracy of photometric redshifts. 

\begin{equation}
\label{eqn:Ze}
\phi_{e} = \frac{1}{N}{\sum_{z\in Z} \llbracket |\Delta z| < e \rrbracket}
\end{equation}
here Z is the set of spectroscopic redshifts, $N$ is the total number of objects with spectroscopic redshift.  $e$ is typically set to be 0.1, 0.2, 0.3 \citep{Bovy2012,Peters2015,Richards2015}. 
 
 The evaluation criteria mentioned earlier directly utilize the difference between the spectroscopic redshift and the estimated redshift, which is referred to as ``non-normalized" and remains unaffected by the magnitude of the spectroscopic redshift. Another evaluation criterion is based on using the normalized difference scaled by the spectroscopic redshift, as shown in Equation \ref{eqn:norm}. 

%$\Delta z_{\rm norm}$ is the normalised $\Delta z$ by the value of redshift:

	\begin{equation}
		\label{eqn:norm}
		\Delta z_{\rm norm} = \frac{\Delta z}{1 + z_{\rm spec}} ,
	\end{equation}	
 here $z_{\rm spec}$ is the spectroscopic redshift.

The fraction of entries where $\Delta z_{\rm norm}$ exceeds a certain threshold compared to the total number is commonly referred to as the outlier rate, $\eta$. As shown in Equation \ref{eqn:outlier}, we typically set this threshold to 0.1. A smaller outlier rate indicates that more of the predicted values are within the predetermined margin of error. This outlier rate evaluation criterion is more suitable for lower photometric redshift evaluations. The outlier rate is defined as:
\begin{equation}
    \label{eqn:outlier}
    \eta_{0.1} = \frac{1}{N}{\sum_{z\in Z} \llbracket|\Delta z_{\rm norm}| > 0.1 \rrbracket},
\end{equation}
where $\llbracket  \rrbracket$ represents an eigenfunction, defined as:
\begin{equation}
    \llbracket x \rrbracket=
    \begin{cases}1,true\\0,false
    \end{cases},
\end{equation}

\begin{figure*}
	\centering
	\begin{tabular}{c}
		\includegraphics[width=.99\linewidth]{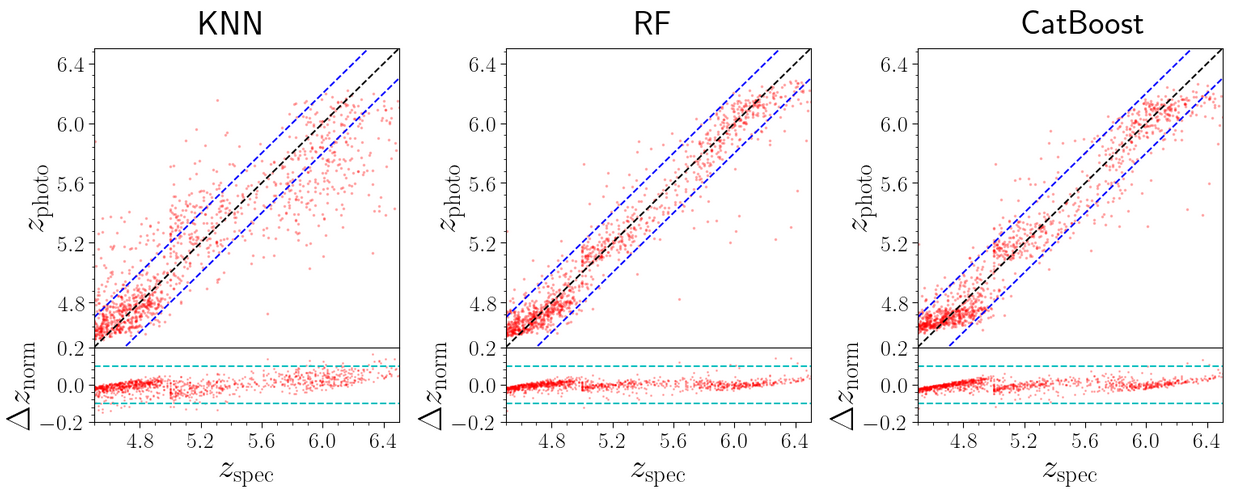} \\
		\includegraphics[width=.99\linewidth]{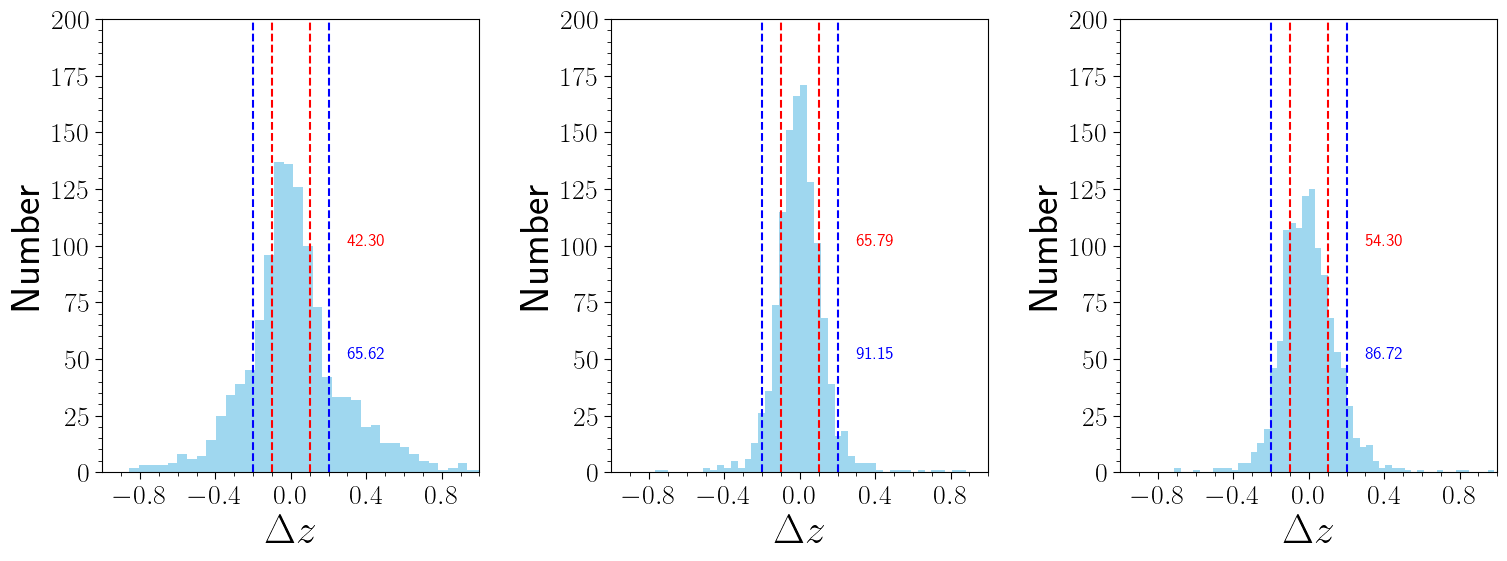} \\
	\end{tabular}
\caption{ The top panel presents the relation between photometric redshift and spectroscopic redshift as well as the distribution of the normalized $\Delta z$ for the three regression algorithms (KNN, RF, CatBoost). The blue dashed line stands for $\Delta z = \pm 0.2$ and the cyan dashed horizontal line represents $\Delta z_{\rm norm} = \pm 0.1$. The bottom panel presents the distribution of the difference between the photometric redshift and the spectroscopic redshift ($\Delta z$). The blue (red) dashed vertical line stands for $\Delta z = \pm 0.2 \ (0.1)$. The numbers in the figure represent the fraction within the boundaries. }
\label{fig:spec-photo}
\end{figure*}

We utilize three widely employed machine learning algorithms (KNN, RF, CatBoost) to construct regression models across two distinct datasets, FeatureSet-mag and FeatureSet-flux, respectively. FeatureSet-mag comprises magnitudes of the {\it g}, {\it r}, {\it z}, {\it W1}, {\it W2}, {\it W}, and {\it grz} bands, alongside the apfluxes in the {\it g}, {\it r}, {\it z}, {\it W1}, and {\it W2} bands. FeatureSet-flux encompasses fluxes of the {\it g}, {\it r}, {\it z}, {\it W1}, {\it W2}, {\it W}, and {\it grz} bands, in addition to the apfluxes in the {\it g}, {\it r}, {\it z}, {\it W1}, and {\it W2} bands. There are only 602 high-redshift quasar samples ($5 < z_{\rm spec} < 6.5$), in order to enhance the accuracy of redshift predictions, we include the mid-z quasars with redshifts greater than 4.5 in the training sample. Similar to the classification models above, we divide the dataset into training, testing, and validation sets and use {\it RandomizedSearchCV} to find the best set of hyperparameters in the model's hyperparameter space.

We have ultimately developed six regression models based on the quasars  with redshifts ranging from 4.5 to 6.5. Table \ref{tab:rgs model scores} presents the $R^2$ (R-squared) and MSE (Mean Squared Error) scores of these six models. $R^2$ and MSE are common regression model evaluation metrics. $R^2$ assesses how well the model explains variable changes (0 to 1, closer to 1 for better fit), while MSE directly reflects prediction accuracy (smaller values for higher accuracy). Among the different training sets, the models trained on FeatureSet-mag outperform those trained on FeatureSet-flux, which also demonstrates the effectiveness of the MICE imputation method we use earlier. When considering different regression algorithms, the KNN model performs significantly worse compared to the other two models. Both the CatBoost and RF models exhibit very close and relatively good scores, with the RF model outperforming the CatBoost model.

\begin{table}[]
\centering
%\resizebox{\columnwidth}{!}
{
\begin{tabular}{|c|ccc|ccc|}
\hline
& \multicolumn{3}{c|}{FeatureSet-mag}                         
& \multicolumn{3}{c|}{FeatureSet-flux}                        \\ \cline{2-7} 
                  & \multicolumn{1}{c|}{KNN}   & \multicolumn{1}{c|}{RF} & CatB & \multicolumn{1}{c|}{KNN}   & \multicolumn{1}{c|}{RF} & CatB \\ \hline
$R^2$             & \multicolumn{1}{c|}{0.714} 
                  & \multicolumn{1}{c|}{\textbf{0.927}} 
                  & 0.924     
                  & \multicolumn{1}{c|}{0.688} 
                  & \multicolumn{1}{c|}{0.912}   
                  & 0.910     
                  \\ \hline
MSE               & \multicolumn{1}{c|}{0.098} 
                  & \multicolumn{1}{c|}{\textbf{0.025}}  
                  & 0.026     
                  & \multicolumn{1}{c|}{0.108} 
                  & \multicolumn{1}{c|}{0.031}  
                  & 0.031     
                  \\ \hline
\end{tabular}
}
\caption{The $R^2$ and MSE score for the six different regression models which predict photometric redshifts.}\label{tab:rgs model scores}
\end{table}

%\begin{figure}[ht!]
%	\includegraphics[width=.99\linewidth]{Bimodal structure.png}
%\caption{The color-color ($z-grz$, $z-W$) diamgram for the known high-redshift quasars. The red dots represent quasars with higher redshifts ($5.6 < z < 6.5$), while blue dots represent quasars with lower redshifts ($5.0 < z < 5.6$). The two shaded regions in the lower panel indicate the density distribution of {\it z-grz} for these two subclasses. The green line, known as the color tracks, represents the distribution of redshifts in the color space for all high-redshift quasars.}\label{fig:Bimodal structure}
%\end{figure}

Figure \ref{fig:spec-photo} illustrates the distribution of spectroscopic redshifts and photometric redshifts for all quasars with redshifts from 4.5 to 6.5 for the three machine learning regression algorithms (KNN, RF, CatBoost) based on FeatureSet-mag, along with the distribution of $\Delta z_{\rm norm}$. According to the traditional evaluation criterion $\phi_{\rm e}$, the KNN, RF, and CatBoost models achieved $\phi_{0.1}$ values of 42.30\%, 66.98\%, and 54.30\%, respectively, representing the proportion of $\Delta z < 0.1$ to the total count. The $\phi_{0.2}$ values for these models are 65.62\%, 90.64\%, 86.72\%. The $\phi_{0.3}$ values are 78.38\%, 96.60\%, and 95.40\%. Judging based on the normalized evaluation criterion $\eta_{0.1}$, with results of 2.89\%, 0.68\%, and 0.51\% for the KNN, RF, and CatBoost models. In the lower panel of the three top subplots, the horizontal cyan lines represent $\Delta z_{\rm norm} = \pm 0.1$. It is clearly seen that there are almost no points falling outside the cyan parallel lines.

All three models show exceptional performance, although the results of the KNN model are slightly inferior to the other two models. In conclusion, it is evident that the RF regression model predicts redshifts with higher precision and accuracy. Therefore, we adopt the random forest regression model to predict the photometric redshifts for the high-redshift quasar candidates which will be discussed in the following.

\section{High-z candidates and verification} \label{sec:High-z candidates and using grism spectra to do some confirmation}

\subsection{Selection criteria}
\label{sec:selection}

Based on the discussions above, we decide to adopt the 11-class random forest classification model on FeatureSet-C and FeatureSet-D. These models are named the ``mag model" and the ``flux model", respectively. Subsequently, we apply these two models separately to the entire Legacy Survey DR9 dataset to obtain the high-redshift quasar candidates. Before that, we have applied several selection criteria to reduce the size of the catalog data from LS DR9 (more than 1 billion entries) without missing too many signals. The selection criteria followed by the reasoning are: 
\begin{enumerate}
    \item $dered\_mag\_g,r,z,W1,W2$ is not null if the features include colors.
    \item $brick\_primary=1$ and $maskbits!=[1,10,12,13]$.
    \item The type is  `PSF'.
    \item $snr\_z>5$, $snr\_{W1}>3$ and $snr\_{W2}>2$.
    \item $dered\_mag\_z > 15$ and $dered\_mag\_z<21.5$.
\end{enumerate}
1) We apply the first criterion to remove sources with missing values because the trained random forest model does not allow missing values, and imputation is challenging due to the large data size and unknown object labels. 2) The second criterion is used to ensure that the quality of the source is minimally affected by uncontrollable factors (such as being on the boundary, near bright sources, etc). 3) Setting the source type to Point Spread Function (PSF) helps eliminate contamination from extended sources like galaxies. 4) The fourth criterion sets the signal-to-noise ratio constraints, as referenced in \cite{yang2023desi}, helping reduce the catalog size without missing too many signals. 5) The last criterion sets the magnitude constraints to help reduce either too bright or too faint sources in the {\it z}-band. 

Although setting the source type to PSF can help us eliminate extended sources significantly, it cannot exclude all galaxies, such as the compact early-type galaxies at intermediate or higher redshift. These galaxies, due to their comparably high redshift and compactness, exhibit the morphology of PSF and have a redder color. We will investigate whether those distant compact galaxies are contaminants for the high-redshift quasars by training a binary RF classifier. For the training sample, we first obtain the galaxy sample from the SDSS DR17 Database\footnote{\href{https://www.sdss4.org/dr17/}{sdss.org/dr17}} \citep{Abdurro2022} with spectral type of ``Galaxy" and with $z \geq 0.2$ (2,207,611 objects in total). Then, we cross-match the galaxy sample to the LS DR9 database  with the above selection criteria to obtain the sample of distant compact galaxies (19,374 objects in total). The sample of high-redshift quasars is the same as the one presented in Table \ref{tab:trainingsample}. 

Similar to the 11-class classification model, we also obtain two parallel feature sets (FeatureSet-C and FeatureSet-D) with the best hyper-parameters determined by {\it RandomizedSearchCV}. The evaluation scores (precision, recall, f1, AGF) of the ``high-z" class for the test set are 0.98, 0.99, 0.98, 0.99, respectively, for the ``binary mag model", and the results for the ``binary flux model" are consistent with those of the ``binary mag model". The corresponding metrics (precision, recall, f1, AGF) of the distant compact galaxies for the test set are 1.00, 1.00, 1.00, 0.99, respectively, for both the ``binary mag model" and the ``binary flux model". These evaluation scores indicate that the distant compact galaxies almost do not contaminate the high-redshift quasars. Moreover, we find that the apflux ratios are indeed very effective to distinguish the high-redshift quasars from the distant compact galaxies.

Furthermore, we apply the trained 11-class classification models to the distant compact galaxy sample, and we found that only 4 galaxies (0.021\%) are predicted as the ``high-z" class by the ``mag model", and only 1 galaxy (0.005\%) is predicted as the ``high-z" class by the ``flux model". In conclusion, the contamination of the distant compact galaxies to the signal of high-redshift quasars is  minimal and negligible. 

\subsection{High-z candidates}

After applying the above selection criteria, we have obtained a total of 140 million sources from LS DR9, which is passed to the two trained random forest classification models to obtain two parallel sets of high-redshift quasar candidates. The random forest classification algorithm provides the predicted probability of each category for each source object, we take the class with the highest predicted probability as the predicted class for each source object. The number of objects for each predicted class using the ``mag model" is shown in the Table \ref{tab:all_classes}. The ``flux model" returns similar results, with 420,208 mid-z quasar candidates and 568,188 high-z quasar candidates, respectively. And there are 216,949 overlapping high-redshift quasar candidates which are both identified by the ``mag model" and the ``flux model". %{\bf The high fraction of overlapping candidates to the total ``mag model" candidates also demonstrates the robustness of the imputation method.}

\begin{table}[h]
	\centering
	\begin{tabular}{lr|lr}
		A & 457,493 		& T 		& 436 \\
		F & 20,026,064 		& vlow-z	& 3,901,910 \\
		G & 580,421 		& low-z 	& 2,844,361 \\
		K & 29,403,366  	& mid-z 	& 399,726 \\
		M & 68,712,381 	    & high-z	& 272,424\\
		L & 109,606	        &  			& 
	\end{tabular}
\caption{\label{tab:all_classes} The number of objects in each category based on the ``mag model" in the entire LS DR9 footprint. We take the class with the highest predicted probability as the predicted category for each source object. }
\end{table}

\begin{figure}[ht!]
\includegraphics[width=.99\linewidth]{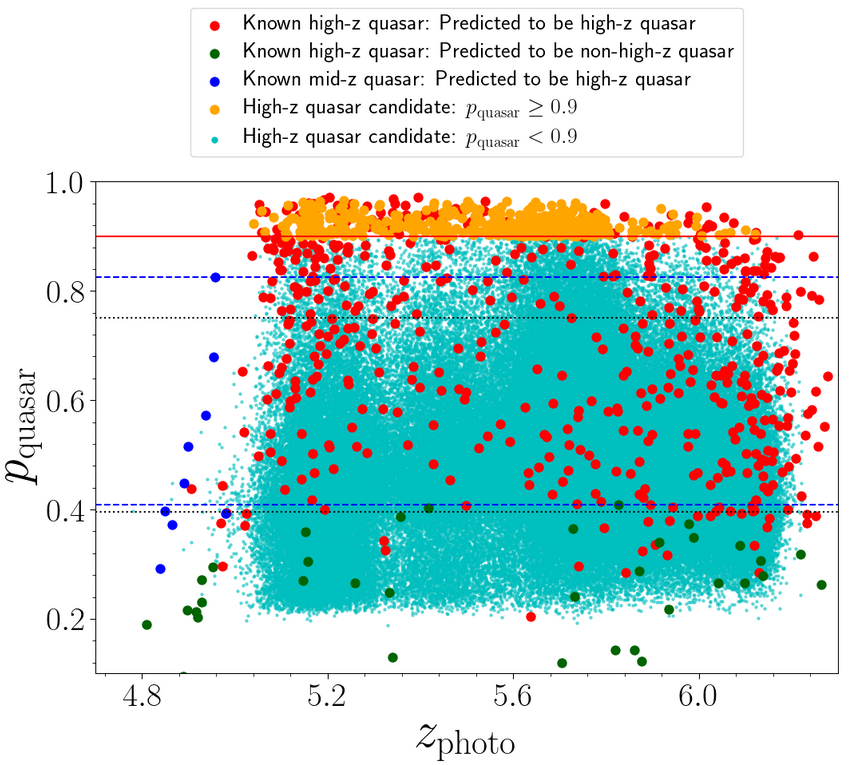}
\caption{ Probability for the high-redshift quasars vs. predicted redshifts for our catalog data. $p_{\rm quasar}$ and $z_{\rm photo}$ are provided by the ``mag model". The orange and cyan dots represent candidates predicted as the high-z class. The red and green dots represent known high-redshift quasars, with green dots indicating sources that are misclassified as other classes. The blue dots represent the known ``mid-z" quasars in the training set that are incorrectly predicted as ``high-z" class. } 
\label{fig:candidates selection}
\end{figure}

To obtain more reliable high-redshift quasar candidates, further screening is necessary. The random forest classification model provides us with the predicted probabilities for each source object. By setting reasonable thresholds on these probabilities, we can select much more reliable high-redshift quasar candidates. Figure \ref{fig:candidates selection} displays the distribution of photometric redshifts for high-redshift quasar candidates and known quasars with the probability of being classified as the high-z class. The orange and cyan dots depict our high-redshift quasar candidates, which are the overlapping candidates obtained from both the ``mag model" and the ``flux model" for the entire LS DR9 footprint. The red and green dots represent spectra-confirmed high-redshift quasars. The photometric redshift for the high-z class is provided by the random forest regression model trained on FeatureSet-mag.

We try first to set a probability threshold such that the precision of the known high-redshift quasars in our predicted results can reach 100\%. To do that, we identify the highest predicted probability among the known high-redshift quasars that are incorrectly predicted as non-high-redshift quasars, and we call it as the first probability threshold, and the values are $p_{\rm thre1,mag} = 0.41$ and $p_{\rm thre1,flux} = 0.40$ for the ``mag model" and the ``flux model", respectively. They are represented by the lower blue and black horizontal lines in Figure \ref{fig:candidates selection}, with the green dots representing known high-redshift quasars that are misclassified to other categories. This threshold helps us remove approximately one-third of the candidates.

The second probability threshold ($p_{\rm thre2}$) is related to the objects which are not ``high-z" class but predicted as the ``high-z" class. Among the objects predicted as the ``high-z" class, we identify the highest probability among those that is incorrectly predicted as the ``high-z" class. They are $p_{\rm thre2,mag} = 0.83$ for the ``mag model" and $p_{\rm thre2,flux} = 0.75$ for the ``flux model", and are represented by the upper blue and black dashed horizontal lines in Figure \ref{fig:candidates selection}. The blue points in Figure \ref{fig:candidates selection} represent the ``mid-z" quasars that have been incorrectly predicted as the ``high-z" class. As has been discussed earlier, the confusion is not a concern because their redshifts are around the boundary. This means that the precision of our trained random forest regression model is actually higher than what is currently being observed.

At this stage, we have obtained a relatively pure candidate set. To further ensure the plausibility of the high-redshift quasar candidates, we also quote the results for higher probability thresholds of 0.90 and 0.95. The number left after each of the probability threshold for the entire LS DR9 footprint is presented in Table \ref{tab:probthres}. For the ``mag model", we have $p_{\rm quasar} \geq p_{\rm thre1,mag}$: 198,339 candidates, $p_{\rm quasar} \geq p_{\rm thre2,mag}$: 2,984 candidates, $p_{\rm quasar} \geq 0.90$: 476 candidates, $p_{\rm quasar} \geq 0.95$: 32 candidates. For the ``flux model", we have $p_{\rm quasar} \geq p_{\rm thre1,flux}$: 473,401 candidates, $p_{\rm quasar} \geq p_{\rm thre2, flux}$: 69,736 candidates, $p_{\rm quasar} \geq 0.90$: 11,012, $p_{\rm quasar} \geq 0.95$: 740. And for the overlapping candidates, we adopt the probability threshold derived from the ``mag model", and we have $p_{\rm quasar} \geq p_{\rm thre1}$: 165,734 candidates, $p_{\rm quasar} \geq p_{\rm thre2}$: 2,984 candidates, $p_{\rm quasar} \geq 0.90$: 476, $p_{\rm quasar} \geq 0.95$: 32. Our catalog includes information of all candidates, and Table \ref{tab:catalog} describes the columns of the data file, which is available in its entirety in machine-readable form. We present the catalogs obtained from the ``mag model", the ``flux model" and the overlapping results.

\begin{table}[ht]
\centering
\begin{center}
\begin{tabular}{|c|c|c|c|c|}
\hline
                      & $p_{thre1}$ & $p_{thre2}$ & $p_{0.90}$ & $p_{0.95}$ \\ \hline
mag candidate         & 198,339     & 2,984       & 476        & 32         \\ \hline
flux candidate        & 473,401     & 69,736      & 11,012     & 740        \\ \hline
overlapping candidate & 165,734     & 2,984       & 476        & 32          \\ \hline
\end{tabular}%
\end{center}
\caption{\label{tab:probthres} The number of high-redsfhit quasar candidates left after each probability threshold for the ``mag model," ``flux model," and the overlapping results. }
\end{table}

\begin{figure}[ht!]
\begin{center}
\includegraphics[width = 0.48 \textwidth]{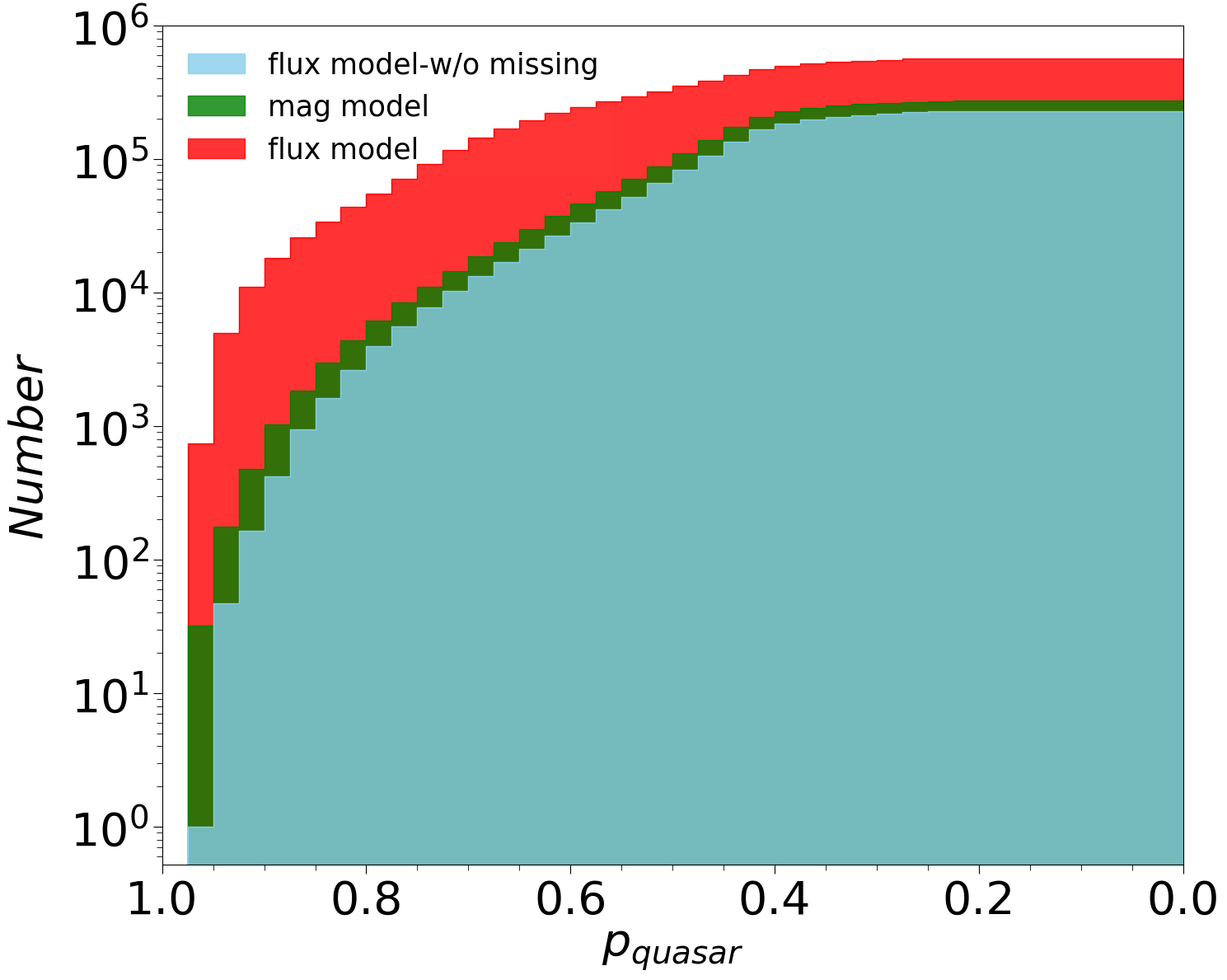}
\end{center}
\caption{
The plot shows a step histogram of the probability of high-z candidates for the ``mag model" (green area) and the ``flux model" (red area). The sky blue area stands for the probability distribution of the candidates without missing values for the ``flux model". The step histogram is obtained by calculating the cumulative number of all data points from the largest probability bin (the leftmost bin) to the end of the current bin, with a width of 0.025 for each bin. 
}
\label{fig:diff}
\end{figure}

We notice that there is a big difference between the ``mag candidate" and the ``flux candidate" in Table \ref{tab:probthres}. This difference becomes more pronounced as the probability threshold increases. Remember that the big advantage of the ``flux model" (FeatureSet-D) is that there is no missing value issue. We suspect that the big difference between the ``mag candidate" and the ``flux candidate" is due to the missing value issue in the LS catalog. To fully understand this issue, we present the probability distributions of the high-redshift quasar candidates from the two models in Figure \ref{fig:diff}. The green area corresponds to the probability distribution of the candidates for the ``mag model". The red area represents the probability distribution of the candidates for the ``flux model".

It is clearly seen that there are many more candidates of the ``flux model" than the ``mag model", especially at the high probability region. It is obvious that the objects with missing values tend to have larger probability to be high-redshift quasars due to the color drop-out, which can be inferred from the training sample (more than half of the current known high-redshift quasars have missing value). If we remove the objects with missing values for the ``flux model", which is represented by the sky blue area. It is obvious that the majority candidates of the ``flux model" at high probability region have missing values, resulting in the big difference between the two models as shown in Table \ref{tab:probthres}. The phenomenon observed in the high-redshift quasar candidates is consistent with that of the training sample.  %Because there exists difference between the probability predicted by the ``mag model" and the ``flux model", we can see the inconsistency of the number of candidates in Table \ref{tab:probthres} and in Figure \ref{fig:diff}.}

Furthermore, by calculating the completeness of known high-redshift quasars in the test set at different thresholds, we provide a reference for the completeness of our high-redshift quasar candidates. The calculation of completeness is similar to recall. At different thresholds, if the probability of the high-z class obtained from the random forest is higher than the threshold, it is considered a correctly classified high-redshift quasar; otherwise, it is deemed a misclassified high-redshift quasar. The completeness of the high-z class is the ratio of correctly classified high-redshift quasars to the total number of high-redshift quasars in the test set. For the ``mag model", the completeness of the known high-redshift quasars in the test set at different thresholds is as follows: 
$p_{\rm quasar} \geq p_{\rm thre1,mag}: 82.20\%$; $p_{\rm quasar} \geq p_{\rm thre2,mag}: 33.05\%$; $p_{\rm quasar} \geq 0.90: 18.64\%$; $p_{\rm quasar} \geq 0.95: 5.09\%$. Similarly, for the ``flux model", the completeness is as follows: $p_{\rm quasar} \geq p_{\rm thre1,flux}: 79.49\%$; $p_{\rm quasar} \geq p_{\rm thre2,flux}: 41.88\%$; $p_{\rm quasar} \geq 0.90: 15.39\%$; $p_{\rm quasar} \geq 0.95: 0.00\%$.

\begin{table}[]
	\centering
	\begin{tabular}{lr}
		\toprule
		Column name                             & {Description } \\
		\toprule
		RA    & R.A. in LS catalog   \\
		DEC     	 & Decl. in LS catalog   \\
		mag\_g    	     & {\it g-} band magnitude in LS catalog   \\
		mag\_r    	     & {\it r-} band magnitude in LS catalog   \\
 		mag\_z    	     & {\it z-} band magnitude in LS catalog   \\                	mag\_W1    	     & {\it W1-} band magnitude in LS catalog   \\  
   		mag\_W2    	     & {\it W2-} band magnitude in LS catalog   \\
          mag\_grz    & constructed {\it grz-} band magnitude \\
  	mag\_W    	   & constructed {\it W-} band magnitude   \\
   	flux\_g    	  & {\it g-} band flux in LS catalog   \\
	flux\_r    	  & {\it r-} band flux in LS catalog   \\
 	flux\_z       & {\it z-} band flux in LS catalog   \\     flux\_W1    	  & {\it W1-} band flux in LS catalog   \\  
   flux\_W2    	   & {\it W2-} band flux in LS catalog   \\
    flux\_grz    & constructed {\it grz-} band flux \\
  	flux\_W    	   & constructed {\it W-} band flux   \\
    HighzProb   	   & Probability for high-z class  \\
    photo-z & High-redshift regression result \\
    spec-z & Redshift determined from spectrum \\

   \toprule
	\end{tabular}

\caption{\label{tab:catalog} List of columns of the dataset consisting of the high-z quasar candidates.}
\end{table}

\subsection{Verification using MUSE}

The Multi Unit Spectroscopic Explorer \citep[MUSE,][]{MUSE2010}, an integral field unit (IFU) mounted on the Very Large Telescope (VLT-{\it Yepun}, UT4), has a field of view (FOV) of 1 arcmin$^2$. The MUSE instrument provides high throughput (35\% end-to-end, including the telescope, at 7000 \AA), moderate spectral resolution (R $\simeq$ 3000 at $\sim$7000 \AA), full optical coverage 4650 $-$ 9300 \AA) spectroscopy
at a spatial sampling scale of 0.2 arcsec across a $1 \times 1$ arcmin$^2$ FOV. The wide wavelength coverage and moderate spectral resolution enable us to identify high-redshift quasars up to redshift of $\sim 6.6$ and suitable for our purpose of high-redshift quasar candidates verification.

So far there are nearly 20,000 publicly available MUSE datacubes, of which the sky coverage is approximately 6 square degrees. Among the entire high-redshift candidates, 21 of them are located within the MUSE field. After spectroscopic verification, 11 of these candidates are high-redshift quasars included in our training set, 3 are known high-redshift quasars but not present in our training sample. 
5 of these candidates are M dwarfs, while the rest two have spectra that are much noisier and are difficult to identify emission/absorption lines, making it hard to determine the type of those sources. By estimating their redshifts using Ly$\alpha$, N {\small V} or C {\small IV} emission lines, we find that these three ``new" high-redshift quasars precisely align with the redshift boundaries we set for high-redshift quasars. Two of them are having redshifts of $z \sim 5$ , and one is near the boundary of 6.5. Because they fall on the redshift boundaries,  we initially miss them when collecting the high-redshift quasars training sample. Since they are not part of our training sample, they provide us a valuable opportunity to validate our model.

\begin{figure}[ht!]
\includegraphics[width=.98\linewidth]{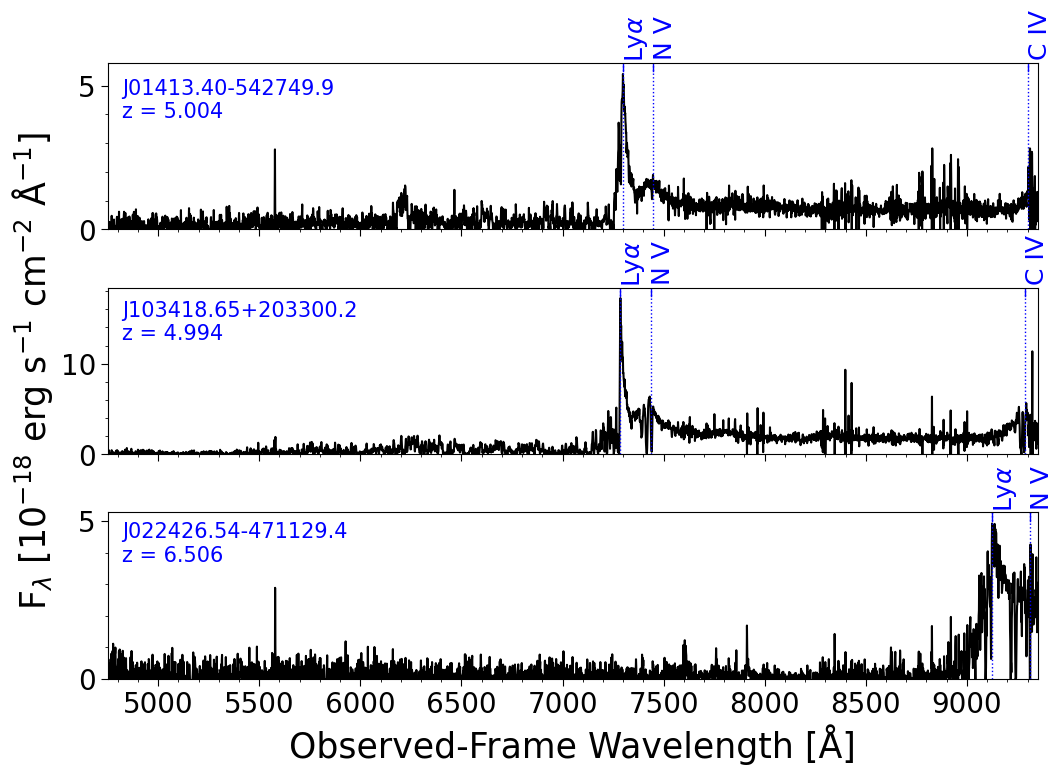}
\caption{The MUSE spectra of these three ``new" high-redshift quasars from our random forest model, with dashed vertical lines denoting the positions of the Ly$\alpha$, N V, and C IV emission lines.} 
\label{fig:sp_quasar_candidate}
\end{figure}

Overall, 14 out of 21 high-redshift quasar candidates from our random forest model has been confirmed to be true high-redshift quasars, reaching a success rate of 66.7\%. If we only focus on the high-redshift quasars not present in the training sample, the success rate is 30\% (3 ``new" high-redshift quasars out of 10 candidates). %Although the success rate using  MUSE data not present in the training sample is not high, it does not imply that our model is not good for a few reasons. First of all, the total sky coverage is only $\sim$ 6 square degrees, which results in a much larger scatter due to such a small sky coverage. Secondly, the data from MUSE observations might be beyond the color-space of the signal in the training sample due to the limited size of high-redshift quasars. Thirdly, the MUSE observations are picked for certain purposes, which might not be representative for high-redshift quasars verification. This issue is highly mitigated by using the DESI-EDR dataset, which will be discussed in the following.} 
Figure \ref{fig:sp_quasar_candidate} shows the MUSE spectra of these three ``new" high-redshift quasar obtained from our random forest model. The details of these three ``new" high-redshift quasars are the following:

$\textbf{J014132.4-542749.9}$. \cite{Belladitta2019} discovered this quasar by cross-matching the first data release of the Dark Energy Survey with the Sidney University Molonglo Survey radio catalog at 0.843 GHz. The {\it z} band magnitude of this quasar is 21.07, %and the spectroscopic redshift calculated from the Ly$\alpha$ emission line is 5.004. T
and the photometric redshift provided by our trained regression model is 5.087. \cite{Belladitta2019} utilized Gaussian profiles based on the Ly$\alpha$, O {\small VI}, N {\small V}, and C {\small IV} emissions to estimate the redshift of this quasar, with the value of 5.000 ± 0.002.

$\textbf{J103418.65+203300.2}$. \cite{Lyke2020} discovered this quasar in the SDSS Quasar Catalog. Its magnitude in the {\it z} band is 19.61. %We estimate the spectroscopic redshift of this quasar to be 4.994 based on the Ly$\alpha$ emission line, and t
The photometric redshift provided by our trained regression model is 5.214. Based on the emission lines of H$\alpha$, H$\beta$, Mg {\small II}, C {\small III}, C {\small IV}, and Ly$\alpha$,  the redshift of this quasar is estimated to be 5.0150 ± 0.0005 \citep{Lyke2020} .

$\textbf{J022426.54–471129.4}$. \cite{Reed2017} discovered this quasar using the European Southern Observatory New Technology Telescope and Gemini South telescopes. The {\it z} band magnitude of this quasar is 20.23, %and the spectroscopic redshift calculated from the Ly$\alpha$ emission line is 6.506. 
and the photometric redshift provided by our trained regression model is 6.139. The significant difference between the photometric and spectroscopic redshifts is due to the scarcity of samples with redshift around 6.5 in our training set. \cite{Reed2017} calculated the redshift of this quasar by fitting its spectrum with a known quasar model, obtaining a value of 6.50 ± 0.01.

\subsection{Verification using DESI}
DESI will collect millions of qusars in its 5-year operation \citep{DESI1, DESI2, DESIover, DESIvalid}, of which thousands of high-redshift quasars will be identified. Currently there are 306 objects at $z > 5$ in the DESI-EDR catalog \citep{DESI-EDR}. We then apply our trained RF classification model to these 306 objects at $z > 5$ in the DESI-EDR catalog to validate and evaluate our model. Before that, we first inspect the spectra of those objects visually. After spectra inspection, we find that only 22 of them are true high-redshift quasars at $z > 5$, 12 of which are in our training sample {and 10 of which are new high-redshift quasars}. The rest 284 objects are not high-redshift quasars after spectra inspection, which might be due to either the incorrect classification or the incorrect redshift determination by the DESI team.

This ensemble of objects at $z >5$ from DESI-EDR provides a perfect labratory to test and validate our random forest classification model. For the ``mag model", 188 out of 306 sources have no missing values, with only 6 out of the 22 true high-redshift quasars found among these 188 sources. 4 out of 188 objects are predicted to be high-redshift quasars, and these 4 objects are among the 6 true high-redshift quasars. The rest 2 true high-redshift quasars are not correctly classified: one is classified as a M dwarf, and the other is classified as a mid-z quasar. This implies that the precision is 1.00 and the recall is 0.67 for the ``mag model". Using the MICE imputation to fill in the missing values for the 118 sources, we find that  17 out of 118 sources are predicted to be high-redshift quasars. 16 out of the 17 candidates are true high-redshift quasars, and the type of the rest 1 source is suspicious because it is too faint without distinct emission or absorption lines. Overall, 21 objects are predicted to be high-redshift quasars, 20 of which are true high-redshift quasars. Therefore, the precision is 0.95 and the recall is 0.91 for the ``mag model".

For the ``flux model", all 306 sources have well-defined features. Among these sources, 21 objects are predicted to be high-redshift quasars, 20 of which are true high-redshift quasars. The rest 1 object is the same as the above one whose type is suspicious. The precision is 0.95, and the recall is 0.91 for the ``flux model". As discussed earlier, there exist a confusion between ``mid-z" and ``high-z" quasars because of the set of arbitrary redshift boundary, which should not be a concern. In this sample, one ``high-z" quasar with true redshift of 5.036 is predicted to be a ``mid-z" quasar, and we should not consider this as an error. This means that the precision and recall will be slightly higher, with a precision and recall of 0.96, 0.96 for both the ``mag model" and the ``flux model". Even if we only focus on the 10 new high-redshift quasars, 9 of which are successfully recovered. It turns out that our random forest classification model can achieve much higher success rate than the DESI target selection. The consistency between the ``mag model" and the ``flux model" further demonstrates the robustness and unbiasness of the imputation method we adopt.

Although the estimation of the precision and recall for the high-redshift quasars using the DESI-EDR database is biased because these objects are not randomly picked from our high-redshift quasar candidates and they are pre-selected by the DESI team, they can still shed a light on the robustness of our machine learning model. On the other hand, the DESI target selection of quasars is mainly for quasars at $0.9 < z < 2.1$ and seldom of these quasars are at $z > 2.1$ \citep{Chaussidon2023}. Moreover, the DESI target selection of high-redshift quasars is still based on the traditional color-cut method that has been adopted in previous high-redshift quasar surveys \citep{wang2016survey, yang2017discovery}. In this sense, these 306 objects extracted from  DESI-EDR database can be viewed as randomly picked objects from our high-redshift quasar candidates. Therefore, if our estimation of the precision and recall for the high-redshift quasars from DESI-EDR database is biased, they are least biased. The high success rate of our high-redshift quasar candidates using the objects at $z>5$ from DESI-EDR further validates our random forest classification model. Therefore, if we pick the high-redshift quasar candidates with high probability ($p_{\rm quasar} > 0.9$ for instance) from our classification model for follow-up spectroscopy observations, we will discover many more high-redshift quasars than expected.

\section{Summary} \label{sec:Summary}

In this work, we obtain a representative training sample which are all spectra-confirmed and it consists of 588 high-redshift quasars and millions of contaminators in the DESI Imaging Legacy Survey DR9 footprint. We then obtain the photometric data for the training sample from Legacy Survey DR9 and WISE survey and constructed various features to train a machine learning model. In addition to the photometric measurements in the {\it g, r, z, W1}, and {\it W2} bands, we also construct the combined photometric measurements of {\it grz} and {\it W}. We  implement the fluxes at different aperture radius provided by the LS DR9, the apfluxes are very effective in distinguishing high-redshift quasars from the extended sources like distant compact galaxies. And we demonstrate that the contamination from distant compact galaxies can be negligible via a binary classification for the first time. In addition, we address the issue of missing values in the known high-redshift quasar dataset. Among the various imputation methods, we find that the predictive results of the model trained with the MICE method are best consistent with the complete dataset.

We compare the performance of several different commonly used machine learning classification algorithms, including KNN, decision tree, random forest, LGBM, GaussianNB. The random forest classification model emerged as the most optimal algorithm. To further enhance the performance of the random forest classification model, we perform feature selections on the training sample. Additionally, we discuss whether the number of classes in the training sample affects the performance of the classification model. Considering the intrinsic characteristics of each class in the training sample, we decide to adopt the 11-class classification model as our final model. We discuss the issue of imbalance sample and find that different algorithms handling the imbalanced data sample do not show significant differences in all metrics compared to those with the imbalanced dataset.

The performance of the random forest with 11-class can reach a high precision of 96.43\%, a high recall of 91.53\% for the high-redshift quasars for the test set, both of which are significantly higher than the previous studies. In addition to the classification model, we also train a regression model on a dataset of quasars with redshift between 4.5 and 6.5 to predict the photometric redshift of the high-redshift candidates. Among the different machine learning regression algorithms (KNN, random forest, CatBoost), we find that the random forest regression model has the best performance, with 99.3\% data points within the range of $\frac{\Delta z}{1 + z_{\rm spec}} < 0.1$ and MSE $\le$ 0.025. %In addition, we discover that known high-redshift quasars may exhibit two subclass in color space, for example in the color space spanned by {\it z-W} and {\it z-grz}. This finding could potentially be related to the bimodal distribution of known high-redshift quasars in terms of their number density as a function of redshift.

We apply several selection criteria to reduce the catalog size of the entire Legacy Survey DR9 catalog. Applying the final 11-class models using random forest algorithm on FeatureSet-C (FeatureSet-D) to the 140 millions sources from LS DR9, we obtain 272,424 (568,188) high-redshift candidates. And there are 216,949 candidates which are classified as high-redshift quasars by both the ``mag model" and the ``flux model". We further narrow down the high-redhisft quasar candidates by setting two cutoff probabilities from the random forest model: the maximum probability of a known high-redshift quasar being misclassified as another class (40.82\%, 40.03\% for the ``mag model" and ``flux model", respectively) and the maximum probability of a contaminator being misclassified as a high-redshift quasar (82.49\%, 75.23\% for the ``mag model" and ``flux model", respectively). By applying these cutoff probabilities to the common high-redhisft quasar candidates identified by both the ``mag model" and the ``flux model", we obtain 165,734 and 2,984 high-redhisft quasar candidates, respectively. There are 476 candidates with probability greater than 90\%, which could be the targets of the highest priority for future spectroscopy followup.

Using MUSE spectra to do the spectra verification of the high-redshift quasar candidates, we find that there are 21 high-redshift quasar candidates in the total $\sim$ 20000 MUSE observations footprint. After inspection, we find that 11 of these candidates are high-redshift quasars already included in our training sample, 3 are known high-redshift quasars but not present in our training sample. Using DESI-EDR spectra, we confirm that 21 out of 22 true high-redshift quasars with correct redshift can be successfully identified for the ``mag model" with missing values imputed and 21 out of 22 true high-redshift quasars can also be successfully classified by the ``flux model", reaching much higher success rate than the DESI target selection.

Our current model still has  room for improvement. The future DESI survey will discover thousands of new high-redshift quasars, which will significantly expand the training sample size. The current photometric measurements available are very limited, therefore introducing new photometric measurements in the near- or mid-infrared could enhance the model performance significantly. The Legacy Survey DR10 currently includes the {\it i} band for $\sim$ half sky coverage of DR9, and the future data release including the {\it y} band is on the way as well. We find that the inclusion of {\it i} band photometric measurement improves the classification performance significantly, which demonstrates that more photometric measurements are critical to improving the machine learning performance. The future imaging surveys such as CSST, RST, EST and LSST, would deliver much deeper images with many more photometric measurements such as the {\it y, J, H} and {\it K} bands, which are beyond the current surveys and would be the frontier for high-redshift quasars searching with redshift even up to 7 or 8. %{\bf The apflux is essential for separating the high-redshift quasars from the extended sources.}

\section{Acknowledgments}

GY and HZ acknowledge financial support from the start-up funding of the Huazhong University of Science and Technology and the National Science Foundation of China grant (No. 12303007). QW acknowledges financial support from the National Science Foundation of China grant (No. 12233007). The authors thank the referees for comments that helped us improve the manuscript. This research was supported by the Munich Institute for Astro-, Particle and BioPhysics (MIAPbP) which is funded by the Deutsche Forschungsgemeinschaft (DFG, German Research Foundation) under Germany's Excellence Strategy EXC-2094 390783311.

The Legacy Surveys consist of three individual and complementary projects: the Dark Energy Camera Legacy Survey (DECaLS; Proposal ID \#2014B-0404; PIs: David Schlegel and Arjun Dey), the Beijing-Arizona Sky Survey (BASS; NOAO Prop. ID \#2015A-0801; PIs: Zhou Xu and Xiaohui Fan), and the Mayall z-band Legacy Survey (MzLS; Prop. ID \#2016A-0453; PI: Arjun Dey). DECaLS, BASS and MzLS together include data obtained, respectively, at the Blanco telescope, Cerro Tololo Inter-American Observatory, NSF's NOIRLab; the Bok telescope, Steward Observatory, University of Arizona; and the Mayall telescope, Kitt Peak National Observatory, NOIRLab. Pipeline processing and analyses of the data are supported by NOIRLab and the Lawrence Berkeley National Laboratory (LBNL). The Legacy Surveys project is honored to be permitted to conduct astronomical research on Iolkam Du'ag (Kitt Peak), a mountain with particular significance to the Tohono O'odham Nation.

NOIRLab is operated by the Association of Universities for Research in Astronomy (AURA) under a cooperative agreement with the National Science Foundation. LBNL is managed by the Regents of the University of California under contract to the U.S. Department of Energy.

This project used data obtained with the Dark Energy Camera (DECam), which was constructed by the Dark Energy Survey (DES) collaboration. Funding for the DES Projects has been provided by the U.S. Department of Energy, the U.S. National Science Foundation, the Ministry of Science and Education of Spain, the Science and Technology Facilities Council of the United Kingdom, the Higher Education Funding Council for England, the National Center for Supercomputing Applications at the University of Illinois at Urbana-Champaign, the Kavli Institute of Cosmological Physics at the University of Chicago, Center for Cosmology and Astro-Particle Physics at the Ohio State University, the Mitchell Institute for Fundamental Physics and Astronomy at Texas A\&M University, Financiadora de Estudos e Projetos, Fundacao Carlos Chagas Filho de Amparo, Financiadora de Estudos e Projetos, Fundacao Carlos Chagas Filho de Amparo a Pesquisa do Estado do Rio de Janeiro, Conselho Nacional de Desenvolvimento Cientifico e Tecnologico and the Ministerio da Ciencia, Tecnologia e Inovacao, the Deutsche Forschungsgemeinschaft and the Collaborating Institutions in the Dark Energy Survey. The Collaborating Institutions are Argonne National Laboratory, the University of California at Santa Cruz, the University of Cambridge, Centro de Investigaciones Energeticas, Medioambientales y Tecnologicas-Madrid, the University of Chicago, University College London, the DES-Brazil Consortium, the University of Edinburgh, the Eidgenossische Technische Hochschule (ETH) Zurich, Fermi National Accelerator Laboratory, the University of Illinois at Urbana-Champaign, the Institut de Ciencies de l'Espai (IEEC/CSIC), the Institut de Fisica d’Altes Energies, Lawrence Berkeley National Laboratory, the Ludwig Maximilians Universitat Munchen and the associated Excellence Cluster Universe, the University of Michigan, NSF’s NOIRLab, the University of Nottingham, the Ohio State University, the University of Pennsylvania, the University of Portsmouth, SLAC National Accelerator Laboratory, Stanford University, the University of Sussex, and Texas A\&M University.

BASS is a key project of the Telescope Access Program (TAP), which has been funded by the National Astronomical Observatories of China, the Chinese Academy of Sciences (the Strategic Priority Research Program “The Emergence of Cosmological Structures” Grant \#XDB09000000), and the Special Fund for Astronomy from the Ministry of Finance. The BASS is also supported by the External Cooperation Program of Chinese Academy of Sciences (Grant \#4A11KYSB20160057), and Chinese National Natural Science Foundation (Grant \#12120101003, \#11433005).

The Legacy Survey team makes use of data products from the Near-Earth Object Wide-field Infrared Survey Explorer (NEOWISE), which is a project of the Jet Propulsion Laboratory/California Institute of Technology. NEOWISE is funded by the National Aeronautics and Space Administration.

The Legacy Surveys imaging of the DESI footprint is supported by the Director, Office of Science, Office of High Energy Physics of the U.S. Department of Energy under Contract No. DE-AC02-05CH1123, by the National Energy Research Scientific Computing Center, a DOE Office of Science User Facility under the same contract; and by the U.S. National Science Foundation, Division of Astronomical Sciences under Contract No. AST-0950945 to NOAO.

This research used data obtained with the Dark Energy Spectroscopic Instrument (DESI). DESI construction and operations is managed by the Lawrence Berkeley National Laboratory. This material is based upon work supported by the U.S. Department of Energy, Office of Science, Office of High-Energy Physics, under Contract No. DE–AC02–05CH11231, and by the National Energy Research Scientific Computing Center, a DOE Office of Science User Facility under the same contract. Additional support for DESI was provided by the U.S. National Science Foundation (NSF), Division of Astronomical Sciences under Contract No. AST-0950945 to the NSF’s National Optical-Infrared Astronomy Research Laboratory; the Science and Technology Facilities Council of the United Kingdom; the Gordon and Betty Moore Foundation; the Heising-Simons Foundation; the French Alternative Energies and Atomic Energy Commission (CEA); the National Council of Science and Technology of Mexico (CONACYT); the Ministry of Science and Innovation of Spain (MICINN), and by the DESI Member Institutions: www.desi.lbl.gov/collaborating-institutions. The DESI collaboration is honored to be permitted to conduct scientific research on Iolkam Du’ag (Kitt Peak), a mountain with particular significance to the Tohono O’odham Nation. Any opinions, findings, and conclusions or recommendations expressed in this material are those of the author(s) and do not necessarily reflect the views of the U.S. National Science Foundation, the U.S. Department of Energy, or any of the listed funding agencies.

Based on observations collected at the European Southern Observatory under ESO programme(s) 109.238W.003, 105.208F.001, and 0104.A-0812(A), and/or data obtained from the ESO Science Archive Facility with DOI(s) under https://doi.org/10.18727/archive/41.

\bibliography{bibliography}

\begin{thebibliography}{}
\expandafter\ifx\csname natexlab\endcsname\relax\def\natexlab#1{#1}\fi
\providecommand{\url}[1]{\href{#1}{#1}}
\providecommand{\dodoi}[1]{doi:~\href{http://doi.org/#1}{\nolinkurl{#1}}}
\providecommand{\doeprint}[1]{\href{http://ascl.net/#1}{\nolinkurl{http://ascl.net/#1}}}
\providecommand{\doarXiv}[1]{\href{https://arxiv.org/abs/#1}{\nolinkurl{https://arxiv.org/abs/#1}}}

\bibitem[{{Abazajian} {et~al.}(2009){Abazajian}, {Adelman-McCarthy},
  {Ag{\"u}eros}, {Allam}, {Allende Prieto}, {An}, {Anderson}, {Anderson},
  {Annis}, {Bahcall}, {Bailer-Jones}, {Barentine}, {Bassett}, {Becker},
  {Beers}, {Bell}, {Belokurov}, {Berlind}, {Berman}, {Bernardi}, {Bickerton},
  {Bizyaev}, {Blakeslee}, {Blanton}, {Bochanski}, {Boroski}, {Brewington},
  {Brinchmann}, {Brinkmann}, {Brunner}, {Budav{\'a}ri}, {Carey}, {Carliles},
  {Carr}, {Castander}, {Cinabro}, {Connolly}, {Csabai}, {Cunha}, {Czarapata},
  {Davenport}, {de Haas}, {Dilday}, {Doi}, {Eisenstein}, {Evans}, {Evans},
  {Fan}, {Friedman}, {Frieman}, {Fukugita}, {G{\"a}nsicke}, {Gates},
  {Gillespie}, {Gilmore}, {Gonzalez}, {Gonzalez}, {Grebel}, {Gunn},
  {Gy{\"o}ry}, {Hall}, {Harding}, {Harris}, {Harvanek}, {Hawley}, {Hayes},
  {Heckman}, {Hendry}, {Hennessy}, {Hindsley}, {Hoblitt}, {Hogan}, {Hogg},
  {Holtzman}, {Hyde}, {Ichikawa}, {Ichikawa}, {Im}, {Ivezi{\'c}}, {Jester},
  {Jiang}, {Johnson}, {Jorgensen}, {Juri{\'c}}, {Kent}, {Kessler}, {Kleinman},
  {Knapp}, {Konishi}, {Kron}, {Krzesinski}, {Kuropatkin}, {Lampeitl},
  {Lebedeva}, {Lee}, {Lee}, {French Leger}, {L{\'e}pine}, {Li}, {Lima}, {Lin},
  {Long}, {Loomis}, {Loveday}, {Lupton}, {Magnier}, {Malanushenko},
  {Malanushenko}, {Mandelbaum}, {Margon}, {Marriner}, {Mart{\'\i}nez-Delgado},
  {Matsubara}, {McGehee}, {McKay}, {Meiksin}, {Morrison}, {Mullally}, {Munn},
  {Murphy}, {Nash}, {Nebot}, {Neilsen}, {Newberg}, {Newman}, {Nichol},
  {Nicinski}, {Nieto-Santisteban}, {Nitta}, {Okamura}, {Oravetz}, {Ostriker},
  {Owen}, {Padmanabhan}, {Pan}, {Park}, {Pauls}, {Peoples}, {Percival}, {Pier},
  {Pope}, {Pourbaix}, {Price}, {Purger}, {Quinn}, {Raddick}, {Re Fiorentin},
  {Richards}, {Richmond}, {Riess}, {Rix}, {Rockosi}, {Sako}, {Schlegel},
  {Schneider}, {Scholz}, {Schreiber}, {Schwope}, {Seljak}, {Sesar}, {Sheldon},
  {Shimasaku}, {Sibley}, {Simmons}, {Sivarani}, {Allyn Smith}, {Smith},
  {Smol{\v{c}}i{\'c}}, {Snedden}, {Stebbins}, {Steinmetz}, {Stoughton},
  {Strauss}, {SubbaRao}, {Suto}, {Szalay}, {Szapudi}, {Szkody}, {Tanaka},
  {Tegmark}, {Teodoro}, {Thakar}, {Tremonti}, {Tucker}, {Uomoto}, {Vanden
  Berk}, {Vandenberg}, {Vidrih}, {Vogeley}, {Voges}, {Vogt}, {Wadadekar},
  {Watters}, {Weinberg}, {West}, {White}, {Wilhite}, {Wonders}, {Yanny},
  {Yocum}, {York}, {Zehavi}, {Zibetti}, \& {Zucker}}]{Abazajian2009}
{Abazajian}, K.~N., {Adelman-McCarthy}, J.~K., {Ag{\"u}eros}, M.~A., {et~al.}
  2009, \apjs, 182, 543, \dodoi{10.1088/0067-0049/182/2/543}

\bibitem[{{Abdurro'uf} {et~al.}(2022){Abdurro'uf}, {Accetta}, {Aerts}, {Silva
  Aguirre}, {Ahumada}, {Ajgaonkar}, {Filiz Ak}, {Alam}, {Allende Prieto},
  {Almeida}, {Anders}, {Anderson}, {Andrews}, {Anguiano}, {Aquino-Ort{\'\i}z},
  {Arag{\'o}n-Salamanca}, {Argudo-Fern{\'a}ndez}, {Ata}, {Aubert},
  {Avila-Reese}, {Badenes}, {Barb{\'a}}, {Barger}, {Barrera-Ballesteros},
  {Beaton}, {Beers}, {Belfiore}, {Bender}, {Bernardi}, {Bershady}, {Beutler},
  {Bidin}, {Bird}, {Bizyaev}, {Blanc}, {Blanton}, {Boardman}, {Bolton},
  {Boquien}, {Borissova}, {Bovy}, {Brandt}, {Brown}, {Brownstein}, {Brusa},
  {Buchner}, {Bundy}, {Burchett}, {Bureau}, {Burgasser}, {Cabang}, {Campbell},
  {Cappellari}, {Carlberg}, {Wanderley}, {Carrera}, {Cash}, {Chen}, {Chen},
  {Cherinka}, {Chiappini}, {Choi}, {Chojnowski}, {Chung}, {Clerc}, {Cohen},
  {Comerford}, {Comparat}, {da Costa}, {Covey}, {Crane}, {Cruz-Gonzalez},
  {Culhane}, {Cunha}, {Dai}, {Damke}, {Darling}, {Davidson}, {Davies},
  {Dawson}, {De Lee}, {Diamond-Stanic}, {Cano-D{\'\i}az}, {S{\'a}nchez},
  {Donor}, {Duckworth}, {Dwelly}, {Eisenstein}, {Elsworth}, {Emsellem},
  {Eracleous}, {Escoffier}, {Fan}, {Farr}, {Feng}, {Fern{\'a}ndez-Trincado},
  {Feuillet}, {Filipp}, {Fillingham}, {Frinchaboy}, {Fromenteau}, {Galbany},
  {Garc{\'\i}a}, {Garc{\'\i}a-Hern{\'a}ndez}, {Ge}, {Geisler}, {Gelfand},
  {G{\'e}ron}, {Gibson}, {Goddy}, {Godoy-Rivera}, {Grabowski}, {Green},
  {Greener}, {Grier}, {Griffith}, {Guo}, {Guy}, {Hadjara}, {Harding},
  {Hasselquist}, {Hayes}, {Hearty}, {Hern{\'a}ndez}, {Hill}, {Hogg},
  {Holtzman}, {Horta}, {Hsieh}, {Hsu}, {Hsu}, {Huber}, {Huertas-Company},
  {Hutchinson}, {Hwang}, {Ibarra-Medel}, {Chitham}, {Ilha}, {Imig}, {Jaekle},
  {Jayasinghe}, {Ji}, {Johnson}, {Jones}, {J{\"o}nsson}, {Katkov}, {Khalatyan},
  {Kinemuchi}, {Kisku}, {Knapen}, {Kneib}, {Kollmeier}, {Kong}, {Kounkel},
  {Kreckel}, {Krishnarao}, {Lacerna}, {Lane}, {Langgin}, {Lavender}, {Law},
  {Lazarz}, {Leung}, {Leung}, {Lewis}, {Li}, {Li}, {Lian}, {Liang}, {Lin},
  {Lin}, {Lin}, {Lintott}, {Long}, {Longa-Pe{\~n}a}, {L{\'o}pez-Cob{\'a}},
  {Lu}, {Lundgren}, {Luo}, {Mackereth}, {de la Macorra}, {Mahadevan},
  {Majewski}, {Manchado}, {Mandeville}, {Maraston}, {Margalef-Bentabol},
  {Masseron}, {Masters}, {Mathur}, {McDermid}, {Mckay}, {Merloni},
  {Merrifield}, {Meszaros}, {Miglio}, {Di Mille}, {Minniti}, {Minsley},
  {Monachesi}, {Moon}, {Mosser}, {Mulchaey}, {Muna}, {Mu{\~n}oz}, {Myers},
  {Myers}, {Nadathur}, {Nair}, {Nandra}, {Neumann}, {Newman}, {Nidever},
  {Nikakhtar}, {Nitschelm}, {O'Connell}, {Garma-Oehmichen}, {Luan Souza de
  Oliveira}, {Olney}, {Oravetz}, {Ortigoza-Urdaneta}, {Osorio}, {Otter},
  {Pace}, {Padilla}, {Pan}, {Pan}, {Parikh}, {Parker}, {Peirani}, {Pe{\~n}a
  Ram{\'\i}rez}, {Penny}, {Percival}, {Perez-Fournon}, {Pinsonneault},
  {Poidevin}, {Poovelil}, {Price-Whelan}, {B{\'a}rbara de Andrade Queiroz},
  {Raddick}, {Ray}, {Rembold}, {Riddle}, {Riffel}, {Riffel}, {Rix}, {Robin},
  {Rodr{\'\i}guez-Puebla}, {Roman-Lopes}, {Rom{\'a}n-Z{\'u}{\~n}iga}, {Rose},
  {Ross}, {Rossi}, {Rubin}, {Salvato}, {S{\'a}nchez}, {S{\'a}nchez-Gallego},
  {Sanderson}, {Santana Rojas}, {Sarceno}, {Sarmiento}, {Sayres}, {Sazonova},
  {Schaefer}, {Schiavon}, {Schlegel}, {Schneider}, {Schultheis}, {Schwope},
  {Serenelli}, {Serna}, {Shao}, {Shapiro}, {Sharma}, {Shen}, {Shetrone}, {Shu},
  {Simon}, {Skrutskie}, {Smethurst}, {Smith}, {Sobeck}, {Spoo}, {Sprague},
  {Stark}, {Stassun}, {Steinmetz}, {Stello}, {Stone-Martinez},
  {Storchi-Bergmann}, {Stringfellow}, {Stutz}, {Su}, {Taghizadeh-Popp},
  {Talbot}, {Tayar}, {Telles}, {Teske}, {Thakar}, {Theissen}, {Tkachenko},
  {Thomas}, {Tojeiro}, {Hernandez Toledo}, {Troup}, {Trump}, {Trussler},
  {Turner}, {Tuttle}, {Unda-Sanzana}, {V{\'a}zquez-Mata}, {Valentini},
  {Valenzuela}, {Vargas-Gonz{\'a}lez}, {Vargas-Maga{\~n}a}, {Alfaro},
  {Villanova}, {Vincenzo}, {Wake}, {Warfield}, {Washington}, {Weaver},
  {Weijmans}, {Weinberg}, {Weiss}, {Westfall}, {Wild}, {Wilde}, {Wilson},
  {Wilson}, {Wilson}, {Wolf}, {Wood-Vasey}, {Yan}, {Zamora}, {Zasowski},
  {Zhang}, {Zhao}, {Zheng}, {Zheng}, \& {Zhu}}]{Abdurro2022}
{Abdurro'uf}, {Accetta}, K., {Aerts}, C., {et~al.} 2022, \apjs, 259, 35,
  \dodoi{10.3847/1538-4365/ac4414}

\bibitem[{{Ahumada} {et~al.}(2020){Ahumada}, {Allende Prieto}, {Almeida},
  {Anders}, {Anderson}, {Andrews}, {Anguiano}, {Arcodia}, {Armengaud},
  {Aubert}, {Avila}, {Avila-Reese}, {Badenes}, {Balland}, {Barger},
  {Barrera-Ballesteros}, {Basu}, {Bautista}, {Beaton}, {Beers}, {Benavides},
  {Bender}, {Bernardi}, {Bershady}, {Beutler}, {Bidin}, {Bird}, {Bizyaev},
  {Blanc}, {Blanton}, {Boquien}, {Borissova}, {Bovy}, {Brandt}, {Brinkmann},
  {Brownstein}, {Bundy}, {Bureau}, {Burgasser}, {Burtin}, {Cano-D{\'\i}az},
  {Capasso}, {Cappellari}, {Carrera}, {Chabanier}, {Chaplin}, {Chapman},
  {Cherinka}, {Chiappini}, {Doohyun Choi}, {Chojnowski}, {Chung}, {Clerc},
  {Coffey}, {Comerford}, {Comparat}, {da Costa}, {Cousinou}, {Covey}, {Crane},
  {Cunha}, {Ilha}, {Dai}, {Damsted}, {Darling}, {Davidson}, {Davies}, {Dawson},
  {De}, {de la Macorra}, {De Lee}, {Queiroz}, {Deconto Machado}, {de la Torre},
  {Dell'Agli}, {du Mas des Bourboux}, {Diamond-Stanic}, {Dillon}, {Donor},
  {Drory}, {Duckworth}, {Dwelly}, {Ebelke}, {Eftekharzadeh}, {Davis Eigenbrot},
  {Elsworth}, {Eracleous}, {Erfanianfar}, {Escoffier}, {Fan}, {Farr},
  {Fern{\'a}ndez-Trincado}, {Feuillet}, {Finoguenov}, {Fofie},
  {Fraser-McKelvie}, {Frinchaboy}, {Fromenteau}, {Fu}, {Galbany}, {Garcia},
  {Garc{\'\i}a-Hern{\'a}ndez}, {Garma Oehmichen}, {Ge}, {Geimba Maia},
  {Geisler}, {Gelfand}, {Goddy}, {Gonzalez-Perez}, {Grabowski}, {Green},
  {Grier}, {Guo}, {Guy}, {Harding}, {Hasselquist}, {Hawken}, {Hayes}, {Hearty},
  {Hekker}, {Hogg}, {Holtzman}, {Horta}, {Hou}, {Hsieh}, {Huber}, {Hunt}, {Ider
  Chitham}, {Imig}, {Jaber}, {Jimenez Angel}, {Johnson}, {Jones},
  {J{\"o}nsson}, {Jullo}, {Kim}, {Kinemuchi}, {Kirkpatrick}, {Kite}, {Klaene},
  {Kneib}, {Kollmeier}, {Kong}, {Kounkel}, {Krishnarao}, {Lacerna}, {Lan},
  {Lane}, {Law}, {Le Goff}, {Leung}, {Lewis}, {Li}, {Lian}, {Lin}, {Long},
  {Longa-Pe{\~n}a}, {Lundgren}, {Lyke}, {Mackereth}, {MacLeod}, {Majewski},
  {Manchado}, {Maraston}, {Martini}, {Masseron}, {Masters}, {Mathur},
  {McDermid}, {Merloni}, {Merrifield}, {M{\'e}sz{\'a}ros}, {Miglio}, {Minniti},
  {Minsley}, {Miyaji}, {Mohammad}, {Mosser}, {Mueller}, {Muna},
  {Mu{\~n}oz-Guti{\'e}rrez}, {Myers}, {Nadathur}, {Nair}, {Nandra}, {Correa do
  Nascimento}, {Nevin}, {Newman}, {Nidever}, {Nitschelm}, {Noterdaeme},
  {O'Connell}, {Olmstead}, {Oravetz}, {Oravetz}, {Osorio}, {Pace}, {Padilla},
  {Palanque-Delabrouille}, {Palicio}, {Pan}, {Pan}, {Parker}, {Paviot},
  {Peirani}, {Ram{\'r}ez}, {Penny}, {Percival}, {Perez-Fournon},
  {P{\'e}rez-R{\`a}fols}, {Petitjean}, {Pieri}, {Pinsonneault}, {Poovelil},
  {Povick}, {Prakash}, {Price-Whelan}, {Raddick}, {Raichoor}, {Ray}, {Rembold},
  {Rezaie}, {Riffel}, {Riffel}, {Rix}, {Robin}, {Roman-Lopes},
  {Rom{\'a}n-Z{\'u}{\~n}iga}, {Rose}, {Ross}, {Rossi}, {Rowlands}, {Rubin},
  {Salvato}, {S{\'a}nchez}, {S{\'a}nchez-Menguiano}, {S{\'a}nchez-Gallego},
  {Sayres}, {Schaefer}, {Schiavon}, {Schimoia}, {Schlafly}, {Schlegel},
  {Schneider}, {Schultheis}, {Schwope}, {Seo}, {Serenelli}, {Shafieloo},
  {Shamsi}, {Shao}, {Shen}, {Shetrone}, {Shirley}, {Silva Aguirre}, {Simon},
  {Skrutskie}, {Slosar}, {Smethurst}, {Sobeck}, {Sodi}, {Souto}, {Stark},
  {Stassun}, {Steinmetz}, {Stello}, {Stermer}, {Storchi-Bergmann},
  {Streblyanska}, {Stringfellow}, {Stutz}, {Su{\'a}rez}, {Sun},
  {Taghizadeh-Popp}, {Talbot}, {Tayar}, {Thakar}, {Theriault}, {Thomas},
  {Thomas}, {Tinker}, {Tojeiro}, {Toledo}, {Tremonti}, {Troup}, {Tuttle},
  {Unda-Sanzana}, {Valentini}, {Vargas-Gonz{\'a}lez}, {Vargas-Maga{\~n}a},
  {V{\'a}zquez-Mata}, {Vivek}, {Wake}, {Wang}, {Weaver}, {Weijmans}, {Wild},
  {Wilson}, {Wilson}, {Wolthuis}, {Wood-Vasey}, {Yan}, {Yang}, {Y{\`e}che},
  {Zamora}, {Zarrouk}, {Zasowski}, {Zhang}, {Zhao}, {Zhao}, {Zheng}, {Zheng},
  {Zhu}, \& {Zou}}]{Ahumada2020}
{Ahumada}, R., {Allende Prieto}, C., {Almeida}, A., {et~al.} 2020, \apjs, 249,
  3, \dodoi{10.3847/1538-4365/ab929e}

\bibitem[{{Aihara} {et~al.}(2011){Aihara}, {Allende Prieto}, {An}, {Anderson},
  {Aubourg}, {Balbinot}, {Beers}, {Berlind}, {Bickerton}, {Bizyaev}, {Blanton},
  {Bochanski}, {Bolton}, {Bovy}, {Brandt}, {Brinkmann}, {Brown}, {Brownstein},
  {Busca}, {Campbell}, {Carr}, {Chen}, {Chiappini}, {Comparat}, {Connolly},
  {Cortes}, {Croft}, {Cuesta}, {da Costa}, {Davenport}, {Dawson}, {Dhital},
  {Ealet}, {Ebelke}, {Edmondson}, {Eisenstein}, {Escoffier}, {Esposito},
  {Evans}, {Fan}, {Femen{\'\i}a Castell{\'a}}, {Font-Ribera}, {Frinchaboy},
  {Ge}, {Gillespie}, {Gilmore}, {Gonz{\'a}lez Hern{\'a}ndez}, {Gott}, {Gould},
  {Grebel}, {Gunn}, {Hamilton}, {Harding}, {Harris}, {Hawley}, {Hearty}, {Ho},
  {Hogg}, {Holtzman}, {Honscheid}, {Inada}, {Ivans}, {Jiang}, {Johnson},
  {Jordan}, {Jordan}, {Kazin}, {Kirkby}, {Klaene}, {Knapp}, {Kneib},
  {Kochanek}, {Koesterke}, {Kollmeier}, {Kron}, {Lampeitl}, {Lang}, {Le Goff},
  {Lee}, {Lin}, {Long}, {Loomis}, {Lucatello}, {Lundgren}, {Lupton}, {Ma},
  {MacDonald}, {Mahadevan}, {Maia}, {Makler}, {Malanushenko}, {Malanushenko},
  {Mandelbaum}, {Maraston}, {Margala}, {Masters}, {McBride}, {McGehee},
  {McGreer}, {M{\'e}nard}, {Miralda-Escud{\'e}}, {Morrison}, {Mullally},
  {Muna}, {Munn}, {Murayama}, {Myers}, {Naugle}, {Neto}, {Nguyen}, {Nichol},
  {O'Connell}, {Ogando}, {Olmstead}, {Oravetz}, {Padmanabhan},
  {Palanque-Delabrouille}, {Pan}, {Pandey}, {P{\^a}ris}, {Percival},
  {Petitjean}, {Pfaffenberger}, {Pforr}, {Phleps}, {Pichon}, {Pieri}, {Prada},
  {Price-Whelan}, {Raddick}, {Ramos}, {Reyl{\'e}}, {Rich}, {Richards}, {Rix},
  {Robin}, {Rocha-Pinto}, {Rockosi}, {Roe}, {Rollinde}, {Ross}, {Ross},
  {Rossetto}, {S{\'a}nchez}, {Sayres}, {Schlegel}, {Schlesinger}, {Schmidt},
  {Schneider}, {Sheldon}, {Shu}, {Simmerer}, {Simmons}, {Sivarani}, {Snedden},
  {Sobeck}, {Steinmetz}, {Strauss}, {Szalay}, {Tanaka}, {Thakar}, {Thomas},
  {Tinker}, {Tofflemire}, {Tojeiro}, {Tremonti}, {Vandenberg}, {Vargas
  Maga{\~n}a}, {Verde}, {Vogt}, {Wake}, {Wang}, {Weaver}, {Weinberg}, {White},
  {White}, {Yanny}, {Yasuda}, {Yeche}, \& {Zehavi}}]{Aihara2011}
{Aihara}, H., {Allende Prieto}, C., {An}, D., {et~al.} 2011, \apjs, 193, 29,
  \dodoi{10.1088/0067-0049/193/2/29}

\bibitem[{{Akeson} {et~al.}(2019){Akeson}, {Armus}, {Bachelet}, {Bailey},
  {Bartusek}, {Bellini}, {Benford}, {Bennett}, {Bhattacharya}, {Bohlin},
  {Boyer}, {Bozza}, {Bryden}, {Calchi Novati}, {Carpenter}, {Casertano},
  {Choi}, {Content}, {Dayal}, {Dressler}, {Dor{\'e}}, {Fall}, {Fan}, {Fang},
  {Filippenko}, {Finkelstein}, {Foley}, {Furlanetto}, {Kalirai}, {Gaudi},
  {Gilbert}, {Girard}, {Grady}, {Greene}, {Guhathakurta}, {Heinrich},
  {Hemmati}, {Hendel}, {Henderson}, {Henning}, {Hirata}, {Ho}, {Huff},
  {Hutter}, {Jansen}, {Jha}, {Johnson}, {Jones}, {Kasdin}, {Kelly}, {Kirshner},
  {Koekemoer}, {Kruk}, {Lewis}, {Macintosh}, {Madau}, {Malhotra}, {Mandel},
  {Massara}, {Masters}, {McEnery}, {McQuinn}, {Melchior}, {Melton},
  {Mennesson}, {Peeples}, {Penny}, {Perlmutter}, {Pisani}, {Plazas}, {Poleski},
  {Postman}, {Ranc}, {Rauscher}, {Rest}, {Roberge}, {Robertson}, {Rodney},
  {Rhoads}, {Rhodes}, {Ryan}, {Sahu}, {Sand}, {Scolnic}, {Seth}, {Shvartzvald},
  {Siellez}, {Smith}, {Spergel}, {Stassun}, {Street}, {Strolger}, {Szalay},
  {Trauger}, {Troxel}, {Turnbull}, {van der Marel}, {von der Linden}, {Wang},
  {Weinberg}, {Williams}, {Windhorst}, {Wollack}, {Wu}, {Yee}, \&
  {Zimmerman}}]{Akeson2019}
{Akeson}, R., {Armus}, L., {Bachelet}, E., {et~al.} 2019, arXiv e-prints,
  arXiv:1902.05569, \dodoi{10.48550/arXiv.1902.05569}

\bibitem[{Akosa(2017)}]{Akosa2017}
Akosa, J.~S. 2017, in , Vol.~12.
\newblock
  \url{http://support.sas.com/resources/papers/proceedings17/0942-2017.pdf}

\bibitem[{Azur {et~al.}(2011)Azur, Stuart, Frangakis, \& Leaf}]{Azur2011}
Azur, M.~J., Stuart, E.~A., Frangakis, C., \& Leaf, P.~J. 2011, International
  Journal of Methods in Psychiatric Research, 20, \dodoi{10.1002/mpr.329}

\bibitem[{{Bacon} {et~al.}(2010){Bacon}, {Accardo}, {Adjali}, {Anwand},
  {Bauer}, {Biswas}, {Blaizot}, {Boudon}, {Brau-Nogue}, {Brinchmann},
  {Caillier}, {Capoani}, {Carollo}, {Contini}, {Couderc}, {Daguis{\'e}},
  {Deiries}, \& {Delabre}}]{MUSE2010}
{Bacon}, R., {Accardo}, M., {Adjali}, L., {et~al.} 2010, in Society of
  Photo-Optical Instrumentation Engineers (SPIE) Conference Series, Vol. 7735,
  Ground-based and Airborne Instrumentation for Astronomy III, ed. I.~S.
  {McLean}, S.~K. {Ramsay}, \& H.~{Takami}, 773508, \dodoi{10.1117/12.856027}

\bibitem[{{Ball} {et~al.}(2007){Ball}, {Brunner}, \& {Myers}}]{Ball2007}
{Ball}, N.~M., {Brunner}, R.~J., \& {Myers}, A.~D. 2007, arXiv e-prints,
  arXiv:0710.4482, \dodoi{10.48550/arXiv.0710.4482}

\bibitem[{Ba{\~n}ados {et~al.}(2016)Ba{\~n}ados, Venemans, Decarli, Farina,
  Mazzucchelli, Walter, Fan, Stern, Schlafly, Chambers,
  {et~al.}}]{banados2016pan}
Ba{\~n}ados, E., Venemans, B., Decarli, R., {et~al.} 2016, The Astrophysical
  Journal Supplement Series, 227, 11

\bibitem[{Barnett {et~al.}(2019)Barnett, Warren, Mortlock, Cuby, Conselice,
  Hewett, Willott, Auricchio, Balaguera-Antol{\'\i}nez, Baldi,
  {et~al.}}]{barnett2019euclid}
Barnett, R., Warren, S., Mortlock, D.~J., {et~al.} 2019, Astronomy \&
  Astrophysics, 631, A85

\bibitem[{Becker {et~al.}(2007)Becker, Rauch, \& Sargent}]{becker2007evolution}
Becker, G.~D., Rauch, M., \& Sargent, W.~L. 2007, The Astrophysical Journal,
  662, 72

\bibitem[{{Belladitta} {et~al.}(2019){Belladitta}, {Moretti}, {Caccianiga},
  {Ghisellini}, {Cicone}, {Sbarrato}, {Ighina}, \& {Pedani}}]{Belladitta2019}
{Belladitta}, S., {Moretti}, A., {Caccianiga}, A., {et~al.} 2019, \aap, 629,
  A68, \dodoi{10.1051/0004-6361/201935965}

\bibitem[{Best {et~al.}(2020)Best, Dupuy, Liu, Siverd, \& Zhang}]{Best2020}
Best, W. M.~J., Dupuy, T.~J., Liu, M.~C., Siverd, R.~J., \& Zhang, Z. 2020,
  \dodoi{10.5281/zenodo.4169085}

\bibitem[{Bosman {et~al.}(2018)Bosman, Fan, Jiang, Reed, Matsuoka, Becker, \&
  Haehnelt}]{bosman2018new}
Bosman, S.~E., Fan, X., Jiang, L., {et~al.} 2018, Monthly Notices of the Royal
  Astronomical Society, 479, 1055

\bibitem[{Bosman {et~al.}(2022)Bosman, Davies, Becker, Keating, Davies, Zhu,
  Eilers, D’Odorico, Bian, Bischetti, Cristiani, Fan, Farina, Haehnelt,
  Hennawi, Kulkarni, Mesinger, Meyer, Onoue, Pallottini, Qin, Ryan-Weber,
  Schindler, Walter, Wang, \& Yang}]{bosman2022hydrogen}
Bosman, S. E.~I., Davies, F.~B., Becker, G.~D., {et~al.} 2022, Monthly Notices
  of the Royal Astronomical Society, 514, 55, \dodoi{10.1093/mnras/stac1046}

\bibitem[{{Bovy} {et~al.}(2011){Bovy}, {Hennawi}, {Hogg}, {Myers},
  {Kirkpatrick}, {Schlegel}, {Ross}, {Sheldon}, {McGreer}, {Schneider}, \&
  {Weaver}}]{Bovy2011}
{Bovy}, J., {Hennawi}, J.~F., {Hogg}, D.~W., {et~al.} 2011, \apj, 729, 141,
  \dodoi{10.1088/0004-637X/729/2/141}

\bibitem[{{Bovy} {et~al.}(2012){Bovy}, {Myers}, {Hennawi}, {Hogg}, {McMahon},
  {Schiminovich}, {Sheldon}, {Brinkmann}, {Schneider}, \& {Weaver}}]{Bovy2012}
{Bovy}, J., {Myers}, A.~D., {Hennawi}, J.~F., {et~al.} 2012, \apj, 749, 41,
  \dodoi{10.1088/0004-637X/749/1/41}

\bibitem[{{Brammer} {et~al.}(2008){Brammer}, {van Dokkum}, \&
  {Coppi}}]{Brammer2008}
{Brammer}, G.~B., {van Dokkum}, P.~G., \& {Coppi}, P. 2008, \apj, 686, 1503,
  \dodoi{10.1086/591786}

\bibitem[{{Carliles} {et~al.}(2010){Carliles}, {Budav{\'a}ri}, {Heinis},
  {Priebe}, \& {Szalay}}]{Carliles2010}
{Carliles}, S., {Budav{\'a}ri}, T., {Heinis}, S., {Priebe}, C., \& {Szalay},
  A.~S. 2010, \apj, 712, 511, \dodoi{10.1088/0004-637X/712/1/511}

\bibitem[{{Carrasco Kind} \& {Brunner}(2013)}]{Carrasco2013}
{Carrasco Kind}, M., \& {Brunner}, R.~J. 2013, \mnras, 432, 1483,
  \dodoi{10.1093/mnras/stt574}

\bibitem[{{Chambers} {et~al.}(2016){Chambers}, {Magnier}, {Metcalfe},
  {Flewelling}, {Huber}, {Waters}, {Denneau}, {Draper}, {Farrow}, {Finkbeiner},
  {Holmberg}, {Koppenhoefer}, {Price}, {Rest}, {Saglia}, {Schlafly}, {Smartt},
  {Sweeney}, {Wainscoat}, {Burgett}, {Chastel}, {Grav}, {Heasley}, {Hodapp},
  {Jedicke}, {Kaiser}, {Kudritzki}, {Luppino}, {Lupton}, {Monet}, {Morgan},
  {Onaka}, {Shiao}, {Stubbs}, {Tonry}, {White}, {Ba{\~n}ados}, {Bell},
  {Bender}, {Bernard}, {Boegner}, {Boffi}, {Botticella}, {Calamida},
  {Casertano}, {Chen}, {Chen}, {Cole}, {Deacon}, {Frenk}, {Fitzsimmons},
  {Gezari}, {Gibbs}, {Goessl}, {Goggia}, {Gourgue}, {Goldman}, {Grant},
  {Grebel}, {Hambly}, {Hasinger}, {Heavens}, {Heckman}, {Henderson}, {Henning},
  {Holman}, {Hopp}, {Ip}, {Isani}, {Jackson}, {Keyes}, {Koekemoer}, {Kotak},
  {Le}, {Liska}, {Long}, {Lucey}, {Liu}, {Martin}, {Masci}, {McLean}, {Mindel},
  {Misra}, {Morganson}, {Murphy}, {Obaika}, {Narayan}, {Nieto-Santisteban},
  {Norberg}, {Peacock}, {Pier}, {Postman}, {Primak}, {Rae}, {Rai}, {Riess},
  {Riffeser}, {Rix}, {R{\"o}ser}, {Russel}, {Rutz}, {Schilbach}, {Schultz},
  {Scolnic}, {Strolger}, {Szalay}, {Seitz}, {Small}, {Smith}, {Soderblom},
  {Taylor}, {Thomson}, {Taylor}, {Thakar}, {Thiel}, {Thilker}, {Unger},
  {Urata}, {Valenti}, {Wagner}, {Walder}, {Walter}, {Watters}, {Werner},
  {Wood-Vasey}, \& {Wyse}}]{Chambers2016}
{Chambers}, K.~C., {Magnier}, E.~A., {Metcalfe}, N., {et~al.} 2016, arXiv
  e-prints, arXiv:1612.05560, \dodoi{10.48550/arXiv.1612.05560}

\bibitem[{{Chaussidon} {et~al.}(2023){Chaussidon}, {Y{\`e}che},
  {Palanque-Delabrouille}, {Alexander}, {Yang}, {Ahlen}, {Bailey}, {Brooks},
  {Cai}, {Chabanier}, {Davis}, {Dawson}, {de laMacorra}, {Dey}, {Dey},
  {Eftekharzadeh}, {Eisenstein}, {Fanning}, {Font-Ribera}, {Gazta{\~n}aga}, {A
  Gontcho}, {Gonzalez-Morales}, {Guy}, {Herrera-Alcantar}, {Honscheid},
  {Ishak}, {Jiang}, {Juneau}, {Kehoe}, {Kisner}, {Kov{\'a}cs}, {Kremin}, {Lan},
  {Landriau}, {Le Guillou}, {Levi}, {Magneville}, {Martini}, {Meisner},
  {Moustakas}, {Mu{\~n}oz-Guti{\'e}rrez}, {Myers}, {Newman}, {Nie}, {Percival},
  {Poppett}, {Prada}, {Raichoor}, {Ravoux}, {Ross}, {Schlafly}, {Schlegel},
  {Tan}, {Tarl{\'e}}, {Zhou}, {Zhou}, \& {Zou}}]{Chaussidon2023}
{Chaussidon}, E., {Y{\`e}che}, C., {Palanque-Delabrouille}, N., {et~al.} 2023,
  \apj, 944, 107, \dodoi{10.3847/1538-4357/acb3c2}

\bibitem[{{Chawla} {et~al.}(2011){Chawla}, {Bowyer}, {Hall}, \&
  {Kegelmeyer}}]{Chawla2011}
{Chawla}, N.~V., {Bowyer}, K.~W., {Hall}, L.~O., \& {Kegelmeyer}, W.~P. 2011,
  arXiv e-prints, arXiv:1106.1813, \dodoi{10.48550/arXiv.1106.1813}

\bibitem[{Chen {et~al.}(2004)Chen, Liaw, \& Breiman}]{Chen2004}
Chen, C., Liaw, A., \& Breiman, L. 2004, Using Random Forest to Learn
  Imbalanced Data, Tech. Rep. 666, University of California, Berkeley 110
  (1-12): 24.
\newblock
  \url{https://statistics.berkeley.edu/sites/default/files/tech-reports/666.pdf}

\bibitem[{{Curran} {et~al.}(2021){Curran}, {Moss}, \& {Perrott}}]{Curran2021}
{Curran}, S.~J., {Moss}, J.~P., \& {Perrott}, Y.~C. 2021, \mnras, 503, 2639,
  \dodoi{10.1093/mnras/stab485}

\bibitem[{{Dark Energy Survey Collaboration} {et~al.}(2016){Dark Energy Survey
  Collaboration}, {Abbott}, {Abdalla}, {Aleksi{\'c}}, {Allam}, {Amara},
  {Bacon}, {Balbinot}, {Banerji}, {Bechtol}, {Benoit-L{\'e}vy}, {Bernstein},
  {Bertin}, {Blazek}, {Bonnett}, {Bridle}, {Brooks}, {Brunner}, {Buckley-Geer},
  {Burke}, {Caminha}, {Capozzi}, {Carlsen}, {Carnero-Rosell}, {Carollo},
  {Carrasco-Kind}, {Carretero}, {Castander}, {Clerkin}, {Collett}, {Conselice},
  {Crocce}, {Cunha}, {D'Andrea}, {da Costa}, {Davis}, {Desai}, {Diehl},
  {Dietrich}, {Dodelson}, {Doel}, {Drlica-Wagner}, {Estrada}, {Etherington},
  {Evrard}, {Fabbri}, {Finley}, {Flaugher}, {Foley}, {Fosalba}, {Frieman},
  {Garc{\'\i}a-Bellido}, {Gaztanaga}, {Gerdes}, {Giannantonio}, {Goldstein},
  {Gruen}, {Gruendl}, {Guarnieri}, {Gutierrez}, {Hartley}, {Honscheid}, {Jain},
  {James}, {Jeltema}, {Jouvel}, {Kessler}, {King}, {Kirk}, {Kron}, {Kuehn},
  {Kuropatkin}, {Lahav}, {Li}, {Lima}, {Lin}, {Maia}, {Makler}, {Manera},
  {Maraston}, {Marshall}, {Martini}, {McMahon}, {Melchior}, {Merson}, {Miller},
  {Miquel}, {Mohr}, {Morice-Atkinson}, {Naidoo}, {Neilsen}, {Nichol}, {Nord},
  {Ogando}, {Ostrovski}, {Palmese}, {Papadopoulos}, {Peiris}, {Peoples},
  {Percival}, {Plazas}, {Reed}, {Refregier}, {Romer}, {Roodman}, {Ross},
  {Rozo}, {Rykoff}, {Sadeh}, {Sako}, {S{\'a}nchez}, {Sanchez}, {Santiago},
  {Scarpine}, {Schubnell}, {Sevilla-Noarbe}, {Sheldon}, {Smith}, {Smith},
  {Soares-Santos}, {Sobreira}, {Soumagnac}, {Suchyta}, {Sullivan}, {Swanson},
  {Tarle}, {Thaler}, {Thomas}, {Thomas}, {Tucker}, {Vieira}, {Vikram},
  {Walker}, {Wechsler}, {Weller}, {Wester}, {Whiteway}, {Wilcox}, {Yanny},
  {Zhang}, \& {Zuntz}}]{desCollaboration2016}
{Dark Energy Survey Collaboration}, {Abbott}, T., {Abdalla}, F.~B., {et~al.}
  2016, \mnras, 460, 1270, \dodoi{10.1093/mnras/stw641}

\bibitem[{{Davies} {et~al.}(2023{\natexlab{a}}){Davies}, {Ryan-Weber},
  {D'Odorico}, {Bosman}, {Meyer}, {Becker}, {Cupani}, {Bischetti}, {Sebastian},
  {Eilers}, {Farina}, {Wang}, {Yang}, \& {Zhu}}]{Davies2023a}
{Davies}, R.~L., {Ryan-Weber}, E., {D'Odorico}, V., {et~al.}
  2023{\natexlab{a}}, \mnras, 521, 289, \dodoi{10.1093/mnras/stac3662}

\bibitem[{{Davies} {et~al.}(2023{\natexlab{b}}){Davies}, {Ryan-Weber},
  {D'Odorico}, {Bosman}, {Meyer}, {Becker}, {Cupani}, {Keating}, {Bischetti},
  {Davies}, {Eilers}, {Farina}, {Haehnelt}, {Pallottini}, \&
  {Zhu}}]{Davies2023b}
---. 2023{\natexlab{b}}, \mnras, 521, 314, \dodoi{10.1093/mnras/stad294}

\bibitem[{{DESI Collaboration} {et~al.}(2016{\natexlab{a}}){DESI
  Collaboration}, {Aghamousa}, {Aguilar}, {Ahlen}, {Alam}, {Allen}, {Allende
  Prieto}, {Annis}, {Bailey}, {Balland}, {Ballester}, {Baltay}, {Beaufore},
  {Bebek}, {Beers}, {Bell}, {Bernal}, {Besuner}, {Beutler}, {Blake}, {Bleuler},
  {Blomqvist}, {Blum}, {Bolton}, {Briceno}, {Brooks}, {Brownstein},
  {Buckley-Geer}, {Burden}, {Burtin}, {Busca}, {Cahn}, {Cai}, {Cardiel-Sas},
  {Carlberg}, {Carton}, {Casas}, {Castander}, {Cervantes-Cota}, {Claybaugh},
  {Close}, {Coker}, {Cole}, {Comparat}, {Cooper}, {Cousinou}, {Crocce}, {Cuby},
  {Cunningham}, {Davis}, {Dawson}, {de la Macorra}, {De Vicente}, {Delubac},
  {Derwent}, {Dey}, {Dhungana}, {Ding}, {Doel}, {Duan}, {Ealet}, {Edelstein},
  {Eftekharzadeh}, {Eisenstein}, {Elliott}, {Escoffier}, {Evatt}, {Fagrelius},
  {Fan}, {Fanning}, {Farahi}, {Farihi}, {Favole}, {Feng}, {Fernandez},
  {Findlay}, {Finkbeiner}, {Fitzpatrick}, {Flaugher}, {Flender}, {Font-Ribera},
  {Forero-Romero}, {Fosalba}, {Frenk}, {Fumagalli}, {Gaensicke}, {Gallo},
  {Garcia-Bellido}, {Gaztanaga}, {Pietro Gentile Fusillo}, {Gerard},
  {Gershkovich}, {Giannantonio}, {Gillet}, {Gonzalez-de-Rivera},
  {Gonzalez-Perez}, {Gott}, {Graur}, {Gutierrez}, {Guy}, {Habib}, {Heetderks},
  {Heetderks}, {Heitmann}, {Hellwing}, {Herrera}, {Ho}, {Holland}, {Honscheid},
  {Huff}, {Hutchinson}, {Huterer}, {Hwang}, {Illa Laguna}, {Ishikawa},
  {Jacobs}, {Jeffrey}, {Jelinsky}, {Jennings}, {Jiang}, {Jimenez}, {Johnson},
  {Joyce}, {Jullo}, {Juneau}, {Kama}, {Karcher}, {Karkar}, {Kehoe}, {Kennamer},
  {Kent}, {Kilbinger}, {Kim}, {Kirkby}, {Kisner}, {Kitanidis}, {Kneib},
  {Koposov}, {Kovacs}, {Koyama}, {Kremin}, {Kron}, {Kronig}, {Kueter-Young},
  {Lacey}, {Lafever}, {Lahav}, {Lambert}, {Lampton}, {Landriau}, {Lang},
  {Lauer}, {Le Goff}, {Le Guillou}, {Le Van Suu}, {Lee}, {Lee}, {Leitner},
  {Lesser}, {Levi}, {L'Huillier}, {Li}, {Liang}, {Lin}, {Linder}, {Loebman},
  {Luki{\'c}}, {Ma}, {MacCrann}, {Magneville}, {Makarem}, {Manera}, {Manser},
  {Marshall}, {Martini}, {Massey}, {Matheson}, {McCauley}, {McDonald},
  {McGreer}, {Meisner}, {Metcalfe}, {Miller}, {Miquel}, {Moustakas}, {Myers},
  {Naik}, {Newman}, {Nichol}, {Nicola}, {Nicolati da Costa}, {Nie}, {Niz},
  {Norberg}, {Nord}, {Norman}, {Nugent}, {O'Brien}, {Oh}, {Olsen}, {Padilla},
  {Padmanabhan}, {Padmanabhan}, {Palanque-Delabrouille}, {Palmese},
  {Pappalardo}, {P{\^a}ris}, {Park}, {Patej}, {Peacock}, {Peiris}, {Peng},
  {Percival}, {Perruchot}, {Pieri}, {Pogge}, {Pollack}, {Poppett}, {Prada},
  {Prakash}, {Probst}, {Rabinowitz}, {Raichoor}, {Ree}, {Refregier}, {Regal},
  {Reid}, {Reil}, {Rezaie}, {Rockosi}, {Roe}, {Ronayette}, {Roodman}, {Ross},
  {Ross}, {Rossi}, {Rozo}, {Ruhlmann-Kleider}, {Rykoff}, {Sabiu}, {Samushia},
  {Sanchez}, {Sanchez}, {Schlegel}, {Schneider}, {Schubnell}, {Secroun},
  {Seljak}, {Seo}, {Serrano}, {Shafieloo}, {Shan}, {Sharples}, {Sholl},
  {Shourt}, {Silber}, {Silva}, {Sirk}, {Slosar}, {Smith}, {Smoot}, {Som},
  {Song}, {Sprayberry}, {Staten}, {Stefanik}, {Tarle}, {Sien Tie}, {Tinker},
  {Tojeiro}, {Valdes}, {Valenzuela}, {Valluri}, {Vargas-Magana}, {Verde},
  {Walker}, {Wang}, {Wang}, {Weaver}, {Weaverdyck}, {Wechsler}, {Weinberg},
  {White}, {Yang}, {Yeche}, {Zhang}, {Zhao}, {Zheng}, {Zhou}, {Zhou}, {Zhu},
  {Zou}, \& {Zu}}]{DESI1}
{DESI Collaboration}, {Aghamousa}, A., {Aguilar}, J., {et~al.}
  2016{\natexlab{a}}, arXiv e-prints, arXiv:1611.00036.
\newblock \doarXiv{1611.00036}

\bibitem[{{DESI Collaboration} {et~al.}(2016{\natexlab{b}}){DESI
  Collaboration}, {Aghamousa}, {Aguilar}, {Ahlen}, {Alam}, {Allen}, {Allende
  Prieto}, {Annis}, {Bailey}, {Balland}, {Ballester}, {Baltay}, {Beaufore},
  {Bebek}, {Beers}, {Bell}, {Bernal}, {Besuner}, {Beutler}, {Blake}, {Bleuler},
  {Blomqvist}, {Blum}, {Bolton}, {Briceno}, {Brooks}, {Brownstein},
  {Buckley-Geer}, {Burden}, {Burtin}, {Busca}, {Cahn}, {Cai}, {Cardiel-Sas},
  {Carlberg}, {Carton}, {Casas}, {Castander}, {Cervantes-Cota}, {Claybaugh},
  {Close}, {Coker}, {Cole}, {Comparat}, {Cooper}, {Cousinou}, {Crocce}, {Cuby},
  {Cunningham}, {Davis}, {Dawson}, {de la Macorra}, {De Vicente}, {Delubac},
  {Derwent}, {Dey}, {Dhungana}, {Ding}, {Doel}, {Duan}, {Ealet}, {Edelstein},
  {Eftekharzadeh}, {Eisenstein}, {Elliott}, {Escoffier}, {Evatt}, {Fagrelius},
  {Fan}, {Fanning}, {Farahi}, {Farihi}, {Favole}, {Feng}, {Fernandez},
  {Findlay}, {Finkbeiner}, {Fitzpatrick}, {Flaugher}, {Flender}, {Font-Ribera},
  {Forero-Romero}, {Fosalba}, {Frenk}, {Fumagalli}, {Gaensicke}, {Gallo},
  {Garcia-Bellido}, {Gaztanaga}, {Pietro Gentile Fusillo}, {Gerard},
  {Gershkovich}, {Giannantonio}, {Gillet}, {Gonzalez-de-Rivera},
  {Gonzalez-Perez}, {Gott}, {Graur}, {Gutierrez}, {Guy}, {Habib}, {Heetderks},
  {Heetderks}, {Heitmann}, {Hellwing}, {Herrera}, {Ho}, {Holland}, {Honscheid},
  {Huff}, {Hutchinson}, {Huterer}, {Hwang}, {Illa Laguna}, {Ishikawa},
  {Jacobs}, {Jeffrey}, {Jelinsky}, {Jennings}, {Jiang}, {Jimenez}, {Johnson},
  {Joyce}, {Jullo}, {Juneau}, {Kama}, {Karcher}, {Karkar}, {Kehoe}, {Kennamer},
  {Kent}, {Kilbinger}, {Kim}, {Kirkby}, {Kisner}, {Kitanidis}, {Kneib},
  {Koposov}, {Kovacs}, {Koyama}, {Kremin}, {Kron}, {Kronig}, {Kueter-Young},
  {Lacey}, {Lafever}, {Lahav}, {Lambert}, {Lampton}, {Landriau}, {Lang},
  {Lauer}, {Le Goff}, {Le Guillou}, {Le Van Suu}, {Lee}, {Lee}, {Leitner},
  {Lesser}, {Levi}, {L'Huillier}, {Li}, {Liang}, {Lin}, {Linder}, {Loebman},
  {Luki{\'c}}, {Ma}, {MacCrann}, {Magneville}, {Makarem}, {Manera}, {Manser},
  {Marshall}, {Martini}, {Massey}, {Matheson}, {McCauley}, {McDonald},
  {McGreer}, {Meisner}, {Metcalfe}, {Miller}, {Miquel}, {Moustakas}, {Myers},
  {Naik}, {Newman}, {Nichol}, {Nicola}, {Nicolati da Costa}, {Nie}, {Niz},
  {Norberg}, {Nord}, {Norman}, {Nugent}, {O'Brien}, {Oh}, {Olsen}, {Padilla},
  {Padmanabhan}, {Padmanabhan}, {Palanque-Delabrouille}, {Palmese},
  {Pappalardo}, {P{\^a}ris}, {Park}, {Patej}, {Peacock}, {Peiris}, {Peng},
  {Percival}, {Perruchot}, {Pieri}, {Pogge}, {Pollack}, {Poppett}, {Prada},
  {Prakash}, {Probst}, {Rabinowitz}, {Raichoor}, {Ree}, {Refregier}, {Regal},
  {Reid}, {Reil}, {Rezaie}, {Rockosi}, {Roe}, {Ronayette}, {Roodman}, {Ross},
  {Ross}, {Rossi}, {Rozo}, {Ruhlmann-Kleider}, {Rykoff}, {Sabiu}, {Samushia},
  {Sanchez}, {Sanchez}, {Schlegel}, {Schneider}, {Schubnell}, {Secroun},
  {Seljak}, {Seo}, {Serrano}, {Shafieloo}, {Shan}, {Sharples}, {Sholl},
  {Shourt}, {Silber}, {Silva}, {Sirk}, {Slosar}, {Smith}, {Smoot}, {Som},
  {Song}, {Sprayberry}, {Staten}, {Stefanik}, {Tarle}, {Sien Tie}, {Tinker},
  {Tojeiro}, {Valdes}, {Valenzuela}, {Valluri}, {Vargas-Magana}, {Verde},
  {Walker}, {Wang}, {Wang}, {Weaver}, {Weaverdyck}, {Wechsler}, {Weinberg},
  {White}, {Yang}, {Yeche}, {Zhang}, {Zhao}, {Zheng}, {Zhou}, {Zhou}, {Zhu},
  {Zou}, \& {Zu}}]{DESI2}
---. 2016{\natexlab{b}}, arXiv e-prints, arXiv:1611.00037.
\newblock \doarXiv{1611.00037}

\bibitem[{{DESI Collaboration} {et~al.}(2022){DESI Collaboration}, {Abareshi},
  {Aguilar}, {Ahlen}, {Alam}, {Alexander}, {Alfarsy}, {Allen}, {Allende
  Prieto}, {Alves}, \& et~al.}]{DESIover}
{DESI Collaboration}, {Abareshi}, B., {Aguilar}, J., {et~al.} 2022, \aj, 164,
  207, \dodoi{10.3847/1538-3881/ac882b}

\bibitem[{{DESI Collaboration} {et~al.}(2023){DESI Collaboration}, {Adame},
  {Aguilar}, {Ahlen}, {Alam}, {Aldering}, {Alexander}, {Alfarsy}, {Allende
  Prieto}, {Alvarez}, {Alves}, {Anand}, {Andrade-Oliveira}, {Armengaud},
  {Asorey}, {Avila}, {Aviles}, {Bailey}, {Balaguera-Antol{\'\i}nez},
  {Ballester}, {Baltay}, {Bault}, {Bautista}, {Behera}, {Beltran}, {BenZvi},
  {Beraldo e Silva}, {Bermejo-Climent}, {Berti}, {Besuner}, {Beutler},
  {Bianchi}, {Blake}, {Blum}, {Bolton}, {Brieden}, {Brodzeller}, {Brooks},
  {Brown}, {Buckley-Geer}, {Burtin}, {Cabayol-Garcia}, {Cai}, {Canning},
  {Cardiel-Sas}, {Carnero Rosell}, {Castander}, {Cervantes-Cota}, {Chabanier},
  {Chaussidon}, {Chaves-Montero}, {Chen}, {Chuang}, {Claybaugh}, {Cole},
  {Cooper}, {Cuceu}, {Davis}, {Dawson}, {de Belsunce}, {de la Cruz}, {de la
  Macorra}, {de Mattia}, {Demina}, {Demirbozan}, {DeRose}, {Dey}, {Dey},
  {Dhungana}, {Ding}, {Ding}, {Doel}, {Doshi}, {Douglass}, {Edge},
  {Eftekharzadeh}, {Eisenstein}, {Elliott}, {Escoffier}, {Fagrelius}, {Fan},
  {Fanning}, {Fawcett}, {Ferraro}, {Ereza}, {Flaugher}, {Font-Ribera},
  {Forero-S{\'a}nchez}, {Forero-Romero}, {Frenk}, {G{\"a}nsicke},
  {Garc{\'\i}a}, {Garc{\'\i}a-Bellido}, {Garcia-Quintero}, {Garrison},
  {Gil-Mar{\'\i}n}, {Golden-Marx}, {Gontcho}, {Gonzalez-Morales},
  {Gonzalez-Perez}, {Gordon}, {Graur}, {Green}, {Gruen}, {Guy}, {Hadzhiyska},
  {Hahn}, {Han}, {Hanif}, {Herrera-Alcantar}, {Honscheid}, {Hou}, {Howlett},
  {Huterer}, {Ir{\v{s}}i{\v{c}}}, {Ishak}, {Jacques}, {Jana}, {Jiang},
  {Jimenez}, {Jing}, {Joudaki}, {Jullo}, {Juneau}, {Kizhuprakkat},
  {Kara{\c{c}}ayl{\i}}, {Karim}, {Kehoe}, {Kent}, {Khederlarian}, {Kim},
  {Kirkby}, {Kisner}, {Kitaura}, {Kneib}, {Koposov}, {Kov{\'a}cs}, {Kremin},
  {Krolewski}, {L'Huillier}, {Lambert}, {Lamman}, {Lan}, {Landriau}, {Lang},
  {Lange}, {Lasker}, {Le Guillou}, {Leauthaud}, {Levi}, {Li}, {Linder},
  {Lyons}, {Magneville}, {Manera}, {Manser}, {Margala}, {Martini}, {McDonald},
  {Medina}, {Medina-Varela}, {Meisner}, {Mena-Fern{\'a}ndez}, {Meneses-Rizo},
  {Mezcua}, {Miquel}, {Montero-Camacho}, {Moon}, {Moore}, {Moustakas},
  {Mueller}, {Mundet}, {Mu{\~n}oz-Guti{\'e}rrez}, {Myers}, {Nadathur},
  {Napolitano}, {Neveux}, {Newman}, {Nie}, {Nikutta}, {Niz}, {Norberg},
  {Noriega}, {Paillas}, {Palanque-Delabrouille}, {Palmese}, {Zhiwei},
  {Parkinson}, {Penmetsa}, {Percival}, {P{\'e}rez-Fern{\'a}ndez},
  {P{\'e}rez-R{\`a}fols}, {Pieri}, {Poppett}, {Porredon}, {Pothier}, {Prada},
  {Pucha}, {Raichoor}, {Ram{\'\i}rez-P{\'e}rez}, {Ramirez-Solano},
  {Rashkovetskyi}, {Ravoux}, {Rocher}, {Rockosi}, {Ross}, {Rossi}, {Ruggeri},
  {Ruhlmann-Kleider}, {Sabiu}, {Said}, {Saintonge}, {Samushia}, {Sanchez},
  {Saulder}, {Schaan}, {Schlafly}, {Schlegel}, {Scholte}, {Schubnell}, {Seo},
  {Shafieloo}, {Sharples}, {Sheu}, {Silber}, {Sinigaglia}, {Siudek}, {Slepian},
  {Smith}, {Sprayberry}, {Stephey}, {Su{\'a}rez-P{\'e}rez}, {Sun}, {Tan},
  {Tarl{\'e}}, {Tojeiro}, {Ure{\~n}a-L{\'o}pez}, {Vaisakh}, {Valcin}, {Valdes},
  {Valluri}, {Vargas-Maga{\~n}a}, {Variu}, {Verde}, {Walther}, {Wang}, {Wang},
  {Weaver}, {Weaverdyck}, {Wechsler}, {White}, {Xie}, {Yang}, {Y{\`e}che},
  {Yu}, {Yuan}, {Zhang}, {Zhang}, {Zhao}, {Zheng}, {Zhou}, {Zhou}, {Zou},
  {Zou}, \& {Zu}}]{DESI-EDR}
{DESI Collaboration}, {Adame}, A.~G., {Aguilar}, J., {et~al.} 2023, arXiv
  e-prints, arXiv:2306.06308, \dodoi{10.48550/arXiv.2306.06308}

\bibitem[{{DESI Collaboration} {et~al.}(2024){DESI Collaboration}, {Adame},
  {Aguilar}, {Ahlen}, {Alam}, {Aldering}, {Alexander}, {Alfarsy}, {Allende
  Prieto}, {Alvarez}, \& et~al.}]{DESIvalid}
---. 2024, \aj, 167, 62, \dodoi{10.3847/1538-3881/ad0b08}

\bibitem[{{Dey} {et~al.}(2019){Dey}, {Schlegel}, {Lang}, {Blum}, {Burleigh},
  {Fan}, {Findlay}, {Finkbeiner}, {Herrera}, {Juneau}, {Landriau}, {Levi},
  {McGreer}, {Meisner}, {Myers}, {Moustakas}, {Nugent}, {Patej}, {Schlafly},
  {Walker}, {Valdes}, {Weaver}, {Y{\`e}che}, {Zou}, {Zhou}, {Abareshi},
  {Abbott}, {Abolfathi}, {Aguilera}, {Alam}, {Allen}, {Alvarez}, {Annis},
  {Ansarinejad}, {Aubert}, {Beechert}, {Bell}, {BenZvi}, {Beutler}, {Bielby},
  {Bolton}, {Brice{\~n}o}, {Buckley-Geer}, {Butler}, {Calamida}, {Carlberg},
  {Carter}, {Casas}, {Castander}, {Choi}, {Comparat}, {Cukanovaite}, {Delubac},
  {DeVries}, {Dey}, {Dhungana}, {Dickinson}, {Ding}, {Donaldson}, {Duan},
  {Duckworth}, {Eftekharzadeh}, {Eisenstein}, {Etourneau}, {Fagrelius},
  {Farihi}, {Fitzpatrick}, {Font-Ribera}, {Fulmer}, {G{\"a}nsicke},
  {Gaztanaga}, {George}, {Gerdes}, {Gontcho}, {Gorgoni}, {Green}, {Guy},
  {Harmer}, {Hernandez}, {Honscheid}, {Huang}, {James}, {Jannuzi}, {Jiang},
  {Joyce}, {Karcher}, {Karkar}, {Kehoe}, {Kneib}, {Kueter-Young}, {Lan},
  {Lauer}, {Le Guillou}, {Le Van Suu}, {Lee}, {Lesser}, {Perreault Levasseur},
  {Li}, {Mann}, {Marshall}, {Mart{\'\i}nez-V{\'a}zquez}, {Martini}, {du Mas des
  Bourboux}, {McManus}, {Meier}, {M{\'e}nard}, {Metcalfe},
  {Mu{\~n}oz-Guti{\'e}rrez}, {Najita}, {Napier}, {Narayan}, {Newman}, {Nie},
  {Nord}, {Norman}, {Olsen}, {Paat}, {Palanque-Delabrouille}, {Peng},
  {Poppett}, {Poremba}, {Prakash}, {Rabinowitz}, {Raichoor}, {Rezaie},
  {Robertson}, {Roe}, {Ross}, {Ross}, {Rudnick}, {Safonova}, {Saha},
  {S{\'a}nchez}, {Savary}, {Schweiker}, {Scott}, {Seo}, {Shan}, {Silva},
  {Slepian}, {Soto}, {Sprayberry}, {Staten}, {Stillman}, {Stupak}, {Summers},
  {Sien Tie}, {Tirado}, {Vargas-Maga{\~n}a}, {Vivas}, {Wechsler}, {Williams},
  {Yang}, {Yang}, {Yapici}, {Zaritsky}, {Zenteno}, {Zhang}, {Zhang}, {Zhou}, \&
  {Zhou}}]{Dey2019OverviewOT}
{Dey}, A., {Schlegel}, D.~J., {Lang}, D., {et~al.} 2019, \aj, 157, 168,
  \dodoi{10.3847/1538-3881/ab089d}

\bibitem[{{Drlica-Wagner} {et~al.}(2021){Drlica-Wagner}, {Carlin}, {Nidever},
  {Ferguson}, {Kuropatkin}, {Adam{\'o}w}, {Cerny}, {Choi}, {Esteves},
  {Mart{\'\i}nez-V{\'a}zquez}, {Mau}, {Miller}, {Mutlu-Pakdil}, {Neilsen},
  {Olsen}, {Pace}, {Riley}, {Sakowska}, {Sand}, {Santana-Silva}, {Tollerud},
  {Tucker}, {Vivas}, {Zaborowski}, {Zenteno}, {Abbott}, {Allam}, {Bechtol},
  {Bell}, {Bell}, {Bilaji}, {Bom}, {Carballo-Bello}, {Crnojevi{\'c}}, {Cioni},
  {Diaz-Ocampo}, {de Boer}, {Erkal}, {Gruendl}, {Hernandez-Lang}, {Hughes},
  {James}, {Johnson}, {Li}, {Mao}, {Mart{\'\i}nez-Delgado}, {Massana},
  {McNanna}, {Morgan}, {Nadler}, {No{\"e}l}, {Palmese}, {Peter}, {Rykoff},
  {S{\'a}nchez}, {Shipp}, {Simon}, {Smercina}, {Soares-Santos}, {Stringfellow},
  {Tavangar}, {van der Marel}, {Walker}, {Wechsler}, {Wu}, {Yanny},
  {Fitzpatrick}, {Huang}, {Jacques}, {Nikutta}, {Scott}, \& {Astro Data
  Lab}}]{delve}
{Drlica-Wagner}, A., {Carlin}, J.~L., {Nidever}, D.~L., {et~al.} 2021, \apjs,
  256, 2, \dodoi{10.3847/1538-4365/ac079d}

\bibitem[{Eilers {et~al.}(2018)Eilers, Davies, \& Hennawi}]{Eilers2018}
Eilers, A.-C., Davies, F.~B., \& Hennawi, J.~F. 2018, The Astrophysical
  Journal, 864, 53

\bibitem[{{Euclid Collaboration} {et~al.}(2022){Euclid Collaboration},
  {Schirmer}, {Jahnke}, {Seidel}, {Aussel}, {Bodendorf}, {Grupp}, {Hormuth},
  {Wachter}, {Appleton}, {Barbier}, {Brinchmann}, {Carrasco}, {Castander},
  {Coupon}, {De Paolis}, {Franco}, {Ganga}, {Hudelot}, {Jullo}, {Lan{\c{c}}on},
  {Nucita}, {Paltani}, {Smadja}, {Strafella}, {Venancio}, {Weiler}, {Amara},
  {Auphan}, {Auricchio}, {Balestra}, {Bender}, {Bonino}, {Branchini},
  {Brescia}, {Capobianco}, {Carbone}, {Carretero}, {Casas}, {Castellano},
  {Cavuoti}, {Cimatti}, {Cledassou}, {Congedo}, {Conselice}, {Conversi},
  {Copin}, {Corcione}, {Costille}, {Courbin}, {Da Silva}, {Degaudenzi},
  {Douspis}, {Dubath}, {Dupac}, {Dusini}, {Ealet}, {Farrens}, {Ferriol},
  {Fosalba}, {Frailis}, {Franceschi}, {Franzetti}, {Fumana}, {Garilli},
  {Gillard}, {Gillis}, {Giocoli}, {Grazian}, {Guzzo}, {Haugan}, {Hoekstra},
  {Holmes}, {Hornstrup}, {K{\"u}mmel}, {Kermiche}, {Kiessling}, {Kilbinger},
  {Kitching}, {Kohley}, {Kunz}, {Kurki-Suonio}, {Laureijs}, {Ligori}, {Lilje},
  {Lloro}, {Maciaszek}, {Maiorano}, {Mansutti}, {Marggraf}, {Markovic},
  {Marulli}, {Massey}, {Maurogordato}, {Mellier}, {Meneghetti}, {Merlin},
  {Meylan}, {Moresco}, {Moscardini}, {Munari}, {Nakajima}, {Nichol}, {Niemi},
  {Padilla}, {Pasian}, {Pedersen}, {Percival}, {Pettorino}, {Pires}, {Poncet},
  {Popa}, {Pozzetti}, {Prieto}, {Raison}, {Rhodes}, {Rix}, {Roncarelli},
  {Rossetti}, {Saglia}, {Sartoris}, {Scaramella}, {Schneider}, {Secroun},
  {Serrano}, {Sirignano}, {Sirri}, {Stanco}, {Tallada-Cresp{\'\i}}, {Taylor},
  {Teplitz}, {Tereno}, {Toledo-Moreo}, {Torradeflot}, {Trifoglio}, {Valentijn},
  {Valenziano}, {Wang}, {Weller}, {Zamorani}, {Zoubian}, {Andreon}, {Bardelli},
  {Boucaud}, {Camera}, {Farinelli}, {Graci{\'a}-Carpio}, {Maino}, {Medinaceli},
  {Mei}, {Morisset}, {Polenta}, {Renzi}, {Romelli}, {Tenti}, {Vassallo},
  {Zacchei}, {Zucca}, {Baccigalupi}, {Balaguera-Antol{\'\i}nez}, {Biviano},
  {Blanchard}, {Borgani}, {Bozzo}, {Burigana}, {Cabanac}, {Cappi}, {Carvalho},
  {Casas}, {Castignani}, {Colodro-Conde}, {Cooray}, {Courtois}, {Crocce},
  {Cuby}, {Davini}, {de la Torre}, {Di Ferdinando}, {Escartin}, {Farina},
  {Ferreira}, {Finelli}, {Fotopoulou}, {Galeotta}, {Garcia-Bellido},
  {Gaztanaga}, {George}, {Gozaliasl}, {Hook}, {Ili{\'c}}, {Kansal},
  {Kashlinsky}, {Keihanen}, {Kirkpatrick}, {Lindholm}, {Mainetti}, {Maoli},
  {Martinelli}, {Martinet}, {Maturi}, {Mauri}, {McCracken}, {Metcalf},
  {Monaco}, {Morgante}, {Nightingale}, {Patrizii}, {Peel}, {Popa}, {Porciani},
  {Potter}, {Reimberg}, {Riccio}, {S{\'a}nchez}, {Sapone}, {Scottez},
  {Sefusatti}, {Teyssier}, {Tutusaus}, {Valieri}, {Valiviita}, {Viel}, \&
  {Hildebrandt}}]{Euclid}
{Euclid Collaboration}, {Schirmer}, M., {Jahnke}, K., {et~al.} 2022, \aap, 662,
  A92, \dodoi{10.1051/0004-6361/202142897}

\bibitem[{{Fan} {et~al.}(2023){Fan}, {Ba{\~n}ados}, \&
  {Simcoe}}]{fan2023quasars}
{Fan}, X., {Ba{\~n}ados}, E., \& {Simcoe}, R.~A. 2023, \araa, 61, 373,
  \dodoi{10.1146/annurev-astro-052920-102455}

\bibitem[{Fan {et~al.}(2006{\natexlab{a}})Fan, Carilli, \&
  Keating}]{fan2006observational}
Fan, X., Carilli, C., \& Keating, B. 2006{\natexlab{a}}, Annu. Rev. Astron.
  Astrophys., 44, 415

\bibitem[{Fan {et~al.}(2006{\natexlab{b}})Fan, Strauss, Becker, White, Gunn,
  Knapp, Richards, Schneider, Brinkmann, \& Fukugita}]{fan2006constraining}
Fan, X., Strauss, M.~A., Becker, R.~H., {et~al.} 2006{\natexlab{b}}, The
  Astronomical Journal, 132, 117

\bibitem[{{Han} {et~al.}(2021){Han}, {Qiao}, {Chen}, {Zhang}, {Zhang}, \&
  {Zhao}}]{Han2021}
{Han}, B., {Qiao}, L.-N., {Chen}, J.-L., {et~al.} 2021, Research in Astronomy
  and Astrophysics, 21, 017, \dodoi{10.1088/1674-4527/21/1/17}

\bibitem[{He {et~al.}(2008)He, Bai, Garcia, \& Li}]{He2008}
He, H., Bai, Y., Garcia, E.~A., \& Li, S. 2008, in 2008 IEEE International
  Joint Conference on Neural Networks (IEEE World Congress on Computational
  Intelligence), 1322--1328, \dodoi{10.1109/IJCNN.2008.4633969}

\bibitem[{{He} \& {Li}(2022)}]{He2022}
{He}, Z., \& {Li}, N. 2022, Research in Astronomy and Astrophysics, 22, 095021,
  \dodoi{10.1088/1674-4527/ac839b}

\bibitem[{{Ivezi{\'c}} {et~al.}(2019){Ivezi{\'c}}, {Kahn}, {Tyson}, {Abel},
  {Acosta}, {Allsman}, {Alonso}, {AlSayyad}, {Anderson}, {Andrew}, {Angel},
  {Angeli}, {Ansari}, {Antilogus}, {Araujo}, {Armstrong}, {Arndt}, {Astier},
  {Aubourg}, {Auza}, {Axelrod}, {Bard}, {Barr}, {Barrau}, {Bartlett}, {Bauer},
  {Bauman}, {Baumont}, {Bechtol}, {Bechtol}, {Becker}, {Becla}, {Beldica},
  {Bellavia}, {Bianco}, {Biswas}, {Blanc}, {Blazek}, {Blandford}, {Bloom},
  {Bogart}, {Bond}, {Booth}, {Borgland}, {Borne}, {Bosch}, {Boutigny},
  {Brackett}, {Bradshaw}, {Brandt}, {Brown}, {Bullock}, {Burchat}, {Burke},
  {Cagnoli}, {Calabrese}, {Callahan}, {Callen}, {Carlin}, {Carlson},
  {Chandrasekharan}, {Charles-Emerson}, {Chesley}, {Cheu}, {Chiang}, {Chiang},
  {Chirino}, {Chow}, {Ciardi}, {Claver}, {Cohen-Tanugi}, {Cockrum}, {Coles},
  {Connolly}, {Cook}, {Cooray}, {Covey}, {Cribbs}, {Cui}, {Cutri}, {Daly},
  {Daniel}, {Daruich}, {Daubard}, {Daues}, {Dawson}, {Delgado}, {Dellapenna},
  {de Peyster}, {de Val-Borro}, {Digel}, {Doherty}, {Dubois},
  {Dubois-Felsmann}, {Durech}, {Economou}, {Eifler}, {Eracleous}, {Emmons},
  {Fausti Neto}, {Ferguson}, {Figueroa}, {Fisher-Levine}, {Focke}, {Foss},
  {Frank}, {Freemon}, {Gangler}, {Gawiser}, {Geary}, {Gee}, {Geha}, {Gessner},
  {Gibson}, {Gilmore}, {Glanzman}, {Glick}, {Goldina}, {Goldstein}, {Goodenow},
  {Graham}, {Gressler}, {Gris}, {Guy}, {Guyonnet}, {Haller}, {Harris},
  {Hascall}, {Haupt}, {Hernandez}, {Herrmann}, {Hileman}, {Hoblitt}, {Hodgson},
  {Hogan}, {Howard}, {Huang}, {Huffer}, {Ingraham}, {Innes}, {Jacoby}, {Jain},
  {Jammes}, {Jee}, {Jenness}, {Jernigan}, {Jevremovi{\'c}}, {Johns}, {Johnson},
  {Johnson}, {Jones}, {Juramy-Gilles}, {Juri{\'c}}, {Kalirai}, {Kallivayalil},
  {Kalmbach}, {Kantor}, {Karst}, {Kasliwal}, {Kelly}, {Kessler}, {Kinnison},
  {Kirkby}, {Knox}, {Kotov}, {Krabbendam}, {Krughoff}, {Kub{\'a}nek},
  {Kuczewski}, {Kulkarni}, {Ku}, {Kurita}, {Lage}, {Lambert}, {Lange},
  {Langton}, {Le Guillou}, {Levine}, {Liang}, {Lim}, {Lintott}, {Long},
  {Lopez}, {Lotz}, {Lupton}, {Lust}, {MacArthur}, {Mahabal}, {Mandelbaum},
  {Markiewicz}, {Marsh}, {Marshall}, {Marshall}, {May}, {McKercher}, {McQueen},
  {Meyers}, {Migliore}, {Miller}, {Mills}, {Miraval}, {Moeyens}, {Moolekamp},
  {Monet}, {Moniez}, {Monkewitz}, {Montgomery}, {Morrison}, {Mueller},
  {Muller}, {Mu{\~n}oz Arancibia}, {Neill}, {Newbry}, {Nief}, {Nomerotski},
  {Nordby}, {O'Connor}, {Oliver}, {Olivier}, {Olsen}, {O'Mullane}, {Ortiz},
  {Osier}, {Owen}, {Pain}, {Palecek}, {Parejko}, {Parsons}, {Pease},
  {Peterson}, {Peterson}, {Petravick}, {Libby Petrick}, {Petry},
  {Pierfederici}, {Pietrowicz}, {Pike}, {Pinto}, {Plante}, {Plate}, {Plutchak},
  {Price}, {Prouza}, {Radeka}, {Rajagopal}, {Rasmussen}, {Regnault}, {Reil},
  {Reiss}, {Reuter}, {Ridgway}, {Riot}, {Ritz}, {Robinson}, {Roby}, {Roodman},
  {Rosing}, {Roucelle}, {Rumore}, {Russo}, {Saha}, {Sassolas}, {Schalk},
  {Schellart}, {Schindler}, {Schmidt}, {Schneider}, {Schneider}, {Schoening},
  {Schumacher}, {Schwamb}, {Sebag}, {Selvy}, {Sembroski}, {Seppala}, {Serio},
  {Serrano}, {Shaw}, {Shipsey}, {Sick}, {Silvestri}, {Slater}, {Smith},
  {Smith}, {Sobhani}, {Soldahl}, {Storrie-Lombardi}, {Stover}, {Strauss},
  {Street}, {Stubbs}, {Sullivan}, {Sweeney}, {Swinbank}, {Szalay}, {Takacs},
  {Tether}, {Thaler}, {Thayer}, {Thomas}, {Thornton}, {Thukral}, {Tice},
  {Trilling}, {Turri}, {Van Berg}, {Vanden Berk}, {Vetter}, {Virieux},
  {Vucina}, {Wahl}, {Walkowicz}, {Walsh}, {Walter}, {Wang}, {Wang}, {Warner},
  {Wiecha}, {Willman}, {Winters}, {Wittman}, {Wolff}, {Wood-Vasey}, {Wu},
  {Xin}, {Yoachim}, \& {Zhan}}]{LSST2019}
{Ivezi{\'c}}, {\v{Z}}., {Kahn}, S.~M., {Tyson}, J.~A., {et~al.} 2019, \apj,
  873, 111, \dodoi{10.3847/1538-4357/ab042c}

\bibitem[{Jiang {et~al.}(2016)Jiang, McGreer, Fan, Strauss, Ba{\~n}ados,
  Becker, Bian, Farnsworth, Shen, Wang, {et~al.}}]{jiang2016final}
Jiang, L., McGreer, I.~D., Fan, X., {et~al.} 2016, The Astrophysical Journal,
  833, 222

\bibitem[{{Jin} {et~al.}(2019){Jin}, {Zhang}, {Zhang}, {Zhao}, {Wu}, \&
  {Fan}}]{Jin2019}
{Jin}, X., {Zhang}, Y., {Zhang}, J., {et~al.} 2019, \mnras, 485, 4539,
  \dodoi{10.1093/mnras/stz680}

\bibitem[{{Khramtsov} {et~al.}(2019{\natexlab{a}}){Khramtsov}, {Sergeyev},
  {Spiniello}, {Tortora}, {Napolitano}, {Agnello}, {Getman}, {de Jong},
  {Kuijken}, {Radovich}, {Shan}, \& {Shulga}}]{Khramtsov2019}
{Khramtsov}, V., {Sergeyev}, A., {Spiniello}, C., {et~al.} 2019{\natexlab{a}},
  \aap, 632, A56, \dodoi{10.1051/0004-6361/201936006}

\bibitem[{{Khramtsov} {et~al.}(2019{\natexlab{b}}){Khramtsov}, {Sergeyev},
  {Spiniello}, {Tortora}, {Napolitano}, {Agnello}, {Getman}, {de Jong},
  {Kuijken}, {Radovich}, {Shan}, \& {Shulga}}]{Khramtsov2019b}
---. 2019{\natexlab{b}}, \aap, 632, A56, \dodoi{10.1051/0004-6361/201936006}

\bibitem[{Kubat \& Matwin(1997)}]{Kubat1997}
Kubat, M., \& Matwin, S. 1997, in Proc. 14th International Conference on
  Machine Learning (Morgan Kaufmann), 179--186.
\newblock \url{http://citeseer.ist.psu.edu/297143.html}

\bibitem[{Li {et~al.}(2022)Li, Zhang, Cui, Fan, Zhao, Wu, Zhang, Han, Xu, Tao,
  {et~al.}}]{Li2022}
Li, C., Zhang, Y., Cui, C., {et~al.} 2022, Monthly Notices of the Royal
  Astronomical Society, 509, 2289

\bibitem[{{LSST Dark Energy Science Collaboration}(2012)}]{LSST}
{LSST Dark Energy Science Collaboration}. 2012, arXiv e-prints,
  arXiv:1211.0310, \dodoi{10.48550/arXiv.1211.0310}

\bibitem[{{Lyke} {et~al.}(2020){Lyke}, {Higley}, {McLane}, {Schurhammer},
  {Myers}, {Ross}, {Dawson}, {Chabanier}, {Martini}, {Busca}, {Mas des
  Bourboux}, {Salvato}, {Streblyanska}, {Zarrouk}, {Burtin}, {Anderson},
  {Bautista}, {Bizyaev}, {Brandt}, {Brinkmann}, {Brownstein}, {Comparat},
  {Green}, {de la Macorra}, {Mu{\~n}oz Guti{\'e}rrez}, {Hou}, {Newman},
  {Palanque-Delabrouille}, {P{\^a}ris}, {Percival}, {Petitjean}, {Rich},
  {Rossi}, {Schneider}, {Smith}, {Vivek}, \& {Weaver}}]{Lyke2020}
{Lyke}, B.~W., {Higley}, A.~N., {McLane}, J.~N., {et~al.} 2020, \apjs, 250, 8,
  \dodoi{10.3847/1538-4365/aba623}

\bibitem[{{Magnier} {et~al.}(2020){Magnier}, {Schlafly}, {Finkbeiner}, {Tonry},
  {Goldman}, {R{\"o}ser}, {Schilbach}, {Casertano}, {Chambers}, {Flewelling},
  {Huber}, {Price}, {Sweeney}, {Waters}, {Denneau}, {Draper}, {Hodapp},
  {Jedicke}, {Kaiser}, {Kudritzki}, {Metcalfe}, {Stubbs}, \&
  {Wainscoat}}]{Magnier2020}
{Magnier}, E.~A., {Schlafly}, E.~F., {Finkbeiner}, D.~P., {et~al.} 2020, \apjs,
  251, 6, \dodoi{10.3847/1538-4365/abb82a}

\bibitem[{{Mainzer} {et~al.}(2011){Mainzer}, {Bauer}, {Grav}, {Masiero},
  {Cutri}, {Dailey}, {Eisenhardt}, {McMillan}, {Wright}, {Walker}, {Jedicke},
  {Spahr}, {Tholen}, {Alles}, {Beck}, {Brandenburg}, {Conrow}, {Evans},
  {Fowler}, {Jarrett}, {Marsh}, {Masci}, {McCallon}, {Wheelock}, {Wittman},
  {Wyatt}, {DeBaun}, {Elliott}, {Elsbury}, {Gautier}, {Gomillion}, {Leisawitz},
  {Maleszewski}, {Micheli}, \& {Wilkins}}]{Mainzer2011}
{Mainzer}, A., {Bauer}, J., {Grav}, T., {et~al.} 2011, \apj, 731, 53,
  \dodoi{10.1088/0004-637X/731/1/53}

\bibitem[{Matsuoka {et~al.}(2019{\natexlab{a}})Matsuoka, Iwasawa, Onoue,
  Kashikawa, Strauss, Lee, Imanishi, Nagao, Akiyama, Asami,
  {et~al.}}]{matsuoka2019subaru}
Matsuoka, Y., Iwasawa, K., Onoue, M., {et~al.} 2019{\natexlab{a}}, The
  Astrophysical Journal, 883, 183

\bibitem[{Matsuoka {et~al.}(2019{\natexlab{b}})Matsuoka, Onoue, Kashikawa,
  Strauss, Iwasawa, Lee, Imanishi, Nagao, Akiyama, Asami,
  {et~al.}}]{matsuoka2019discovery}
Matsuoka, Y., Onoue, M., Kashikawa, N., {et~al.} 2019{\natexlab{b}}, The
  Astrophysical Journal Letters, 872, L2

\bibitem[{Mazzucchelli {et~al.}(2017)Mazzucchelli, Ba{\~n}ados, Venemans,
  Decarli, Farina, Walter, Eilers, Rix, Simcoe, Stern,
  {et~al.}}]{mazzucchelli2017physical}
Mazzucchelli, C., Ba{\~n}ados, E., Venemans, B., {et~al.} 2017, The
  Astrophysical Journal, 849, 91

\bibitem[{McGreer {et~al.}(2018)McGreer, Fan, Jiang, \& Cai}]{mcgreer2018faint}
McGreer, I.~D., Fan, X., Jiang, L., \& Cai, Z. 2018, The Astronomical Journal,
  155, 131

\bibitem[{Mortlock {et~al.}(2011)Mortlock, Warren, Venemans, Patel, Hewett,
  McMahon, Simpson, Theuns, Gonz{\'a}les-Solares, Adamson,
  {et~al.}}]{mortlock2011luminous}
Mortlock, D.~J., Warren, S.~J., Venemans, B.~P., {et~al.} 2011, Nature, 474,
  616

\bibitem[{Nakoneczny {et~al.}(2019)Nakoneczny, Bilicki, Solarz, Pollo, Maddox,
  Spiniello, Brescia, \& Napolitano}]{nakoneczny2019catalog}
Nakoneczny, S., Bilicki, M., Solarz, A., {et~al.} 2019, Astronomy \&
  Astrophysics, 624, A13

\bibitem[{{Nakoneczny} {et~al.}(2021){Nakoneczny}, {Bilicki}, {Pollo},
  {Asgari}, {Dvornik}, {Erben}, {Giblin}, {Heymans}, {Hildebrandt},
  {Kannawadi}, {Kuijken}, {Napolitano}, \& {Valentijn}}]{Nakoneczny2021}
{Nakoneczny}, S.~J., {Bilicki}, M., {Pollo}, A., {et~al.} 2021, \aap, 649, A81,
  \dodoi{10.1051/0004-6361/202039684}

\bibitem[{Pentericci {et~al.}(2002)Pentericci, Fan, Rix, Strauss, Narayanan,
  Richards, Schneider, Krolik, Heckman, Brinkmann, Lamb, \&
  Szokoly}]{Pentericci2002}
Pentericci, L., Fan, X., Rix, H., {et~al.} 2002, Astronomical Journal, 123,
  2151, \dodoi{10.1086/340077}

\bibitem[{{Peters} {et~al.}(2015){Peters}, {Richards}, {Myers}, {Strauss},
  {Schmidt}, {Ivezi{\'c}}, {Ross}, {MacLeod}, \& {Riegel}}]{Peters2015}
{Peters}, C.~M., {Richards}, G.~T., {Myers}, A.~D., {et~al.} 2015, \apj, 811,
  95, \dodoi{10.1088/0004-637X/811/2/95}

\bibitem[{{Reed} {et~al.}(2017){Reed}, {McMahon}, {Martini}, {Banerji},
  {Auger}, {Hewett}, {Koposov}, {Gibbons}, {Gonzalez-Solares}, {Ostrovski},
  {Tie}, {Abdalla}, {Allam}, {Benoit-L{\'e}vy}, {Bertin}, {Brooks},
  {Buckley-Geer}, {Burke}, {Carnero Rosell}, {Carrasco Kind}, {Carretero}, {da
  Costa}, {DePoy}, {Desai}, {Diehl}, {Doel}, {Evrard}, {Finley}, {Flaugher},
  {Fosalba}, {Frieman}, {Garc{\'\i}a-Bellido}, {Gaztanaga}, {Goldstein},
  {Gruen}, {Gruendl}, {Gutierrez}, {James}, {Kuehn}, {Kuropatkin}, {Lahav},
  {Lima}, {Maia}, {Marshall}, {Melchior}, {Miller}, {Miquel}, {Nord}, {Ogando},
  {Plazas}, {Romer}, {Sanchez}, {Scarpine}, {Schubnell}, {Sevilla-Noarbe},
  {Smith}, {Sobreira}, {Suchyta}, {Swanson}, {Tarle}, {Tucker}, {Walker}, \&
  {Wester}}]{Reed2017}
{Reed}, S.~L., {McMahon}, R.~G., {Martini}, P., {et~al.} 2017, \mnras, 468,
  4702, \dodoi{10.1093/mnras/stx728}

\bibitem[{Reed {et~al.}(2019)Reed, Banerji, Becker, Hewett, Martini, McMahon,
  Pons, Rauch, Abbott, Allam, {et~al.}}]{reed2019three}
Reed, S.~L., Banerji, M., Becker, G., {et~al.} 2019, Monthly Notices of the
  Royal Astronomical Society, 487, 1874

\bibitem[{{Richards} {et~al.}(2009){Richards}, {Myers}, {Gray}, {Riegel},
  {Nichol}, {Brunner}, {Szalay}, {Schneider}, \& {Anderson}}]{Richards2009}
{Richards}, G.~T., {Myers}, A.~D., {Gray}, A.~G., {et~al.} 2009, \apjs, 180,
  67, \dodoi{10.1088/0067-0049/180/1/67}

\bibitem[{{Richards} {et~al.}(2015){Richards}, {Myers}, {Peters}, {Krawczyk},
  {Chase}, {Ross}, {Fan}, {Jiang}, {Lacy}, {McGreer}, {Trump}, \&
  {Riegel}}]{Richards2015}
{Richards}, G.~T., {Myers}, A.~D., {Peters}, C.~M., {et~al.} 2015, \apjs, 219,
  39, \dodoi{10.1088/0067-0049/219/2/39}

\bibitem[{{Ross} \& {Cross}(2020)}]{Ross&Cross2020}
{Ross}, N.~P., \& {Cross}, N. J.~G. 2020, \mnras, 494, 789,
  \dodoi{10.1093/mnras/staa544}

\bibitem[{Schindler {et~al.}(2017)Schindler, Fan, McGreer, Yang, Wu, Jiang, \&
  Green}]{schindler2017extremely}
Schindler, J.-T., Fan, X., McGreer, I.~D., {et~al.} 2017, The Astrophysical
  Journal, 851, 13

\bibitem[{{Spergel} {et~al.}(2015){Spergel}, {Gehrels}, {Baltay}, {Bennett},
  {Breckinridge}, {Donahue}, {Dressler}, {Gaudi}, {Greene}, {Guyon}, {Hirata},
  {Kalirai}, {Kasdin}, {Macintosh}, {Moos}, {Perlmutter}, {Postman},
  {Rauscher}, {Rhodes}, {Wang}, {Weinberg}, {Benford}, {Hudson}, {Jeong},
  {Mellier}, {Traub}, {Yamada}, {Capak}, {Colbert}, {Masters}, {Penny},
  {Savransky}, {Stern}, {Zimmerman}, {Barry}, {Bartusek}, {Carpenter}, {Cheng},
  {Content}, {Dekens}, {Demers}, {Grady}, {Jackson}, {Kuan}, {Kruk}, {Melton},
  {Nemati}, {Parvin}, {Poberezhskiy}, {Peddie}, {Ruffa}, {Wallace}, {Whipple},
  {Wollack}, \& {Zhao}}]{RST2015}
{Spergel}, D., {Gehrels}, N., {Baltay}, C., {et~al.} 2015, arXiv e-prints,
  arXiv:1503.03757, \dodoi{10.48550/arXiv.1503.03757}

\bibitem[{Venemans {et~al.}(2015)Venemans, Ba{\~n}ados, Decarli, Farina,
  Walter, Chambers, Fan, Rix, Schlafly, McMahon,
  {et~al.}}]{venemans2015identification}
Venemans, B., Ba{\~n}ados, E., Decarli, R., {et~al.} 2015, The Astrophysical
  Journal Letters, 801, L11

\bibitem[{Volonteri(2012)}]{volonteri2012formation}
Volonteri, M. 2012, Science, 337, 544

\bibitem[{Wang {et~al.}(2016)Wang, Wu, Fan, Yang, Yi, Bian, McGreer, Yang, Ai,
  Dong, {et~al.}}]{wang2016survey}
Wang, F., Wu, X.-B., Fan, X., {et~al.} 2016, The Astrophysical Journal, 819, 24

\bibitem[{Wang {et~al.}(2018)Wang, Yang, Fan, Yue, Wu, Schindler, Bian, Li,
  Farina, Ba{\~n}ados, {et~al.}}]{wang2018discovery}
Wang, F., Yang, J., Fan, X., {et~al.} 2018, The Astrophysical Journal Letters,
  869, L9

\bibitem[{Wang {et~al.}(2019)Wang, Yang, Fan, Wu, Yue, Li, Bian, Jiang,
  Ba{\~n}ados, Schindler, {et~al.}}]{wang2019exploring}
---. 2019, The Astrophysical Journal, 884, 30

\bibitem[{Warren {et~al.}(1987)Warren, Hewett, Irwin, McMahon, Bridgeland,
  Bunclark, \& Kibblewhite}]{warren1987first}
Warren, S., Hewett, P., Irwin, M., {et~al.} 1987, Nature, 325, 131

\bibitem[{{Wenzl} {et~al.}(2021){Wenzl}, {Schindler}, {Fan}, {Andika},
  {Ba{\~n}ados}, {Decarli}, {Jahnke}, {Mazzucchelli}, {Onoue}, {Venemans},
  {Walter}, \& {Yang}}]{Wenzl2021}
{Wenzl}, L., {Schindler}, J.-T., {Fan}, X., {et~al.} 2021, \aj, 162, 72,
  \dodoi{10.3847/1538-3881/ac0254}

\bibitem[{{Williams} {et~al.}(2004){Williams}, {Olszewski}, {Lesser}, \&
  {Burge}}]{Williams2004}
{Williams}, G.~G., {Olszewski}, E., {Lesser}, M.~P., \& {Burge}, J.~H. 2004, in
  Society of Photo-Optical Instrumentation Engineers (SPIE) Conference Series,
  Vol. 5492, Ground-based Instrumentation for Astronomy, ed. A.~F.~M.
  {Moorwood} \& M.~{Iye}, 787--798, \dodoi{10.1117/12.552189}

\bibitem[{Willott {et~al.}(2010)Willott, Delorme, Reyl{\'e}, Albert, Bergeron,
  Crampton, Delfosse, Forveille, Hutchings, McLure,
  {et~al.}}]{willott2010canada}
Willott, C.~J., Delorme, P., Reyl{\'e}, C., {et~al.} 2010, The Astronomical
  Journal, 139, 906

\bibitem[{{Wright} {et~al.}(2010){Wright}, {Eisenhardt}, {Mainzer}, {Ressler},
  {Cutri}, {Jarrett}, {Kirkpatrick}, {Padgett}, {McMillan}, {Skrutskie},
  {Stanford}, {Cohen}, {Walker}, {Mather}, {Leisawitz}, {Gautier}, {McLean},
  {Benford}, {Lonsdale}, {Blain}, {Mendez}, {Irace}, {Duval}, {Liu}, {Royer},
  {Heinrichsen}, {Howard}, {Shannon}, {Kendall}, {Walsh}, {Larsen}, {Cardon},
  {Schick}, {Schwalm}, {Abid}, {Fabinsky}, {Naes}, \& {Tsai}}]{Wright2010}
{Wright}, E.~L., {Eisenhardt}, P. R.~M., {Mainzer}, A.~K., {et~al.} 2010, \aj,
  140, 1868, \dodoi{10.1088/0004-6256/140/6/1868}

\bibitem[{Wu {et~al.}(2015)Wu, Wang, Fan, Yi, Zuo, Bian, Jiang, McGreer, Wang,
  Yang, {et~al.}}]{wu2015ultraluminous}
Wu, X.-B., Wang, F., Fan, X., {et~al.} 2015, Nature, 518, 512

\bibitem[{Yang {et~al.}(2017)Yang, Fan, Wu, Wang, Bian, Yang, McGreer, Yi,
  Jiang, Green, {et~al.}}]{yang2017discovery}
Yang, J., Fan, X., Wu, X.-B., {et~al.} 2017, The Astronomical Journal, 153, 184

\bibitem[{Yang {et~al.}(2019)Yang, Wang, Fan, Wu, Bian, Banados, Yue,
  Schindler, Yang, Jiang, {et~al.}}]{yang2019filling}
Yang, J., Wang, F., Fan, X., {et~al.} 2019, The Astrophysical Journal, 871, 199

\bibitem[{{Yang} {et~al.}(2019){Yang}, {Wang}, {Fan}, {Yue}, {Wu}, {Li},
  {Bian}, {Jiang}, {Ba{\~n}ados}, \& {Beletsky}}]{yang2019exploring}
{Yang}, J., {Wang}, F., {Fan}, X., {et~al.} 2019, \aj, 157, 236,
  \dodoi{10.3847/1538-3881/ab1be1}

\bibitem[{{Yang} {et~al.}(2020){Yang}, {Wang}, {Fan}, {Hennawi}, {Davies},
  {Yue}, {Eilers}, {Farina}, {Wu}, {Bian}, {Pacucci}, \&
  {Lee}}]{yang2020measurements}
---. 2020, \apj, 904, 26, \dodoi{10.3847/1538-4357/abbc1b}

\bibitem[{{Yang} {et~al.}(2021){Yang}, {Wang}, {Fan}, {Barth}, {Hennawi},
  {Nanni}, {Bian}, {Davies}, {Farina}, {Schindler}, {Ba{\~n}ados}, {Decarli},
  {Eilers}, {Green}, {Guo}, {Jiang}, {Li}, {Venemans}, {Walter}, {Wu}, \&
  {Yue}}]{yang2021probing}
---. 2021, \apj, 923, 262, \dodoi{10.3847/1538-4357/ac2b32}

\bibitem[{{Yang} {et~al.}(2023){Yang}, {Fan}, {Gupta}, {Myers},
  {Palanque-Delabrouille}, {Wang}, {Y{\`e}che}, {Aguilar}, {Ahlen},
  {Alexander}, {Brooks}, {Dawson}, {de la Macorra}, {Dey}, {Dhungana},
  {Fanning}, {Font-Ribera}, {Gontcho}, {Guy}, {Honscheid}, {Juneau}, {Kisner},
  {Kremin}, {Le Guillou}, {Levi}, {Magneville}, {Martini}, {Meisner}, {Miquel},
  {Moustakas}, {Nie}, {Percival}, {Poppett}, {Prada}, {Schlafly}, {Tarl{\'e}},
  {Vargas Magana}, {Weaver}, {Wechsler}, {Zhou}, {Zhou}, \&
  {Zou}}]{yang2023desi}
{Yang}, J., {Fan}, X., {Gupta}, A., {et~al.} 2023, \apjs, 269, 27,
  \dodoi{10.3847/1538-4365/acf99b}

\bibitem[{Y{\`e}che {et~al.}(2020)Y{\`e}che, Palanque-Delabrouille, Claveau,
  Brooks, Chaussidon, Davis, Dawson, Dey, Duan, Eftekharzadeh,
  {et~al.}}]{yeche2020preliminary}
Y{\`e}che, C., Palanque-Delabrouille, N., Claveau, C.-A., {et~al.} 2020,
  Research Notes of the AAS, 4, 179

\bibitem[{{York} {et~al.}(2000){York}, {Adelman}, {Anderson}, {Anderson},
  {Annis}, {Bahcall}, {Bakken}, {Barkhouser}, {Bastian}, {Berman}, {Boroski},
  {Bracker}, {Briegel}, {Briggs}, {Brinkmann}, {Brunner}, {Burles}, {Carey},
  {Carr}, {Castander}, {Chen}, {Colestock}, {Connolly}, {Crocker}, {Csabai},
  {Czarapata}, {Davis}, {Doi}, {Dombeck}, {Eisenstein}, {Ellman}, {Elms},
  {Evans}, {Fan}, {Federwitz}, {Fiscelli}, {Friedman}, {Frieman}, {Fukugita},
  {Gillespie}, {Gunn}, {Gurbani}, {de Haas}, {Haldeman}, {Harris}, {Hayes},
  {Heckman}, {Hennessy}, {Hindsley}, {Holm}, {Holmgren}, {Huang}, {Hull},
  {Husby}, {Ichikawa}, {Ichikawa}, {Ivezi{\'c}}, {Kent}, {Kim}, {Kinney},
  {Klaene}, {Kleinman}, {Kleinman}, {Knapp}, {Korienek}, {Kron}, {Kunszt},
  {Lamb}, {Lee}, {Leger}, {Limmongkol}, {Lindenmeyer}, {Long}, {Loomis},
  {Loveday}, {Lucinio}, {Lupton}, {MacKinnon}, {Mannery}, {Mantsch}, {Margon},
  {McGehee}, {McKay}, {Meiksin}, {Merelli}, {Monet}, {Munn}, {Narayanan},
  {Nash}, {Neilsen}, {Neswold}, {Newberg}, {Nichol}, {Nicinski}, {Nonino},
  {Okada}, {Okamura}, {Ostriker}, {Owen}, {Pauls}, {Peoples}, {Peterson},
  {Petravick}, {Pier}, {Pope}, {Pordes}, {Prosapio}, {Rechenmacher}, {Quinn},
  {Richards}, {Richmond}, {Rivetta}, {Rockosi}, {Ruthmansdorfer}, {Sandford},
  {Schlegel}, {Schneider}, {Sekiguchi}, {Sergey}, {Shimasaku}, {Siegmund},
  {Smee}, {Smith}, {Snedden}, {Stone}, {Stoughton}, {Strauss}, {Stubbs},
  {SubbaRao}, {Szalay}, {Szapudi}, {Szokoly}, {Thakar}, {Tremonti}, {Tucker},
  {Uomoto}, {Vanden Berk}, {Vogeley}, {Waddell}, {Wang}, {Watanabe},
  {Weinberg}, {Yanny}, {Yasuda}, \& {SDSS Collaboration}}]{York2000}
{York}, D.~G., {Adelman}, J., {Anderson}, John~E., J., {et~al.} 2000, \aj, 120,
  1579, \dodoi{10.1086/301513}

\bibitem[{{Zeraatgari} {et~al.}(2024){Zeraatgari}, {Hafezianzadeh}, {Zhang},
  {Mei}, {Ayubinia}, {Mosallanezhad}, \& {Zhang}}]{Zeraatgari2024}
{Zeraatgari}, F.~Z., {Hafezianzadeh}, F., {Zhang}, Y., {et~al.} 2024, \mnras,
  527, 4677, \dodoi{10.1093/mnras/stad3436}

\bibitem[{{Zhan}(2011)}]{Zhan2011}
{Zhan}, H. 2011, Scientia Sinica Physica, Mechanica \& Astronomica, 41, 1441,
  \dodoi{10.1360/132011-961}

\bibitem[{{Zhang} {et~al.}(2023){Zhang}, {Behroozi}, {Volonteri}, {Silk},
  {Fan}, {Hopkins}, {Yang}, \& {Aird}}]{Zhang2023}
{Zhang}, H., {Behroozi}, P., {Volonteri}, M., {et~al.} 2023, \mnras, 518, 2123,
  \dodoi{10.1093/mnras/stac2633}

\bibitem[{{Zhang} {et~al.}(2024){Zhang}, {Behroozi}, {Volonteri}, {Silk},
  {Fan}, {Aird}, {Yang}, {Wang}, {Tee}, \& {Hopkins}}]{Zhang2024}
---. 2024, \mnras, 531, 4974, \dodoi{10.1093/mnras/stae1447}

\bibitem[{Zhang \& Mani(2003)}]{Mani2003}
Zhang, J., \& Mani, I. 2003, in {Proceedings of the ICML'2003 Workshop on
  Learning from Imbalanced Datasets}

\bibitem[{Zhang {et~al.}(2013)Zhang, Ma, Peng, Zhao, \& Wu}]{Zhang2013}
Zhang, Y., Ma, H., Peng, N., Zhao, Y., \& Wu, X.-b. 2013, The Astronomical
  Journal, 146, 22

\bibitem[{{Zhang} {et~al.}(2019){Zhang}, {Zhang}, {Jin}, \& {Zhao}}]{Zhang2019}
{Zhang}, Y.-X., {Zhang}, J.-Y., {Jin}, X., \& {Zhao}, Y.-H. 2019, Research in
  Astronomy and Astrophysics, 19, 175, \dodoi{10.1088/1674-4527/19/12/175}

\bibitem[{{Zou} {et~al.}(2021){Zou}, {Jiang}, {Shen}, {Wu}, {Ba{\~n}ados},
  {Fan}, {Ho}, {Riechers}, {Venemans}, {Vestergaard}, {Walter}, {Wang},
  {Willott}, {Joshi}, {Wu}, \& {Yang}}]{Zou2021}
{Zou}, S., {Jiang}, L., {Shen}, Y., {et~al.} 2021, \apj, 906, 32,
  \dodoi{10.3847/1538-4357/abc6ff}

\bibitem[{{Zou} {et~al.}(2024){Zou}, {Cai}, {Wang}, {Fan}, {Champagne},
  {Hennawi}, {Schindler}, {Farina}, {Yang}, {Inayoshi}, {Banados}, {Bosman},
  {Li}, {Lin}, {Wu}, {Sun}, {Guo}, {Kulkarni}, {Habouzit}, {Charlot},
  {Chevallard}, {Connor}, {Eilers}, {Jiang}, {Jin}, {Kakiichi}, {Li}, {Meyer},
  {Walter}, \& {Zhang}}]{Zou2024}
{Zou}, S., {Cai}, Z., {Wang}, F., {et~al.} 2024, arXiv e-prints,
  arXiv:2402.00113, \dodoi{10.48550/arXiv.2402.00113}

\end{thebibliography}

\end{document}